\documentclass[iop]{emulateapj}

\usepackage{amsmath,natbib,graphicx}
\usepackage{epsf}
\usepackage{epstopdf}
\usepackage{epsfig}
\usepackage{color}
\DeclareGraphicsExtensions{.jpg,.pdf,.png,.eps,.ps}
\usepackage{scrextend}
\usepackage{hyperref}

\newcommand{\I}{~{\sc i}}
\newcommand{\II}{~{\sc ii}}
\newcommand{\III}{~{\sc iii}}

\shorttitle{}

\begin{document}

\title{Spatially resolved BPT mapping of nearby Seyfert 2 galaxies}

\author{Jingzhe Ma$^{1}$, W. Peter Maksym$^{1}$, G. Fabbiano$^{1}$, Martin Elvis$^{1}$, Thaisa Storchi-Bergmann$^{2}$, Margarita Karovska$^{1}$, Junfeng Wang$^{3}$, and Andrea Travascio$^{1}$}

\altaffiltext{1}{Center for Astrophysics $|$ Harvard \& Smithsonian, 60 Garden St, Cambridge, MA 02138, USA; \href{jingzhe.ma@cfa.harvard.edu}{jingzhe.ma@cfa.harvard.edu}}
\altaffiltext{2}{Departamento de Astronomia, Universidade Federal do Rio Grande do Sul, IF, CP 15051, 91501-970 Porto Alegre, RS, Brazil}
\altaffiltext{3}{Department of Astronomy, Physics Building, Xiamen University, Xiamen, Fujian, 361005, China}


\begin{abstract}

We present spatially resolved BPT mapping of the extended narrow line regions (ENLRs) of seven nearby Seyfert 2 galaxies, using {\it HST} narrow band filter imaging. We construct the BPT diagrams using $\leq$ 0.1$\arcsec$ resolution emission line images of [O\III]$\lambda$5007, H$\alpha$, [S\II]$\lambda$$\lambda$6717,6731, and H$\beta$. By mapping these diagnostic lines according to the BPT classification, we dissect the ENLR into Seyfert, LINER, and star-forming regions. The nucleus and ionization cones are dominated by Seyfert-type emission, which can be interpreted as predominantly photoionization by the active galactic nucleus (AGN). The Seyfert nucleus and ionization cones transition to and are surrounded by a LINER cocoon, extending up to $\sim$ 250 pc in thickness. The ubiquity of the LINER cocoon in Seyfert 2 galaxies suggests that the circumnuclear regions are not necessarily Seyfert-type, and LINER activity plays an important role in Seyfert 2 galaxies. We demonstrate that spatially resolved diagnostics are crucial to understanding the excitation mechanisms in different regions and the AGN-host galaxy interactions.

\end{abstract}


\section{Introduction}

How super massive black holes interact with and impact their host galaxies via feedback processes is critical to understanding galaxy evolution (e.g., \citealt{Hopkins2006,Fabian2012,Kormendy2013,Heckman2014}). Active galactic nuclei (AGN) can exert positive or negative feedback via either radiative processes, in which energetic AGN radiation photoionize and heat gas in the host galaxy, or via mechanical processes such as jets and outflows.    

The narrow line region (NLR) is a ubiquitous component of AGN and is important for studying the interaction of the AGN with its host galaxy. The size of the NLR was first thought to be sub-kpc scales, but further observations reveal more extended ionized gas up to several kpc or more, referred to as extended narrow-line regions or ENLRs (e.g., \citealt{Heckman1981,Unger1987,Keel2012,Liu2013,Sun2018,Storchi2018}). The NLRs or ENLRs often form in a conical or biconial shape, i.e., ionization cones. Detailed studies of the ENLRs in nearby galaxies are necessary to understand the feedback processes, as the ENLRs can be spatially resolved given the advantage of angular scales of nearby galaxies.

Emission line diagnostics are powerful probes of the physical conditions and processes occurring in galaxies. The Baldwin-Phillips-Terlevich (BPT) diagnostic diagrams, which employ strong optical emission line ratios [N\II]/H$\alpha$ (N-BPT), [S\II]/H$\alpha$ (S-BPT), [O\I]/H$\alpha$ (O-BPT) versus [O\III]/H$\beta$, have been widely used to separate populations of AGN from star formation and classify galaxy spectra dominated by different activities, i.e., Seyfert-like, LINER-like, or star formation driven \citep{Baldwin1981,Veilleux1987,Kewley2001,Kauffmann2003,Kewley2006}. LINERs were first defined by \cite{Heckman1980} as low-ionization nuclear emission-line regions. LINER-type emission was later found to be not necessarily confined to the nuclear region and also seen in extended gas out to kpc scales (e.g., \citealt{Phillips1986, Ho2014,Belfiore2016}). LINER-type emission has been found to be associated with a variety of ionization mechanisms, including shock excitation (e.g., \citealt{Heckman1980,Dopita1995,Ho2014}), low luminosity AGN activity (e.g., \citealt{Storchi1997,Ho2001,Ulvestad2001}), and an evolved stellar population like post-asymptotic giant branch stars  (e.g., \citealt{Binette1994,Kehrig2012,Belfiore2016}). 

BPT mapping, which utilizes spatially resolved line maps of the critical diagnostic emission lines in the BPT diagram, can resolve galaxies into regions dominated by different ionizing processes. The advent of the new generation integral field unit (IFU) spectroscopy such as VLT/MUSE (e.g., \citealt{Cresci2015}) and AAOmega-SPIRAL (e.g., \citealt{Sharp2010}), and large IFU surveys like SDSS-IV/MANGA (e.g., \citealt{Bundy2015,Belfiore2015,Belfiore2016,Pilyugin2020}) and the Siding Spring Southern Seyfert Spectroscopic Snapshot Survey (S7; e.g., \citealt{Dopita2015,Davies2016,Davies2017}) demonstrates that a single galaxy may simultaneously display Seyfert, LINER, and star-forming characteristics depending upon the local conditions. However, these ground-based IFUs cannot reach the $\leq0.1\arcsec$ resolution of {\it HST}, which can reliably resolve the innermost regions of the nucleus on spatial scales of $\sim$ 10 pc. 

We conducted a pilot {\it HST} BPT mapping project of the nearby Compton-thick Seyfert 2 galaxy NGC 3393, using continuum-subtracted narrow band filter imaging \citep{Maksym2016}. We demonstrate that the ENLR of NGC 3393 is predominantly Seyfert-like with a distinct LINER cocoon surrounding the nuclear bicone structure. This is the first study that clearly reveals a LINER cocoon at $\sim$10 pc resolution and resolves the morphologies of the Seyfert and LINER regions. Using the VLT/MUSE IFU with a spatial FWHM resolution of 0.88$\arcsec$, \cite{Cresci2015} also show in another Seyfert 2 galaxy NGC 5643 that the nucleus and the ionization cones are dominated by Seyfert activity and enveloped by a LINER cocoon. To test the generality of LINER cocoons and resolve where Seyfert activity ends and LINER begins, we conducted a survey of nearby Seyfert galaxies with {\it HST} narrow band filter imaging.

In this paper, we report the first sample results from this {\it HST} BPT mapping survey on seven nearby Type 2 Seyfert galaxies not classified as LINERs. Seyfert 2 galaxies have the advantage that the host galaxy is not outshined by the central point source (unresolved nucleus) such that the extended, diffuse NLR or ENLR can be easily revealed and studied in detail. We focus on the {\it HST} BPT mapping results of the Seyfert 2 sample in this work. We will present a comprehensive analysis, combining multi-wavelength data, of the individual sources in future papers. 

In Section \ref{obs}, we describe the sample selection, {\it HST} observations, and data reduction of the narrow band filter imaging. Section \ref{BPT} introduces the BPT mapping methodology. We present the continuum-subtracted line maps and resolved BPT maps of individual sources in Section \ref{results}. We discuss the results and implications in Section \ref{discussion}. Section \ref{conclusions} summarizes our results and conclusions. 

Throughout the work, we adopt a concordance $\Lambda$CDM cosmological model with $H_0$ = 70 km s$^{-1}$Mpc$^{-1}$, $\Omega$$_{\Lambda}$ = 0.7, and $\Omega$$_{\rm m}$ = 0.3. Redshifts were taken from the NASA/IPAC Extragalactic Database (NED) unless otherwise noted.

\section{HST Observations and data reduction}
\label{obs}

\begin{table*}
\centering
\caption{{\it HST} Observation Log}
\begin{tabular}{lccccccc}
\hline\hline
Sourcename & $z$   &  Instrument/Chip &Filter  &    Exposure (s)  &            Date           & PID           & Note\\  
 \hline
NGC 1386     &0.00290& WFPC2/PC1  &  F502N    &  800            &  1997 June 28      & 6419   & [O\III]$\lambda$5007  \\
                       &            & WFPC2/PC1   & F547M    &  80              &  1997 June 28      & 6419  & [O\III] or H$\beta$ continuum\\
                       &            & WFPC2/PC1  & F658N     & 800              &  1997 June 28      & 6419  &H$\alpha$ + [N\II]\\
                       &             & WFPC2/PC1  & F791W   & 80                 &  1997 June 28      & 6419  &H$\alpha$ + [N\II] continuum\\
                       &	    & WFC3/UVIS2  & F673N   &  962              &  2018 Jan 5         &15350  & [S\II]$\lambda$6716,6731\\
                       &	    & WFC3/UVIS2  & F645N  &  242               & 2018 Jan 5         &15350     &   [S\II] continuum\\
                       &            &  WFC3/UVIS2 & F487N  &  962               & 2018 Jan 5         &15350     & H$\beta$\\
\hline
NGC 5643      &0.00399& WFPC2/PC1  & F502N  & 700                & 1995 June 2       &  5411   &  [O\III]$\lambda$5007  \\
		      &            & WFPC2/PC1  & F547M  & 50                   & 1995 June 2       &  5411   & O\III] or H$\beta$ continuum\\
		      &            & WFPC2/PC1  & F658N  &  800                & 1995 June 2       &  5411   &H$\alpha$ + [N\II]\\
		      &            & WFPC2/PC1  &F814W  & 50                    &  1995 June 2       &  5411   &H$\alpha$ + [N\II] continuum\\
		      &            &  WFC3/UVIS2 & F673N  &  976                & 2017 Dec 27    & 15350  & [S\II]$\lambda$6716,6731\\
		      &            &WFC3/UVIS2 &  F645N   & 256                & 2017 Dec 27    & 15350  &[S\II] continuum\\
		      &            &WFC3/UVIS2 & F487N   &  976                &  2017 Dec 27    & 15350  & H$\beta$\\
\hline
NGC 2273     &	0.00614&WFPC2/WF2  &  FR533N   &  1600            &  1997 Feb 5      & 6419   & [O\III]$\lambda$5007  \\
(Mrk 620)       &            &WFPC2/WF2  &  F547M  &  200                   &  1997 Feb 5      & 6419   & O\III] or H$\beta$ continuum\\
                      &            &WFPC2/PC1  &  FR680P15  &  1400              &  1997 Feb 5      & 6419   &H$\alpha$ + [N\II]\\
                      &            &WFPC2/PC1  &  F791W  &  200                &  1997 Feb 5      & 6419   &H$\alpha$ + [N\II] continuum\\
		     &            &  WFC3/UVIS2 & F673N  &  1034               & 2018 Jan 4    & 15350  & [S\II]$\lambda$6716,6731\\
		     &            &  WFC3/UVIS2 & F645N &  314                & 2018 Jan 4    & 15350  & [S\II] continuum\\
		     &            &  WFC3/UVIS2 & F487N &  1034                & 2018 Jan 4    & 15350  &H$\beta$\\
\hline
NGC 3081     &	0.00799&WFPC2/WF2 & FR533N & 400              & 1997 Feb 5    & 6419  & 	[O\III]$\lambda$5007  \\
                     &            &WFPC2/WF2 & F547M & 80                   & 1997 Feb 5    & 6419  & 	O\III] or H$\beta$ continuum\\
                      &             & WFPC2/PC1 & FR680P15 & 400         & 1997 Feb 5    & 6419  &H$\alpha$ + [N\II]\\
                     &             & WFPC2/PC1 & F791W & 80                   & 1997 Feb 5    & 6419  &H$\alpha$ + [N\II] continuum\\                                          
		    &            &  WFC3/UVIS2 & F673N  &  948              & 2018 May 11    & 15350  & [S\II]$\lambda$6716,6731\\
		    &            &  WFC3/UVIS2 & F645N  &  228              & 2018 May 11    & 15350  & [S\II] continuum\\
		    &            &  WFC3/UVIS-QUAD & FQ492N  &  948 & 2018 May 11    & 15350  & H$\beta$\\				        
 \hline			    	    				    		    
Mrk 573         &0.0172  & WFPC2/WF2 & FR533N  &  600              &   1995 Nov 12     &6332& [O\III]$\lambda$5007\\    
(UGC01214)  &            &   WFPC2/WF4 & FR533N &   280             &   1995 Nov 12      &6332&[O\III] or H$\beta$ continuum\\
                      &            &   WFPC2/WF2 & FR680N &    600             &    1995 Nov 12    & 6332&H$\alpha$ + [N\II]\\
                      &            &   WFPC2/WF3 & FR680N &  280            &    1995 Nov 12       &6332 &H$\alpha$ + [N\II] continuum\\
                     &            &   ACS/WFC1    & FR656N  &    772            &   2017 Dec 11       &15350  &[S\II]$\lambda$6716,6731\\
                     &            &   ACS/WFC1    & FR656N  &  212            &   2017 Dec 11       & 15350 &[S\II] continuum\\
                     &            &   ACS/WFC1    & FR505N  &   772            &   2017 Dec 11         &15350 & H$\beta$\\
 \hline
 NGC 7212&	0.0266	& WFPC2/WF2 & FR533N  &  600              &   1995 Sep 26     &6332& [O\III]$\lambda$5007\\    
		  &            &   WFPC2/WF4 & FR533N &   280             &   1995 Sep 26      &6332&[O\III] or H$\beta$ continuum\\
                      &            &   WFPC2/WF2 & FR680N &    598             &    1995 Sep 26    & 6332&H$\alpha$ + [N\II]\\
                      &            &   WFPC2/WF3 & FR680N &  280            &    1995 Sep 26      &6332 &H$\alpha$ + [N\II] continuum\\
                     &            &   ACS/WFC1    & FR716N  &    760            &   2017 Nov 3       &15350  &[S\II]$\lambda$6716,6731\\
                     &            &   ACS/WFC1    & FR656N  &  200            &   2017 Nov 3       & 15350 &[S\II] continuum\\
                     &            &   ACS/WFC1    & FR505N  &   760            &   2017 Nov 3         &15350 & H$\beta$\\
\hline
 NGC 7674&	0.02903   & WFPC2/PC1 & FR533NP15  &  520              &   1999 Nov 8     &8259& [O\III]$\lambda$5007\\    
(Mrk 533)		  &            &   WFPC2/PC1 & F547M &   200             &   1999 Nov 8      &8259&[O\III] or H$\beta$ continuum\\
                      &            &   WFPC2/WF2 & FR680N &    400             &    1999 Nov 8    & 8259&H$\alpha$ + [N\II]\\
                      &            &   WFPC2/PC1 & F814W &  200            &    1999 Nov 8      &8259 &H$\alpha$ + [N\II] continuum\\
                     &            &   ACS/WFC1    & FR716N  &    750            &   2017 Nov 2       &15350  &[S\II]$\lambda$6716,6731\\
                     &            &   ACS/WFC1    & FR656N  &  190            &   2017 Nov 2       & 15350 &[S\II] continuum\\
                     &            &   ACS/WFC1    & F502N  &   750            &   2017 Nov 2         &15350 & H$\beta$\\

  \hline
\end{tabular}
\label{table1}
\end{table*}

The sources were drawn from two samples of nearby Seyfert galaxies, \cite{Fischer2013} and \cite{Schmitt2003}, many of which have archival {\it HST} narrow band imaging data of [O\III] and/or H$\alpha$+[N\II]. To increase the efficiency of our BPT mapping project, we only chose the Seyfert galaxies that have existing [O\III] and H$\alpha$ narrow band data. A new follow-up {\it HST} narrow band imaging program was conducted by our team (P.I. Maksym) to complement the set of optical emission lines in the BPT diagram, i.e., [S\II] and H$\beta$. The seven sources in this work, which span a redshift range of 0.002 $<$ $z$ $<$ 0.03, are all Type 2 Seyfert galaxies and all have post-COSTAR narrow band imaging data. Two sources (Mrk 573 and NGC 1386), and the pilot source NGC 3393 in \cite{Maksym2016} are part of our {\it CHandra} Survey of Extended Emission-line Regions in nearby Seyfert galaxies (CHEERS; \citealt{Wang2010}).

\subsection{WFPC2 observations and data reduction}
\label{WFPC2}


\begin{table*}
\centering
\caption{Observed emission line fluxes}
\begin{tabular}{lcccccccc}
\hline\hline
Sourcename   &  $F_{\rm [O III]}^{\rm int}$   &  $F_{\rm [O III]}^{\rm nuc}$ &  $F_{\rm H\alpha}^{\rm int}$   &  $F_{\rm H\alpha}^{\rm nuc}$ &$A_{\rm int}$ &$A_{\rm nuc}$   \\
                       & (erg cm$^{-2}$ s$^{-1}$)      & (erg cm$^{-2}$ s$^{-1}$)       &  (erg cm$^{-2}$ s$^{-1}$)      &  (erg cm$^{-2}$ s$^{-1}$) &(arcsec)            & (arcsec) \\
\hline                  
NGC 1386      &(1.27 $\pm$ 0.30)$\times$10$^{-12}$& (4.72 $\pm$ 0.10)$\times$10$^{-13}$& (4.98 $\pm$ 0.61)$\times$10$^{-13}$&(1.38 $\pm$ 0.17)$\times$10$^{-13}$ & 4&0.4\\
NGC 5643     &(2.14 $\pm$ 0.22)$\times$10$^{-12}$&(8.62 $\pm$ 0.27)$\times$10$^{-13}$&(5.63 $\pm$ 0.79)$\times$10$^{-13}$&(2.25 $\pm$ 0.24)$\times$10$^{-13}$& 4.8&0.35\\
NGC 2273     & (3.42 $\pm$ 0.45)$\times$10$^{-13}$&  (1.24 $\pm$ 0.35)$\times$10$^{-13}$& (2.22 $\pm$ 0.25)$\times$10$^{-13}$& (5.72 $\pm$ 0.35)$\times$10$^{-14}$&2.5 &0.4\\
NGC 3081     & (6.49 $\pm$ 0.35)$\times$10$^{-13}$&(1.88 $\pm$ 0.09)$\times$10$^{-13}$ &(2.28 $\pm$ 0.38)$\times$10$^{-13}$&(5.38 $\pm$ 0.37)$\times$10$^{-14}$& 2&0.32\\
Mrk 573        &(1.41 $\pm$ 0.08)$\times$10$^{-12}$ & (3.42 $\pm$ 0.11)$\times$10$^{-13}$ &(4.53 $\pm$ 0.41)$\times$10$^{-13}$ & (1.01 $\pm$ 0.09)$\times$10$^{-13}$ &5 & 0.5\\
NGC 7212    &(8.22 $\pm$ 0.42)$\times$10$^{-13}$&(3.03 $\pm$ 0.14)$\times$10$^{-13}$& (2.90 $\pm$ 0.28)$\times$10$^{-13}$&(9.09 $\pm$ 0.72)$\times$10$^{-14}$&3 &0.3\\
NGC 7674    &(4.74 $\pm$ 0.27)$\times$10$^{-13}$&(3.59 $\pm$ 0.16)$\times$10$^{-13}$&(2.67 $\pm$ 0.26)$\times$10$^{-13}$&(1.76 $\pm$ 0.08)$\times$10$^{-13}$&1.3 & 0.35\\
\hline
   &  $F_{\rm [S II]}^{\rm int}$   &  $F_{\rm [S II]}^{\rm nuc}$ &  $F_{\rm H\beta}^{\rm int}$   &  $F_{\rm H\beta}^{\rm nuc}$ &$A_{\rm int}$ &$A_{\rm nuc}$  \\

\hline                  
NGC 1386      &(7.08 $\pm$ 0.14)$\times$10$^{-13}$ & (6.35 $\pm$ 0.12)$\times$10$^{-14}$&(4.53 $\pm$ 0.17)$\times$10$^{-13}$ &(3.80 $\pm$ 0.21)$\times$10$^{-14}$& 4&0.4\\
NGC 5643     &(4.39 $\pm$ 0.09)$\times$10$^{-13}$&(7.32 $\pm$ 0.13)$\times$10$^{-14}$&(3.07 $\pm$ 0.15)$\times$10$^{-13}$&(4.61 $\pm$ 0.18)$\times$10$^{-14}$&4.8 &0.35\\
NGC 2273     & (3.19 $\pm$ 0.09)$\times$10$^{-13}$& (4.69 $\pm$ 0.12)$\times$10$^{-14}$& (8.64 $\pm$ 0.35)$\times$10$^{-14}$& (1.01 $\pm$ 0.22)$\times$10$^{-14}$& 2.5&0.4\\
NGC 3081     &(1.95 $\pm$ 0.05)$\times$10$^{-13}$& (2.76 $\pm$ 0.10)$\times$10$^{-14}$& (1.94 $\pm$ 0.11)$\times$10$^{-13}$&(3.12 $\pm$ 0.20)$\times$10$^{-14}$&2 &0.32\\
Mrk 573        &(4.10 $\pm$ 0.10)$\times$10$^{-13}$ & (5.62 $\pm$ 0.17)$\times$10$^{-14}$ &(3.55 $\pm$ 0.25)$\times$10$^{-13}$ & (6.15 $\pm$ 0.18)$\times$10$^{-14}$ &5 & 0.5\\
NGC 7212    &(1.99 $\pm$ 0.07)$\times$10$^{-13}$&(4.37 $\pm$ 0.12)$\times$10$^{-14}$&(1.76 $\pm$ 0.10)$\times$10$^{-13}$&(3.38 $\pm$ 0.22)$\times$10$^{-14}$&3 &0.3\\
NGC 7674    &(1.67 $\pm$ 0.05)$\times$10$^{-13}$&(8.15 $\pm$ 0.15)$\times$10$^{-14}$&(1.08 $\pm$ 0.11)$\times$10$^{-13}$&(5.18 $\pm$ 0.15)$\times$10$^{-14}$&1.3 &0.35\\

  \hline
\end{tabular}
\tablecomments{$F^{\rm int}$ is the integrated line flux of the NRL/ENLR measured in a circular aperture with a radius of $A_{\rm int}$. $F^{\rm nuc}$ is the integrated line flux of the nuclear region measured in a circular aperture with a radius of $A_{\rm nuc}$. }
\label{table2}
\end{table*}

The [O\III] $\lambda$5007 and H$\alpha$ $\lambda$6563 lines were observed with the Wide Field and Planetary Camera 2 (WFPC2) on board {\it HST} by various programs (PIDs: 5411, 6332, 6419, 8259). WFPC2 consists of 4 chips, PC1, WF2, WF3, and WF4. PC1 has a pixel scale of 0.0455$\arcsec$ and WF chips have a pixel scale of 0.0996$\arcsec$ \citep{Gonzaga2010}. Depending on the redshifts, the [O\III] and H$\alpha$ line imaging were taken with the narrow band filters if the lines can be covered. Otherwise, linear ramp filters (LRFs) were used, where the central wavelength was tuned to the wavelength of the redshifted lines. Corresponding continuum images were taken with medium or wide band filters or LRFs. For the LRFs, the transmissions and passbands are position-dependent, resulting in an upper limit of 14$\arcsec$ on the field of view or the size of an emission line region. Exposure times vary from 400 s to 1600 s for the emission lines and 50 s to 280 s for the corresponding continua. All observations were split into two integrations to allow cosmic ray rejection except for NGC 5643's [O\III] and H$\alpha$ continuum observations, where only a single exposure was executed. The details of the observations, e.g., instrument/chip, filter, exposure time, observation date, and program ID, are given in Table \ref{table1}. 

Data reduction started with reprocessing all the images with {\it astrodrizzle} in the software package {\it Drizzlepac}\footnote{https://www.stsci.edu/scientific-community/software/drizzlepac} \citep{Gonzaga2012}. We set the final image pixel scale to 0.04$\arcsec$ to match WFC3 or 0.05$\arcsec$ to match ACS (see Section \ref{obs2}), as BPT mapping requires all the four emission line images having the exactly same pixel scales, and also because the native WFPC2 images are undersampled. Individual frames were combined to remove cosmic rays and corrected for geometric distortions. In cases where a few cosmic rays or hot pixels remain after running {\it astrodrizzle} or in the case of a single exposure, we manually removed them with the Pyraf tool {\it imedit}. We also rotated all the images such that North is up and East is left during the {\it astrodrizzle} run. 

After the initial processing with {\it astrodrizzle}, we subtracted the background, which was determined by the median value of line-free regions, for both emission line and continuum images. We then multiplied these images by the flux calibration keyword PHOTFLAM and the FWHM of the filter bandwidth 2.355$\times$PHOTBW to convert the image units to flux erg cm$^{-2}$ s$^{-1}$ pixel$^{-1}$. The emission line and corresponding continuum images were aligned according to the positions of the nucleus and stars in the common field. Continuum-free emission line images were generated by subtracting a scaled continuum image from the on-band image. The scaling of the continuum images was estimated from the FWHM of the on-band filter bandpass, and was adjusted when necessary to avoid over-subtraction or under-subtraction. 

Due to the relatively large filter bandwidth and close proximity in wavelengths, the contribution of [N\II] $\lambda$$\lambda$6548, 6584 to the H$\alpha$ filter is significant (typically 30\% - 50\%) and must be subtracted. However, the local [N\II]/H$\alpha$ ratio is subject to spatial variation in the intrinsic [N\II]/H$\alpha$ ratio over the entire emission line region \citep{Tsvetanov1992,Ferruit1999}, the wavelength shifts due to velocity gradients \citep{Whittle1988}, and the position on the chips (e.g., position-dependent transmission of the LRFs). To account for the uncertainties in [N\II] subtraction, we first used a typical [N\II] fractional contribution of 40\% as a constant rescaling factor and applied to the [N\II]+H$\alpha$ images to obtain H$\alpha$ only images. Then we varied the [N\II] contribution fraction and tested the effects on the final BPT maps (Section \ref{BPT}).

The final continuum-subtracted emission line maps were used to measure the integrated line fluxes. We measured the integrated fluxes of both the entire emission line region as well as the nuclear region. The line flux measurements and corresponding circular aperture sizes are listed in Table \ref{table2}. The apertures are centered on the nucleus, the location of which is determined as the peak position of [O\III].

\subsection{WFC3 or ACS observations and data reduction}
\label{obs2}

All the [S\II] $\lambda$$\lambda$6716, 6731 and H$\beta$ $\lambda$4861 imaging data in this work are from our {\it HST} program 15350 (P.I. Maksym). We obtained the [S\II] and H$\beta$ narrow line and the [S\II] continuum images with either WFC3/UVIS or ACS/WFC, depending upon the redshift and available filters. No additional observations were taken for the H$\beta$ continua as the existing [O\III] continuum images can be used for H$\beta$ as well. Exposure times vary from 750 s to 1034 s for the lines and from 190 s to 314 s for the [S\II] continua. WFC3 images have a native pixel scale of 0.04$\arcsec$ \citep{Gennaro2018} and ACS has a pixel scale of 0.05$\arcsec$ \citep{Lucas2018}. The WFC3 and ACS images display faint charge trails following hot pixels, cosmic rays, and bright stars in the direction of CCD readout, which need charge transfer efficiency (CTE) correction. For all the filters except the LRFs, we used the CTE-corrected images (*flc.fits) by the {\it HST} pipeline. The {\it HST} pipeline is not capable of correcting CTE for the LRFs though. 

We first reprocessed the images with {\it astrodrizzle}, setting the final image scales to the native pixel sizes of WFC3 or ACS. Subsequent processing of [S\II] and H$\beta$ is the same as the procedures described in Section \ref{WFPC2}. No contamination due to strong emission lines is expected in these filters. The line flux measurements on the continuum-subtracted line maps are given in Table \ref{table2}.

\section{BPT mapping}
\label{BPT}

We employ the S-BPT diagram, log([O\III]/H$\beta$) versus log([S\II]/H$\alpha$), defined in \cite{Kewley2006} to create spatially resolved BPT maps of these galaxies. In the S-BPT diagram, galaxies dominated by star formation, Seyfert activity, and LINER populate distinct regions. We map each pixel in the emission line maps according to its position on the BPT diagram, i.e., its BPT classification. Then we can investigate the excitation mechanisms in different regions of these galaxies and their implications. 

To do this, we first need to register the four emission line maps. Relative registration was done by aligning the nucleus and common features in the four line maps. We divided pairs of emission line maps from Section \ref{obs}, [O\III]/H$\beta$ and [S\II]/H$\alpha$, and placed the line flux ratio of each pixel as a data point on the BPT diagram. Only pixels with flux values above 3$\sigma$ in all line maps are used, as this is the required emission line significance in the BPT diagram \citep{Kewley2006}. Pixels in different regions are color-coded as red (Seyfert), yellow (LINER), and blue (H\II{} region). The transition zone where Seyfert and LINER classifications overlap is color-coded as green. We show the results of BPT mapping for individual sources in Section \ref{results}.

\section{Results}
\label{results}

\subsection{Mrk 573}

\begin{figure*} 
\centering
\includegraphics[width=8cm]{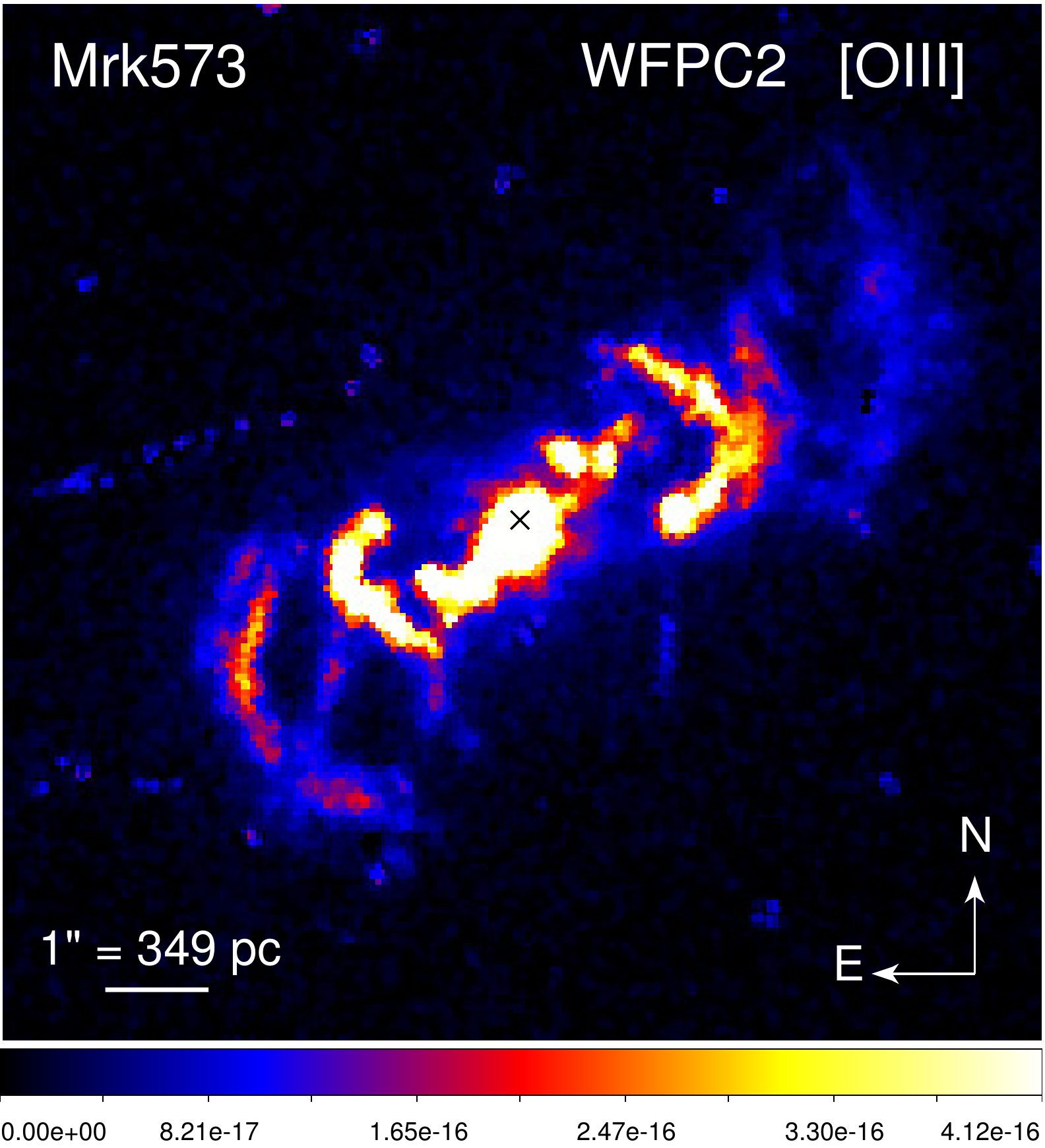}
\includegraphics[width=8cm]{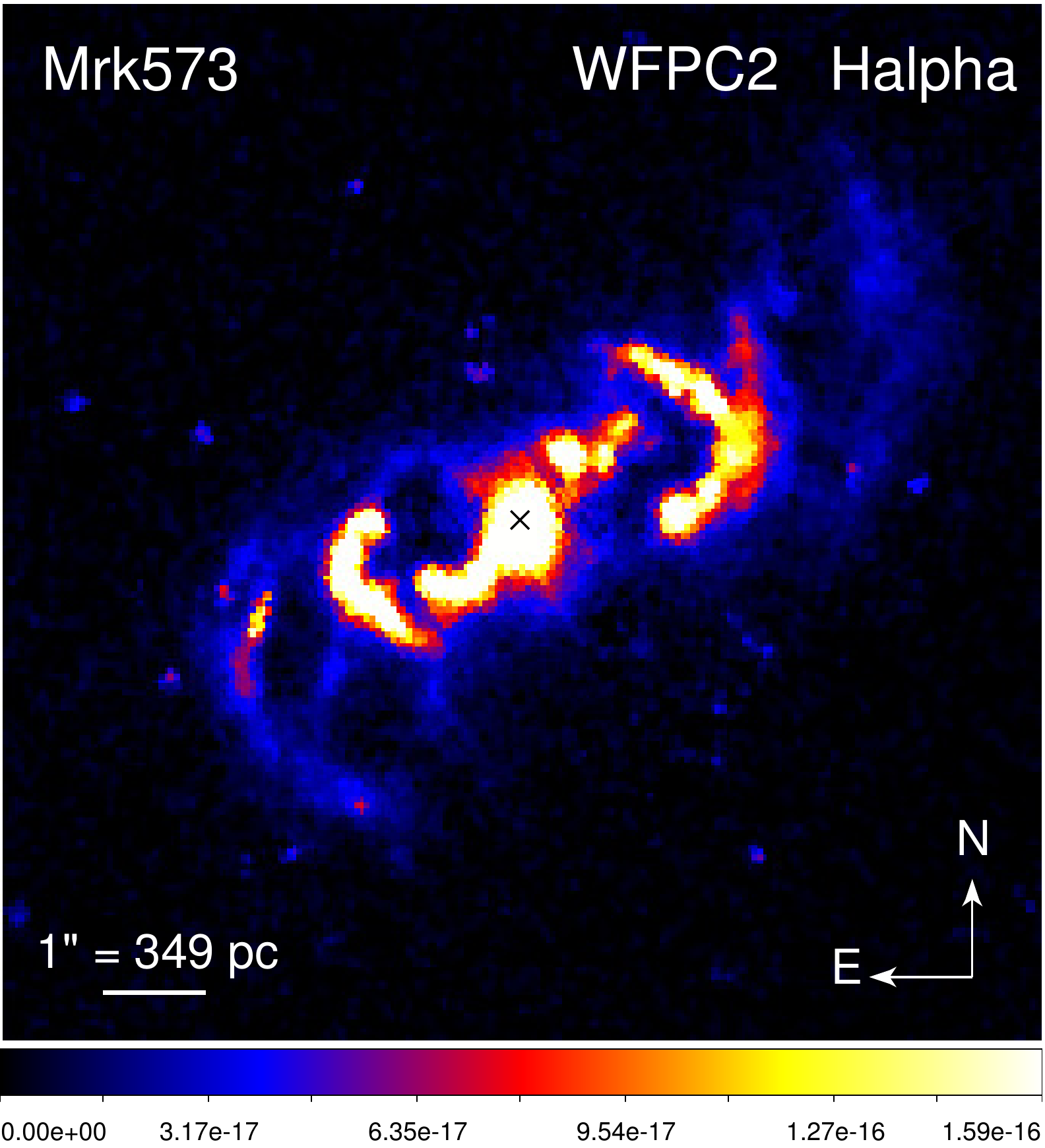}
\includegraphics[width=8cm]{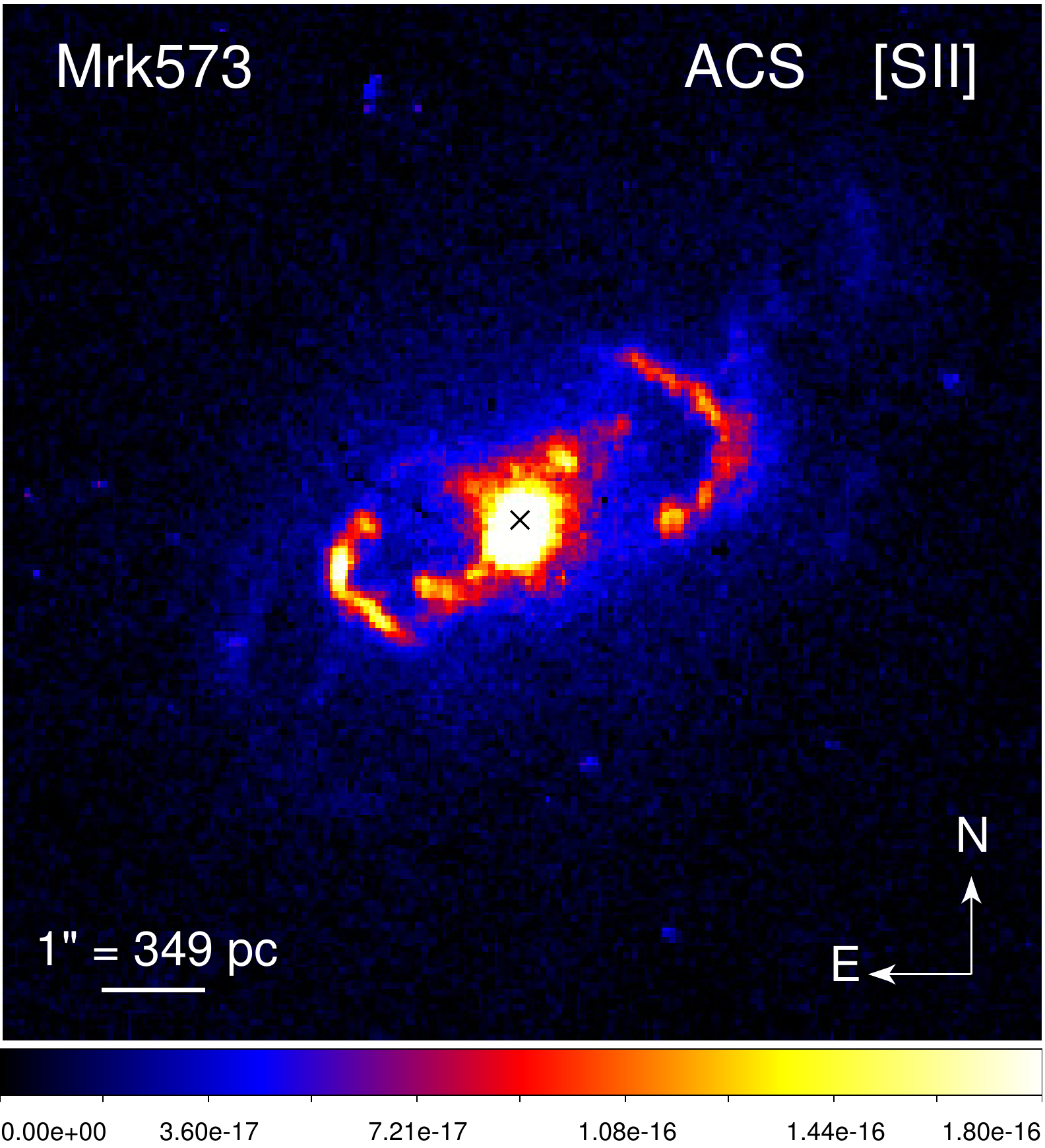}
\includegraphics[width=8cm]{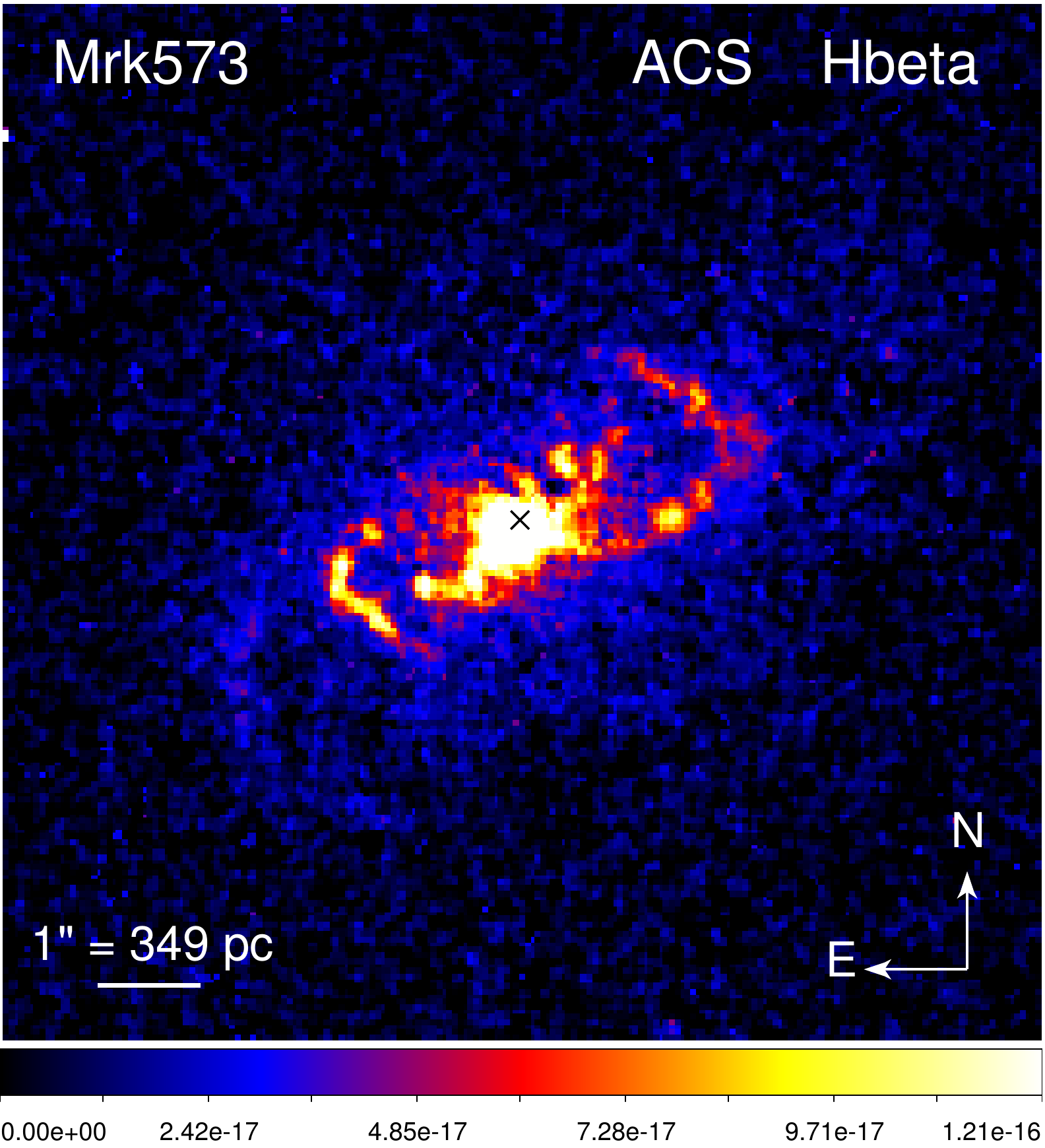}
\caption{Mrk 573 10$\arcsec$$\times$10$\arcsec$ continuum-subtracted emission line maps centered on the nucleus (black cross). All the images have a pixel scale of 0.05$\arcsec$. The colorbar denotes the surface brightness in the units of erg cm$^{-2}$ s$^{-1}$ pixel$^{-1}$.}
\label{Mrk573_linemap}
\end{figure*}

\begin{figure*} 
\centering
\includegraphics[width=9.5cm]{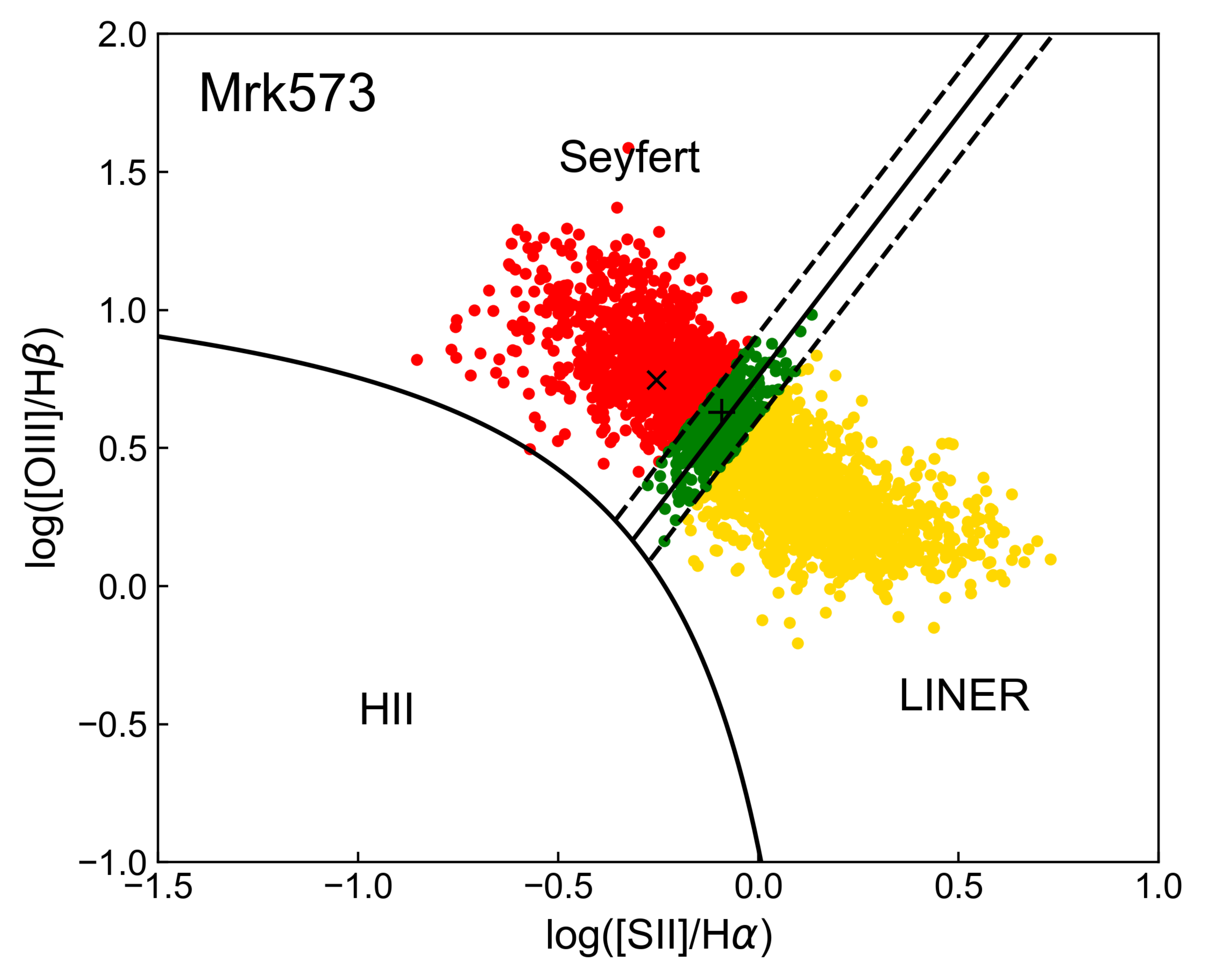}  
\includegraphics[width=7.7cm]{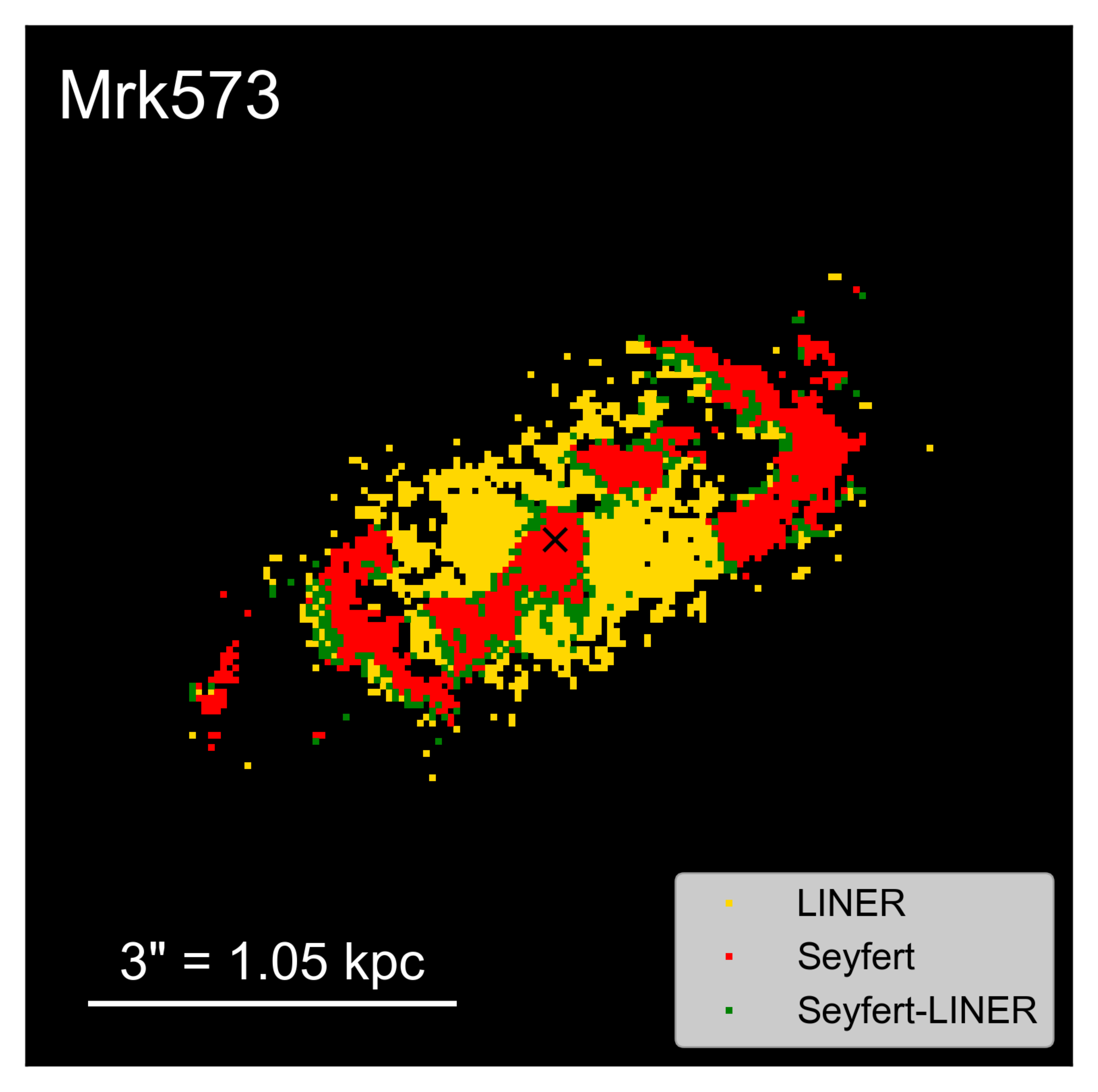} 
\caption{{\bf Left}: BPT diagram of Mrk 573. The dividing lines/curves between different excitation mechanisms are defined in \cite{Kewley2006}. Red corresponds to Seyfert-like activity, yellow denotes LINER-like activity, and the transition zone is coded as green. Blue represents line ratios typical of H\II{} regions. Only pixels detected above 3$\sigma$ in all lines are used. H\II{} pixels are excluded by this criterion. The black cross marks the line flux ratios measured for the nuclear region ($r \leq 0.5\arcsec$). {\bf Right}: Spatially resolved BPT map (8.5$\arcsec$ $\times$ 8.5$\arcsec$) with each pixel color-coded according to the BPT type as shown in the left panel. The black cross marks the nucleus. Black pixels have at least one line with $F_{\rm line}$ $<$ 3$\sigma$. The 3$\arcsec$ white bar represents the diameter of the SDSS 3$\arcsec$ fiber. We calculate the BPT line ratios ('+' in the left panel) using the integrated line fluxes within a 3$\arcsec$ diameter circular aperture centered on the nucleus to mimic SDSS observations. }
\label{Mrk573_BPT}
\end{figure*}

Mrk 573 ($z$ = 0.0172, $D$ = 74.6 Mpc, 1$\arcsec$ = 349 pc) is known for its extended emission line region with rich features in the ionization bicone (e.g., \citealt{Falcke1998,Ferruit1999}). There are two sets of arcs and multiple emission knots in between. A triple radio source is coincident with the nucleus and the inner emission knots (e.g., \citealt{Ulvestad1984,Falcke1998}). 

Figure \ref{Mrk573_linemap} shows the continuum-subtracted [O\III]$\lambda$5007 and H$\alpha$ line maps observed with WFPC2, and the [S\II]$\lambda$$\lambda$6716, 6731 and H$\beta$ line maps taken with ACS in the inner $1.8$ kpc of Mrk 573. [S\II] and H$\beta$ are relatively fainter than [O\III] and H$\alpha$, and the outer arcs are almost invisible. Using these line maps, we created the BPT diagram (left panel) and the spatially resolved BPT image (right panel) with each pixel color-coded according to its BPT type as shown in Figure \ref{Mrk573_BPT}. The boundaries of different excitation mechanisms are defined in \cite{Kewley2006}. Red corresponds to Seyfert-like activity, yellow denotes LINER-like activity, and the transition zone is coded as green. Blue represents line ratios typical of H\II{} regions. Only pixels detected above 3$\sigma$ in all lines are used. H\II{} pixels are excluded by this criterion, and most pixels that meet this criterion are located within the inter arcs. The spatially resolved, color-coded map clearly demonstrates that the nucleus and the ionization bicone are dominated by Seyfert-like activity, closely wrapped by a thin layer ($\sim$ 20 pc) of Seyfert-LINER transition, and surrounded by LINER-type emission ($\sim$ 250 pc in thickness). This structure is highly consistent with the recently identified LINER ``cocoon" feature revealed in our pilot {\it HST} BPT mapping of NGC 3393 \citep{Maksym2016}, although the LINER region in Mrk 573 appears thicker than in NGC 3393. Nevertheless, both sources are limited by the depths of the observations and therefore the full extent of the LINER cocoon is yet to be revealed by deep observations. In this work, we have also identified a thin layer of Seyfert-LINER transition and showed the smooth transition from Seyfert to LINER emission for the first time.  

\cite{Revalski2018} also constructed spatially resolved BPT diagrams in 1D though along the NLR of Mrk 573 using emission line ratios measured from the spectra taken with the {\it HST}/STIS long-slit and ground-based Dual Imaging Spectrograph at the Apache Point Observatory. For the {\it HST}/STIS spectrum, they measured line ratios at each spatial distance (spatial resolution = 0.1$\arcsec$) from the nucleus along the 0.2$\arcsec$ wide slit. The slit position of -71.2$^{\circ}$ spatially samples the bright nuclear emission and inner and outer arcs, but does not pass through the bright emission line knots. The ground-based 2$\arcsec$ wide slit covers most of the NLR features, and they also measured line ratios at each distance (spatial resolution = 0.4$\arcsec$) within 2$\arcsec$ of the nucleus. Their N-BPT, S-BPT, and O-BPT diagrams show that all their data points lie in the Seyfert AGN regime. There are a few points near the Seyfert-LINER border and they interpret the emission as gas exposed to an absorbed or filtered ionizing continuum. In addition, they also performed multi-component photoionization modeling and the results are consistent with pure AGN ionization from the central source. So they conclude that they do not see any evidence for shock ionization in those regions. However, a single spatial dimension cannot simultaneously capture the radial and azimuthal variations in emission. They do not see LINER type most likely because the line ratios are measured along 1D, where both Seyfert and LINER regions within the slit width can be mixed and unresolved and dominated by Seyfert type line ratios. We will further discuss the origin of LINER activity and relation to shock excitation in Section \ref{discussion}.

We have tested the effect on the resulting resolved BPT map due to uncertainties in subtracting the [N\II] contribution from the H$\alpha$ filter. We used the [N\II]/H$\alpha$ line ratio measured at difference distances from the high quality {\it HST}/STIS spectrum in \cite{Revalski2018} as a reference. Only a small number of pixels are affected (e.g., BPT type changed) when varying the fractional [N\II] contribution from the maximum to mininum, and our conclusions are unchanged.

\subsection{NGC 1386}

\begin{figure*} 
\centering
\includegraphics[width=8cm]{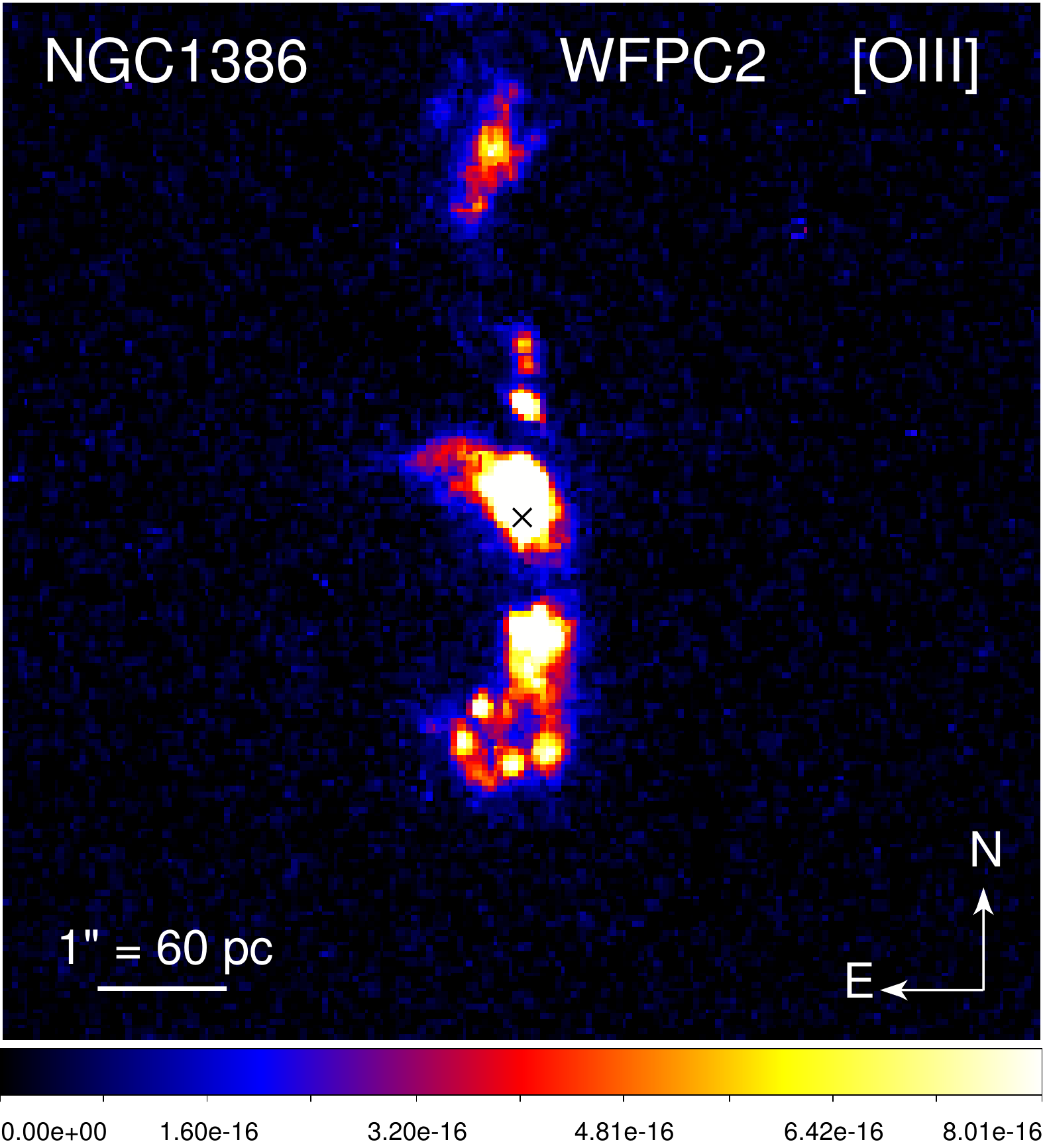}
\includegraphics[width=8cm]{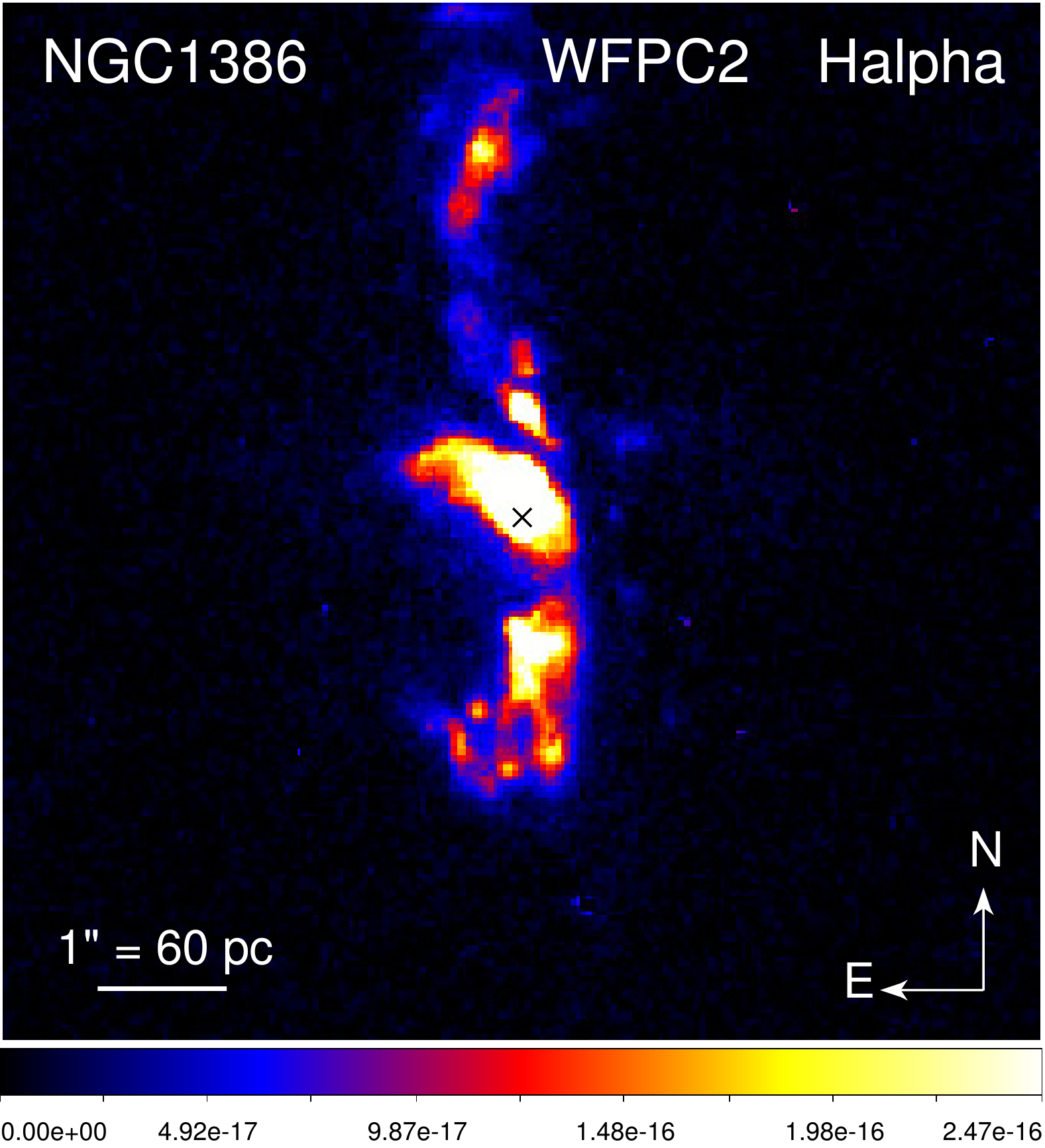}
\includegraphics[width=8cm]{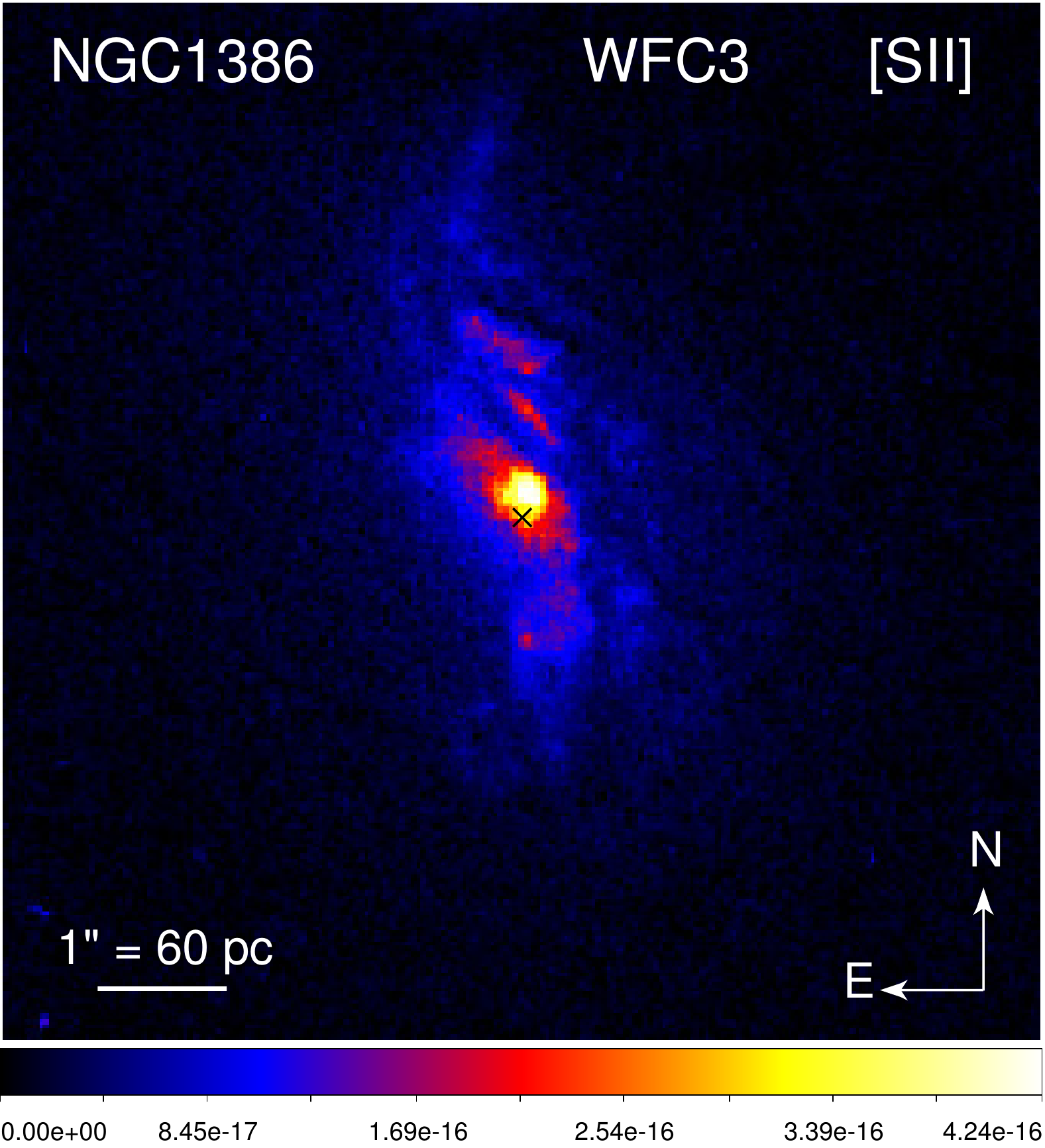}
\includegraphics[width=8cm]{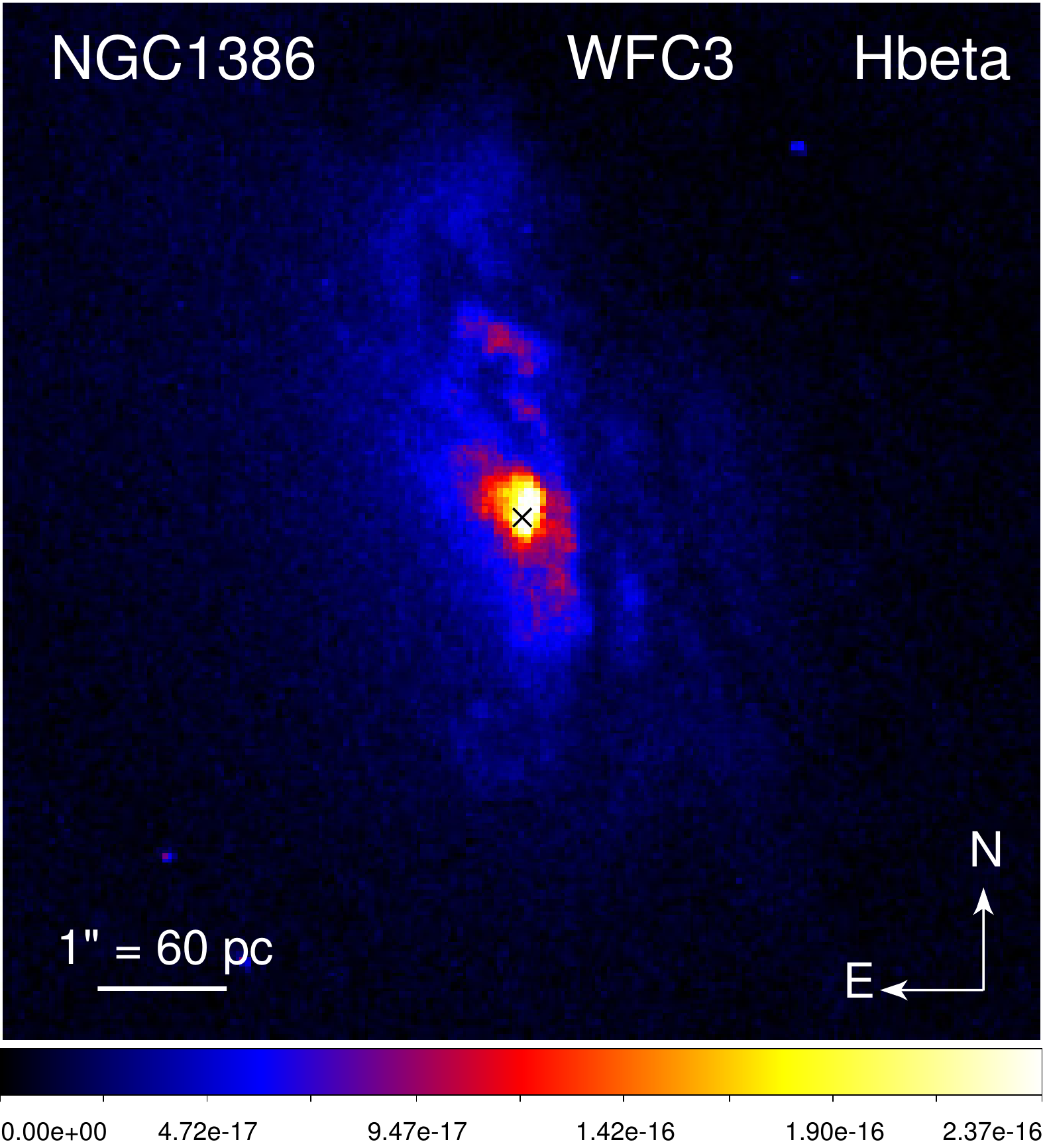}
\caption{NGC 1386 8$\arcsec$ $\times$ 8$\arcsec$ continuum-subtracted emission line maps centered on the nucleus (black cross). All the images have a pixel scale of 0.04$\arcsec$. The colorbar denotes the surface brightness in the units of erg cm$^{-2}$ s$^{-1}$ pixel$^{-1}$.}
\label{NGC1386_linemap}
\end{figure*}

NGC 1386 ($z$ = 0.0029, $D$ = 12.4 Mpc, 1$\arcsec$ = 60 pc) is one of the closest Seyfert galaxies, located in (or in the foreground of) the Fornax Cluster (e.g., \citealt{Storchi1996b,Rossa2000}). The extended [O\III] and H$\alpha$ emission line region (Figure \ref{NGC1386_linemap}) is characterized by an ionization bicone with several blobs distributed along the N-S direction, extending over a region of $\sim$ 6$\arcsec$ (360 pc) (e.g., \citealt{Ferruit2000,Schmitt2003}). The emission along the E-W direction is much narrower, extended by only $\sim$ 1.5$\arcsec$ (90 pc). A ring of small emission blobs is located $\sim$2$\arcsec$ south of the nucleus. The [S\II] and H$\beta$ emission displays a slightly different morphology in the WFC3 images (Figure \ref{NGC1386_linemap}). The far north and south components are much less prominent than those in [O\III] and H$\alpha$. Multiple strong, sharp dust lanes are present in the circumnuclear region. The [S\II] and H$\beta$ image scales are stretched to show the dust features. The morphology of the NLR is strongly shaped by the dust features. \cite{Mundell2009} compared the [O\III] emission with VLA radio data at 8.4 GHz. The radio source generally follows the direction of [O\III], although the radio and optical components do not appear to be directly associated (Figure 5 in \citealt{Mundell2009}).

\begin{figure*} 
\centering
\includegraphics[width=9.5cm]{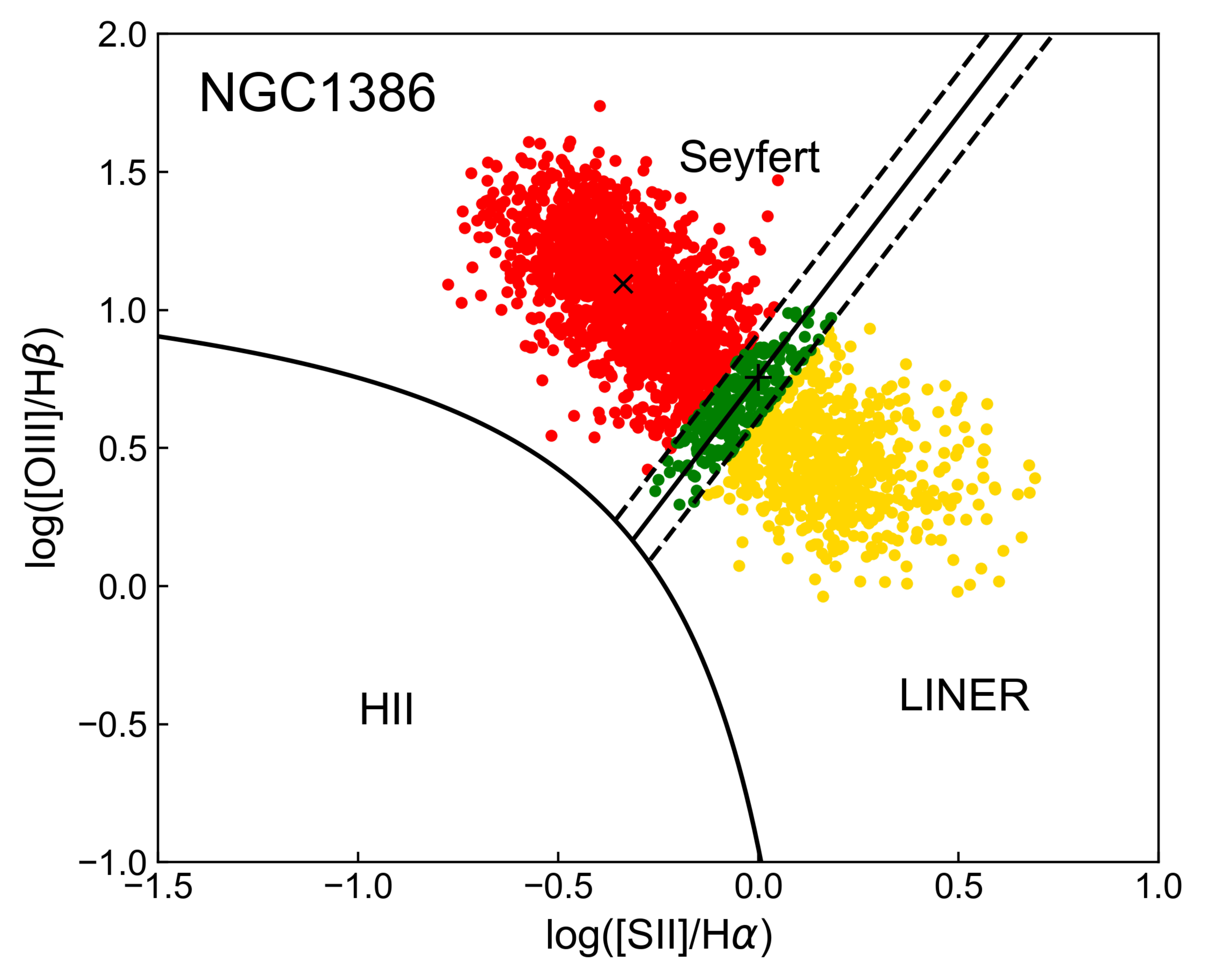} 
\includegraphics[width=7.7cm]{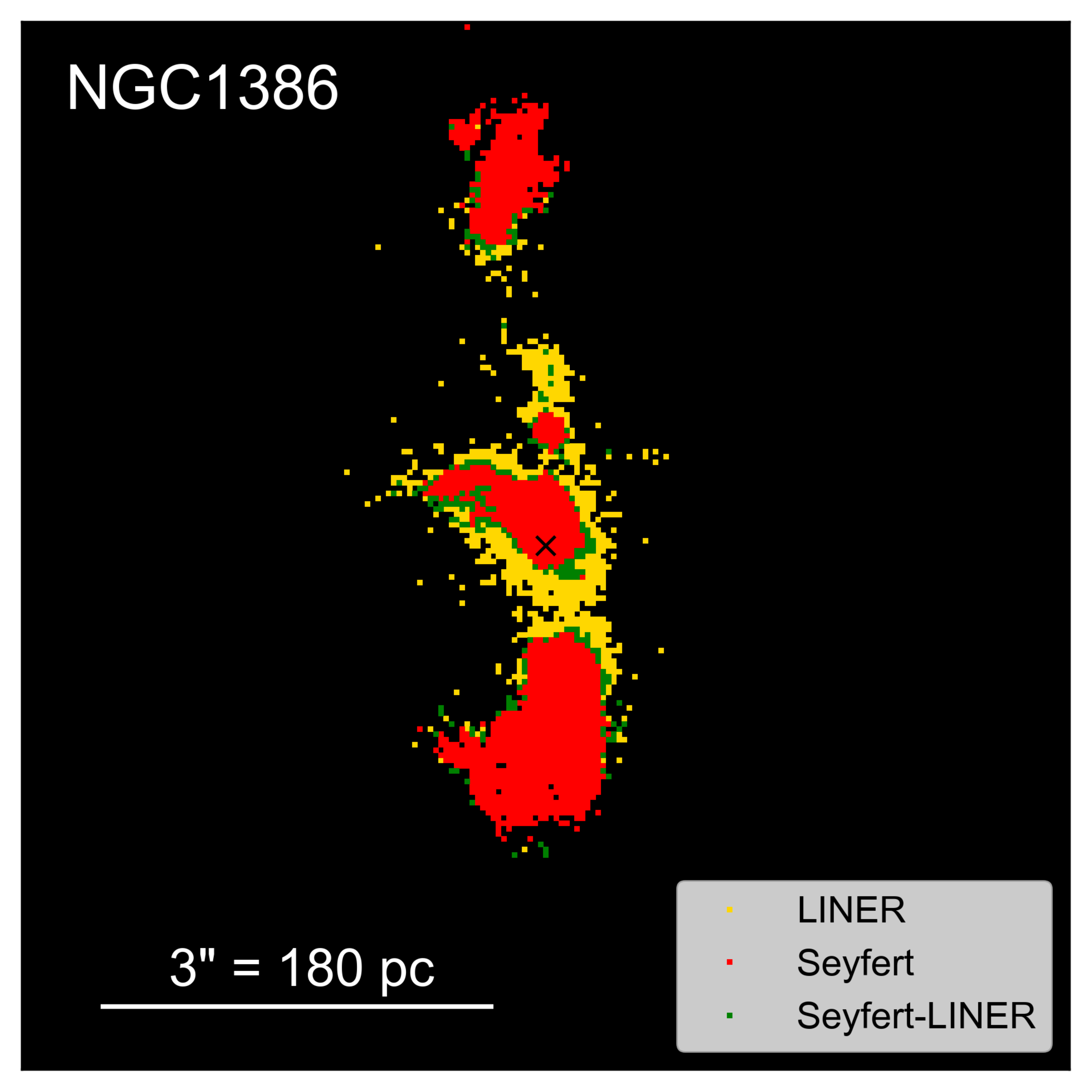} 
\caption{{\bf Left}: BPT diagram of NGC 1386. The dividing lines/curves between different excitation mechanisms are defined in \cite{Kewley2006}. Red corresponds to Seyfert-like activity, yellow denotes LINER-like activity, and the transition zone is coded as green. Blue represents line ratios typical of H\II{} regions. Only pixels detected above 3$\sigma$ in all lines are used. H\II{} pixels are excluded by this criterion. The black cross marks the line flux ratios measured for the nuclear region ($r \leq 0.4\arcsec$). {\bf Right}: Spatially resolved BPT map (8$\arcsec$ $\times$ 8$\arcsec$) with each pixel color-coded according to the BPT type as shown in the left panel. The black cross marks the nucleus. Black pixels have at least one line with $F_{\rm line}$ $<$ 3$\sigma$. The 3$\arcsec$ white bar represents the diameter of the SDSS 3$\arcsec$ fiber. We calculate the BPT line ratios ('+' in the left panel) using the integrated line fluxes within a 3$\arcsec$ diameter circular aperture centered on the nucleus to mimic SDSS observations. }
\label{NGC1386_BPT}
\end{figure*}

Using the continuum-subtracted line maps in Figure \ref{NGC1386_linemap}, we calculated the line ratios for each pixel and created the BPT diagram and the spatially resolved map color-coded by the BPT type (Figure \ref{NGC1386_BPT}). The bright emission blobs show red Seyfert-type activity and transition to yellow LINER regions. A LINER ``cocoon" entirely encloses the central emission line region and partially surrounds the northern and southern components. Using the VLT/MUSE IFU, \cite{Mingozzi2019} conducted BPT mapping of the disk and outflow components in NGC 1386, where they show a more complete picture of the LINER emission, permeating in the disk.

\subsection{NGC 3081}

\begin{figure*} 
\centering
\includegraphics[width=8cm]{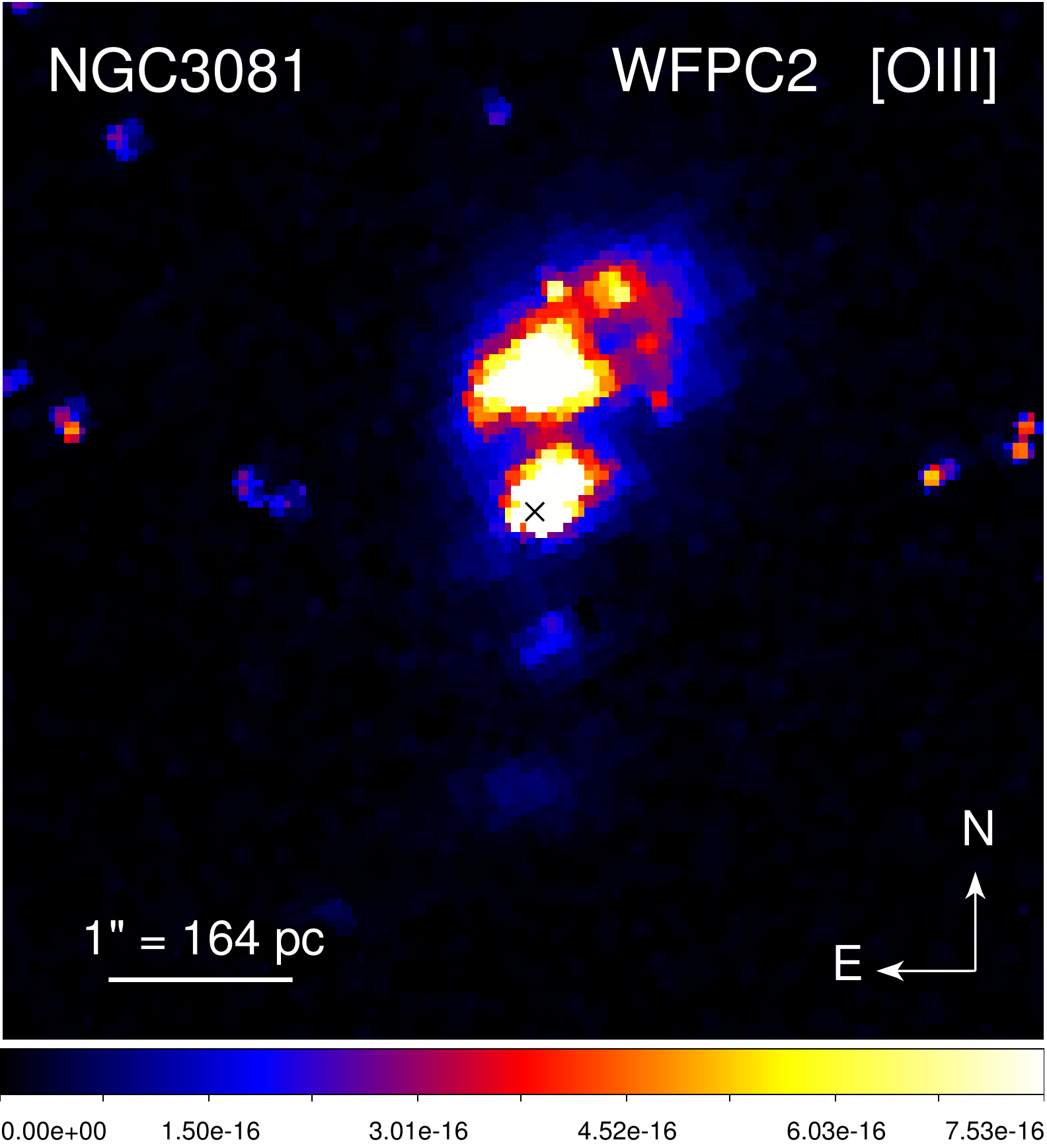}
\includegraphics[width=8cm]{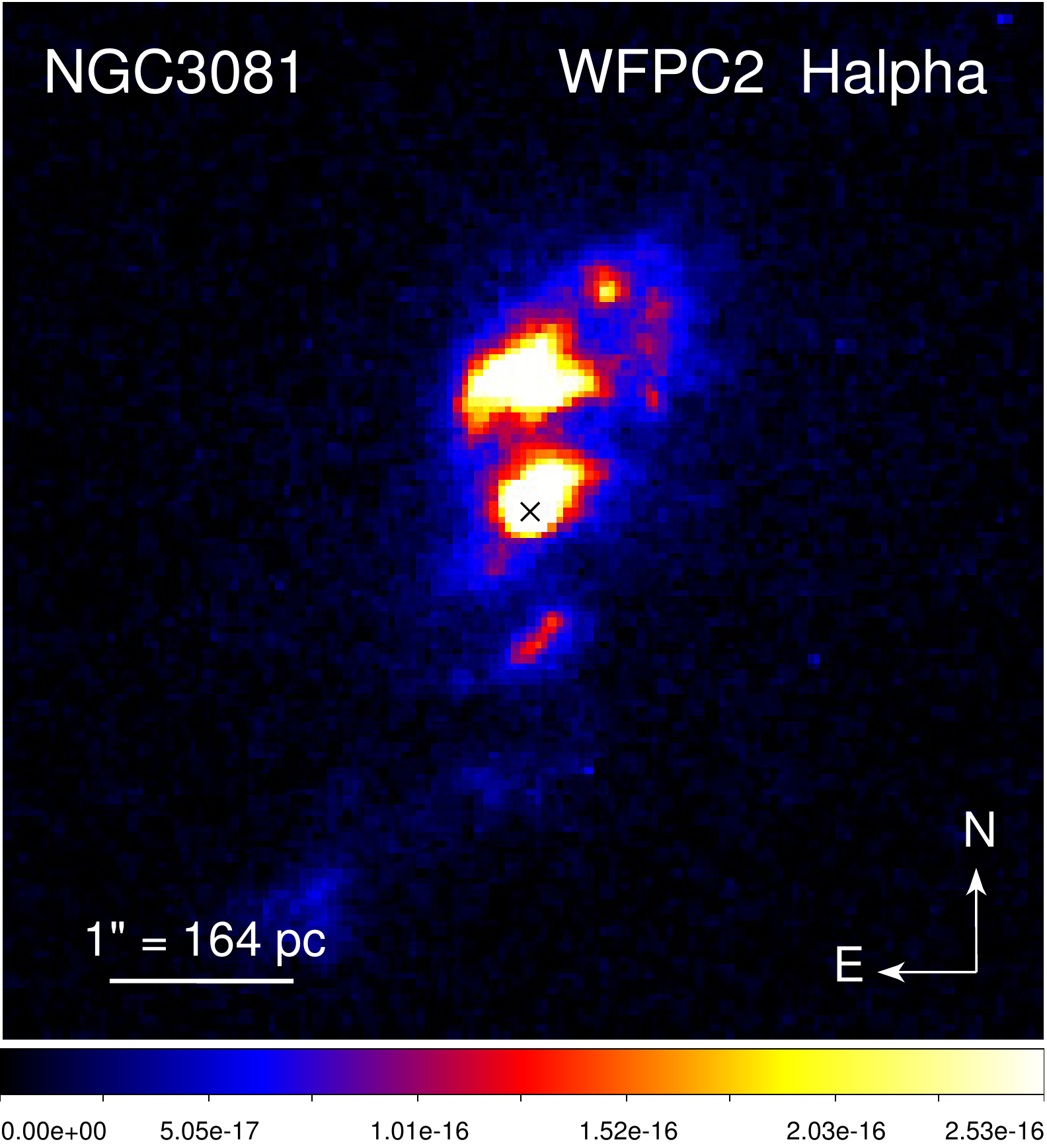}
\includegraphics[width=8cm]{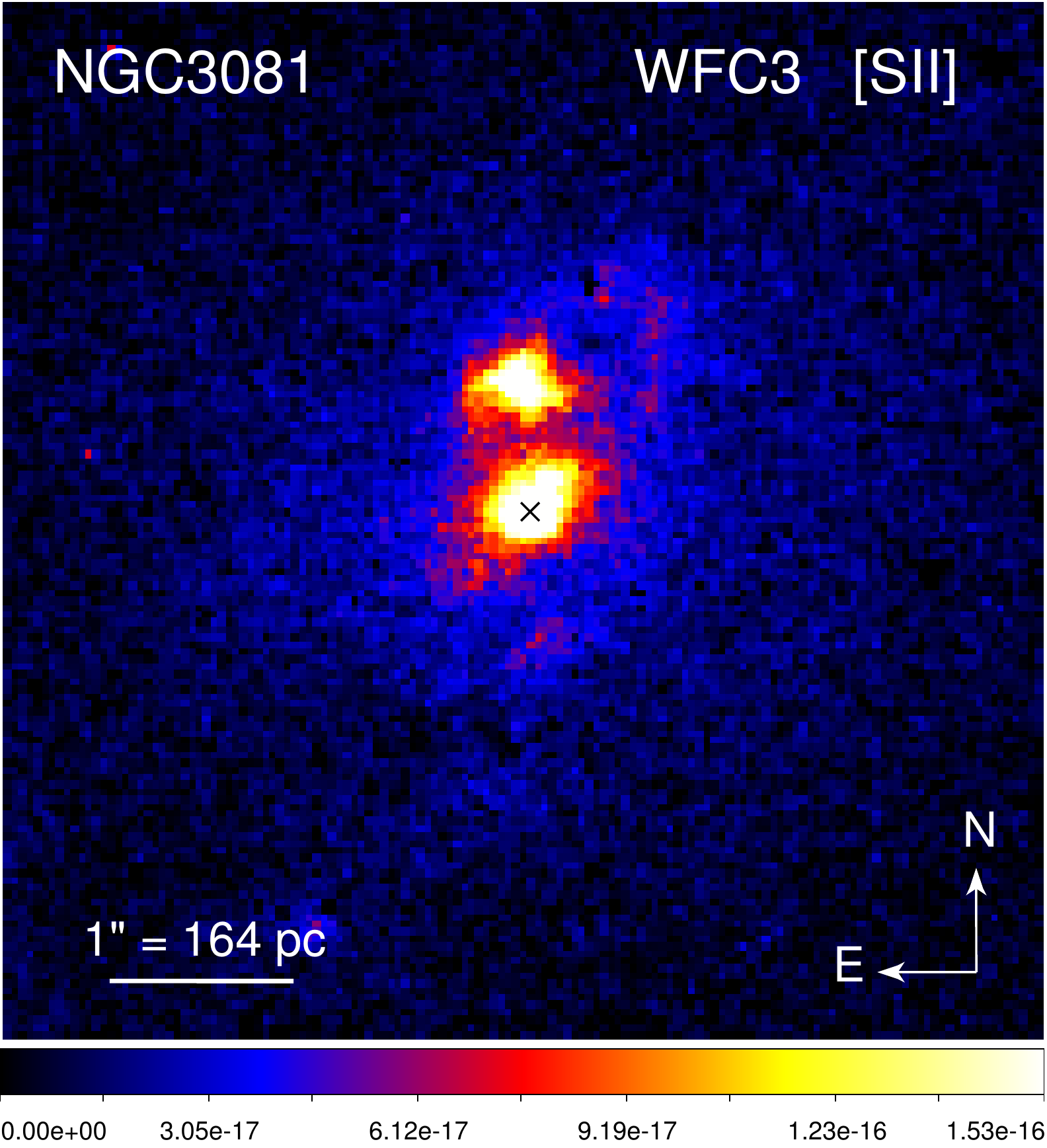}
\includegraphics[width=8cm]{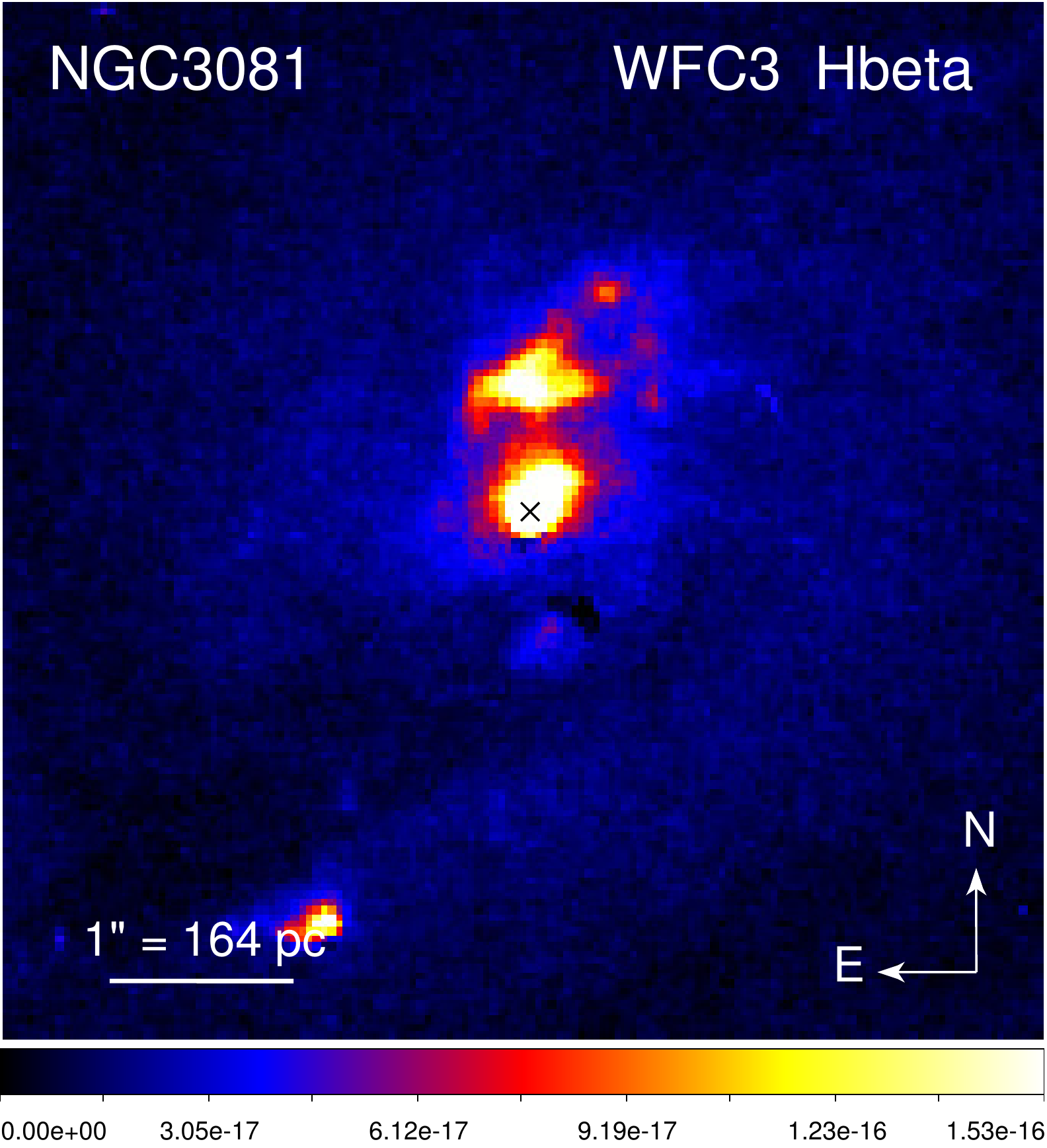}
\caption{NGC 3081 5.6$\arcsec$$\times$5.6$\arcsec$ continuum-subtracted emission line maps centered on the nucleus (black cross). All the images have a pixel scale of 0.04$\arcsec$. The colorbar denotes the surface brightness in the units of erg cm$^{-2}$ s$^{-1}$ pixel$^{-1}$. }
\label{NGC3081_linemap}
\end{figure*}

NGC 3081 ($z$ = 0.00799, $D$ = 34.4 Mpc, 1$\arcsec$ = 164 pc) is a well-known barred spiral galaxy with multiple rings (e.g., \citealt{Ferruit2000,Buta2004}). Extended ionized gas is found within $\sim$ 1 kpc of the nucleus \citep{Durret1986,Ferruit2000}. Figure \ref{NGC3081_linemap} shows the continuum-subtracted line maps within the central 1 kpc $\times$ 1 kpc region. The line-emitting region mainly consists of a central component, a northern component, and a fainter southern component, all distributed along the N-S direction forming an ionization bicone. A loop feature is attached to the northern component in the north-west direction, $\sim$60$^{\circ}$ away from the bicone axis. Since the [O\III] and H$\alpha$ images were taken with different chips on WFPC2, the loop is better resolved into small hot spots in H$\alpha$ on PC1, which has a higher resolution than [O\III] on WF2. The loop feature also appears in the H$\beta$ image and is less obvious in [S\II]. Such a feature has also been discovered in other Seyfert galaxies, e.g., IC 5063 (Maksym et al. in prep). 

\begin{figure*} 
\centering
\includegraphics[width=9.5cm]{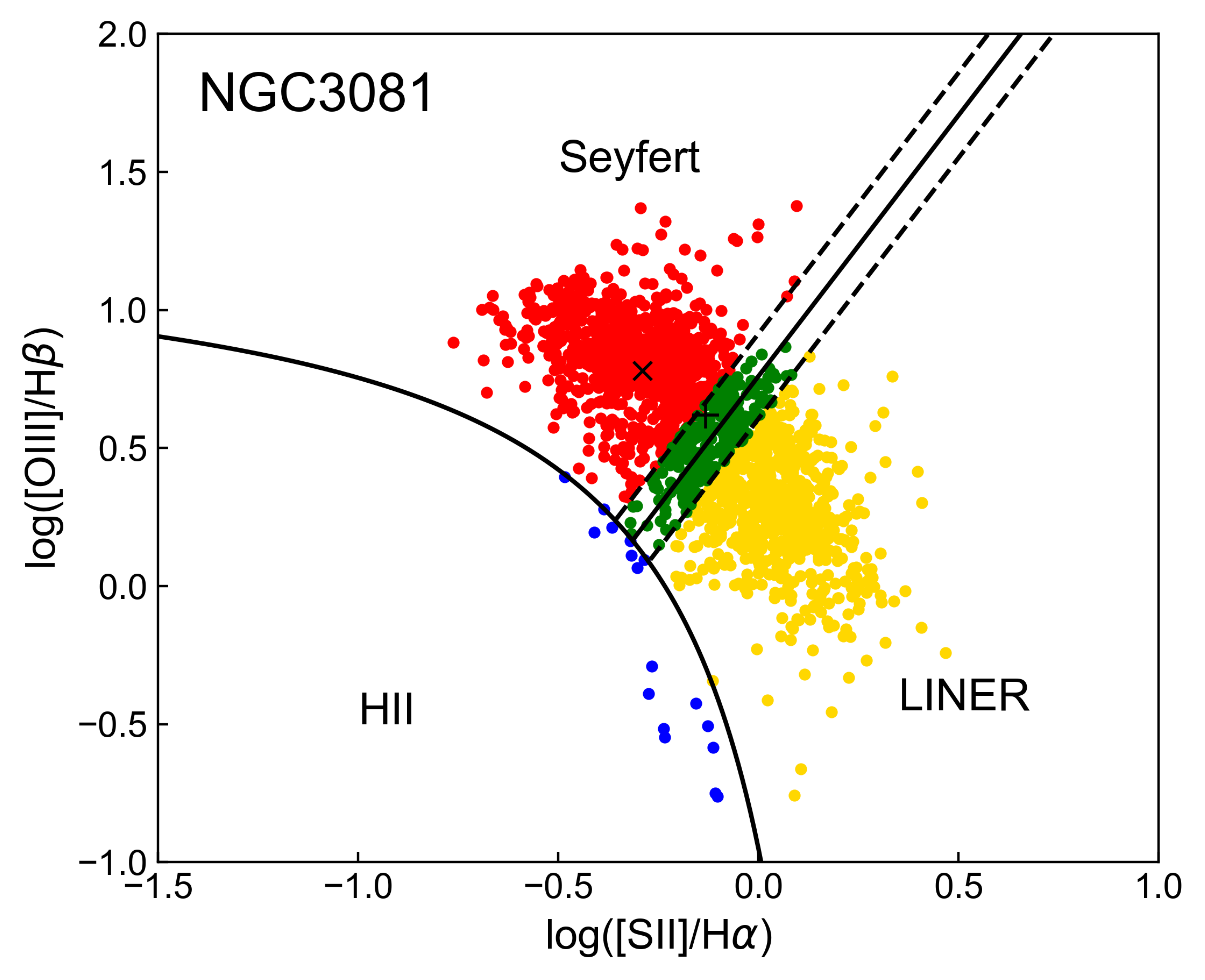} 
\includegraphics[width=7.7cm]{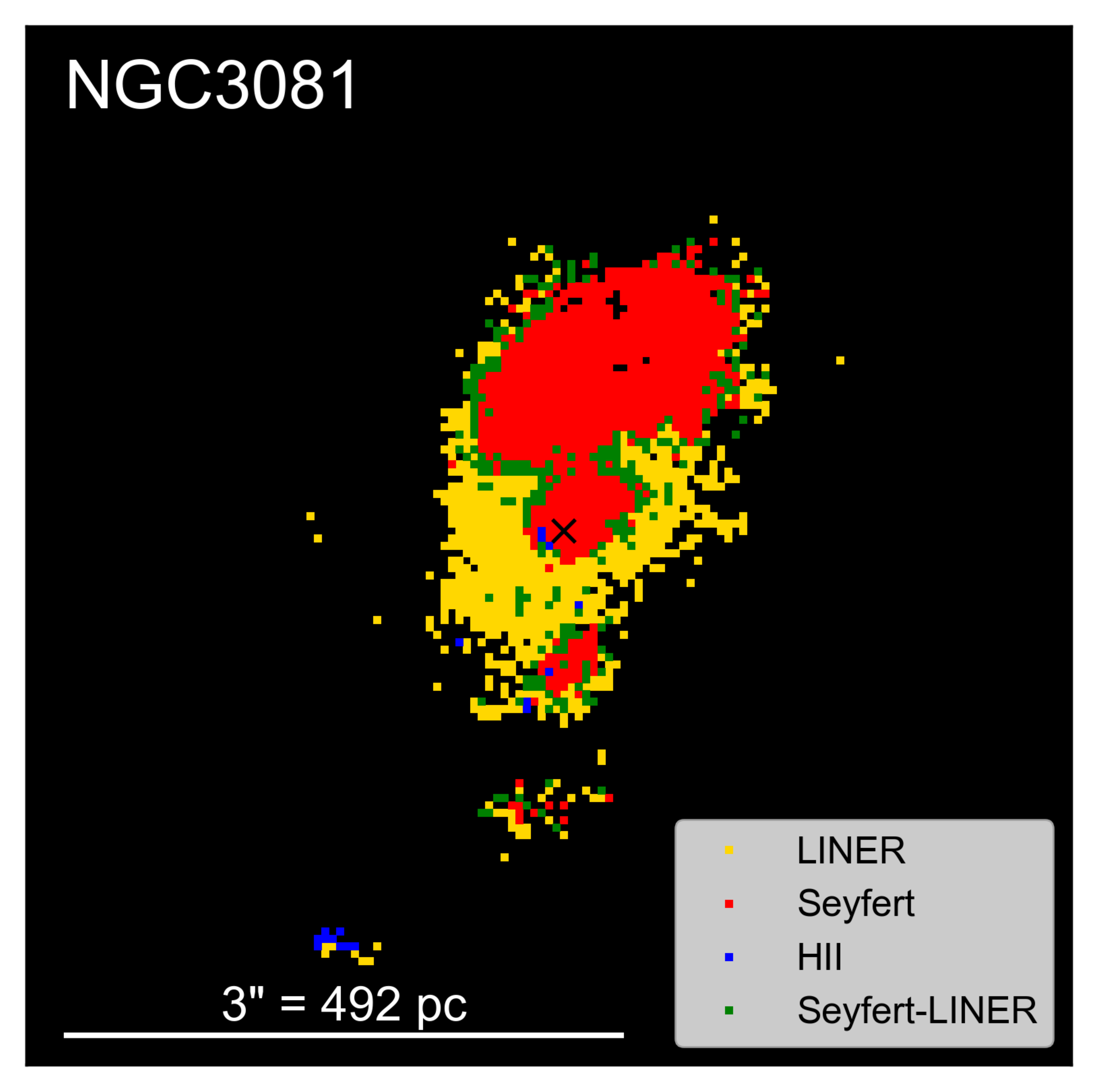}
\caption{{\bf Left}: BPT diagram of NGC 3081. The dividing lines/curves between different excitation mechanisms are defined in \cite{Kewley2006}. Red corresponds to Seyfert-like activity, yellow denotes LINER-like activity, and the transition zone is coded as green. Blue represents line ratios typical of H\II{} regions. Only pixels detected above 3$\sigma$ in all lines are used. Most H\II{} pixels are excluded by this criterion. The black cross marks the line flux ratios measured for the nuclear region ($r \leq 0.32\arcsec$). {\bf Right}: Spatially resolved BPT map (5.6$\arcsec$ $\times$ 5.6$\arcsec$) with each pixel color-coded according to the BPT type as shown in the left panel. The black cross marks the nucleus. Black pixels have at least one line with $F_{\rm line}$ $<$ 3$\sigma$. The 3$\arcsec$ white bar represents the diameter of the SDSS 3$\arcsec$ fiber. We calculate the BPT line ratios ('+' in the left panel) using the integrated line fluxes within a 3$\arcsec$ diameter circular aperture centered on the nucleus to mimic SDSS observations. }
\label{NGC3081_BPT}
\end{figure*}

Based on the continuum-subtracted line maps, we conducted BPT mapping and the results are shown in Figure \ref{NGC3081_BPT}. The nucleus and the ionization cone show red Seyfert activity and are well enclosed by a LINER cocoon through a thin, green layer of Seyfert-LINER transition. A small, blue H\II{} region, where pixels meet the 3$\sigma$ criterion, is located 2.5$\arcsec$ from the nucleus in the southeast direction, as can be seen in the H$\alpha$ and H$\beta$ line maps in Figure \ref{NGC3081_linemap}. It is part of a star-forming region in the circumnuclear spiral arms \citep{Buta2004}.

\subsection{NGC 7212}

\begin{figure*} 
\centering
\includegraphics[width=8cm]{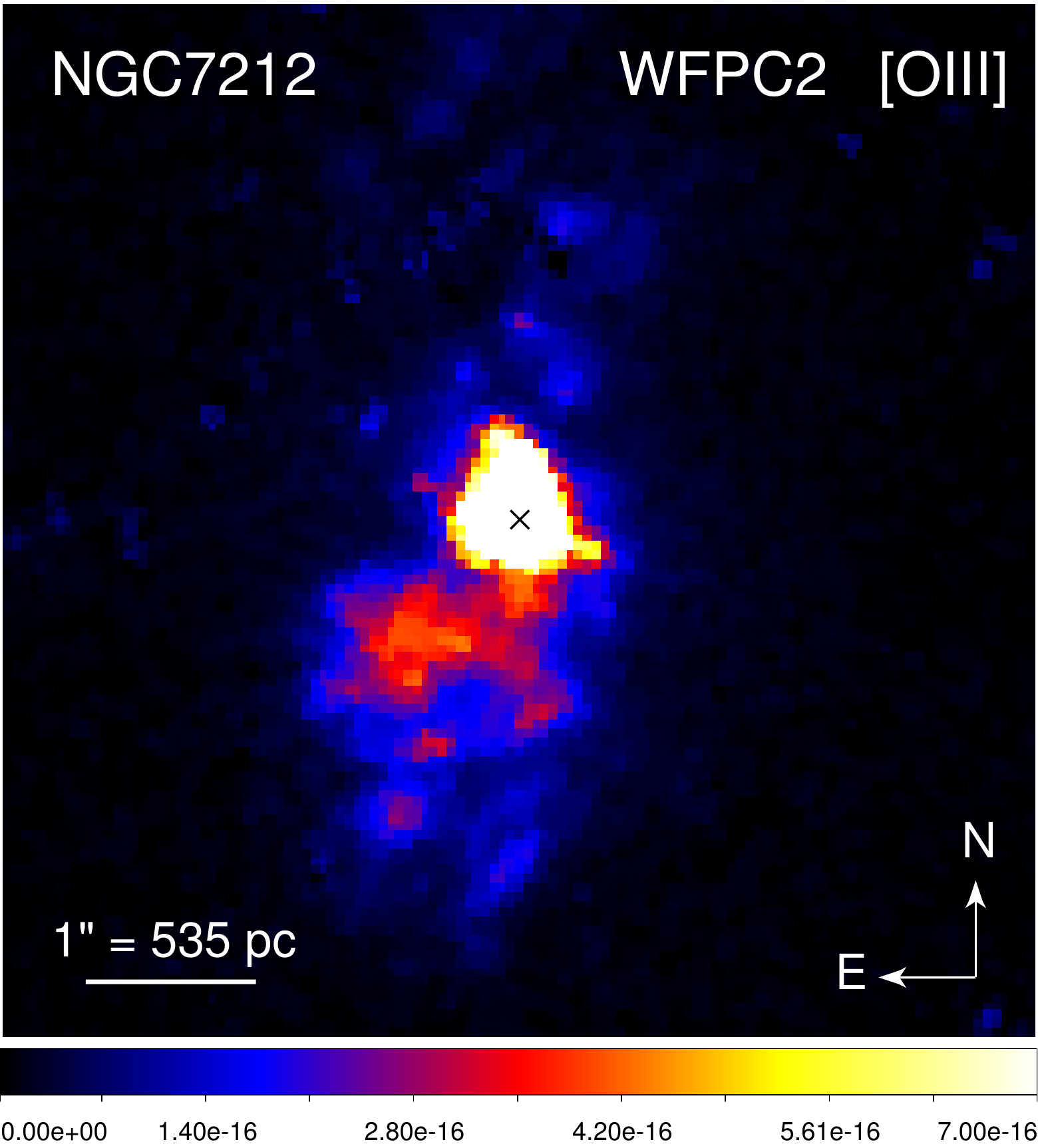}
\includegraphics[width=8cm]{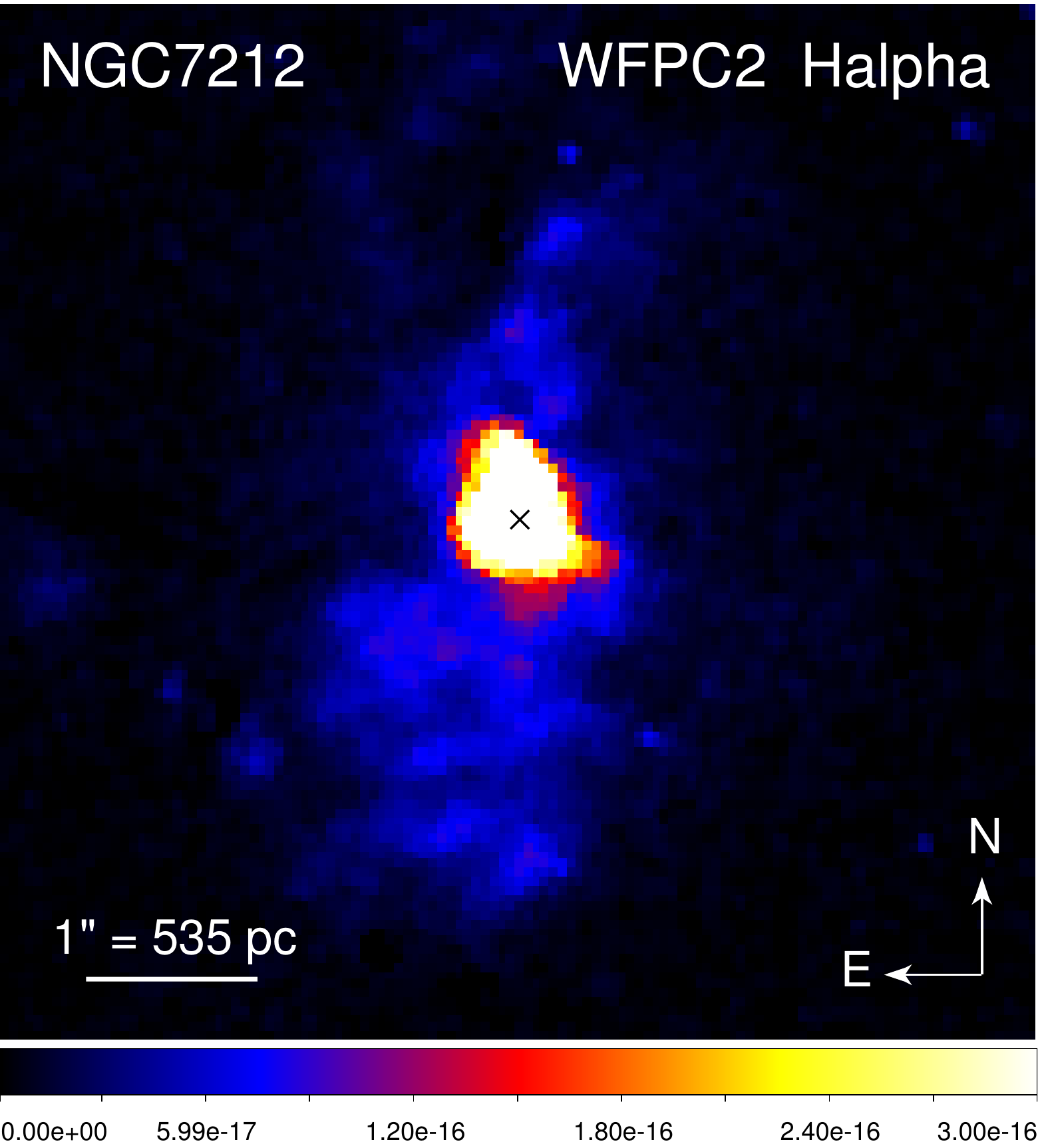}
\includegraphics[width=8cm]{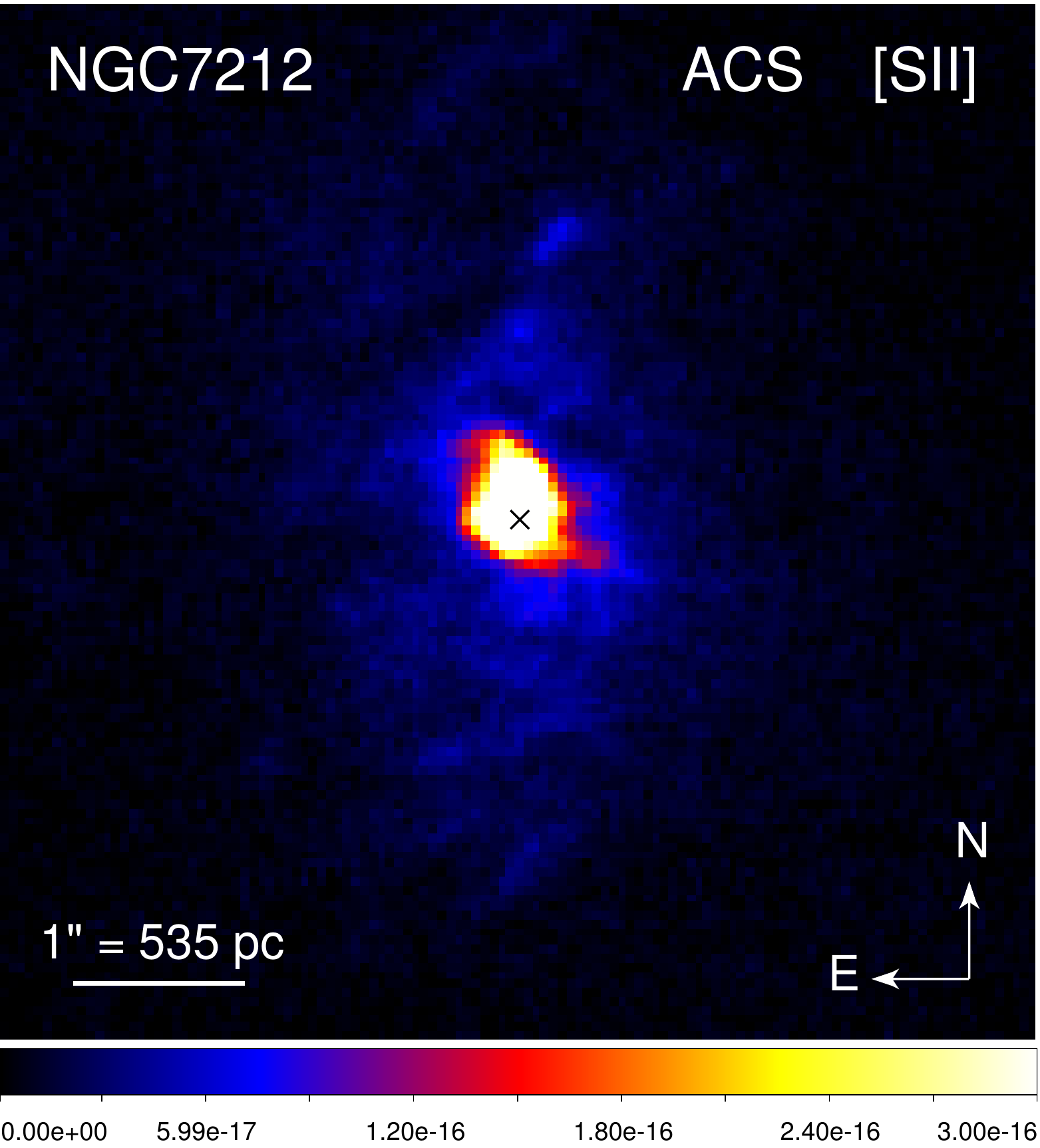}
\includegraphics[width=8cm]{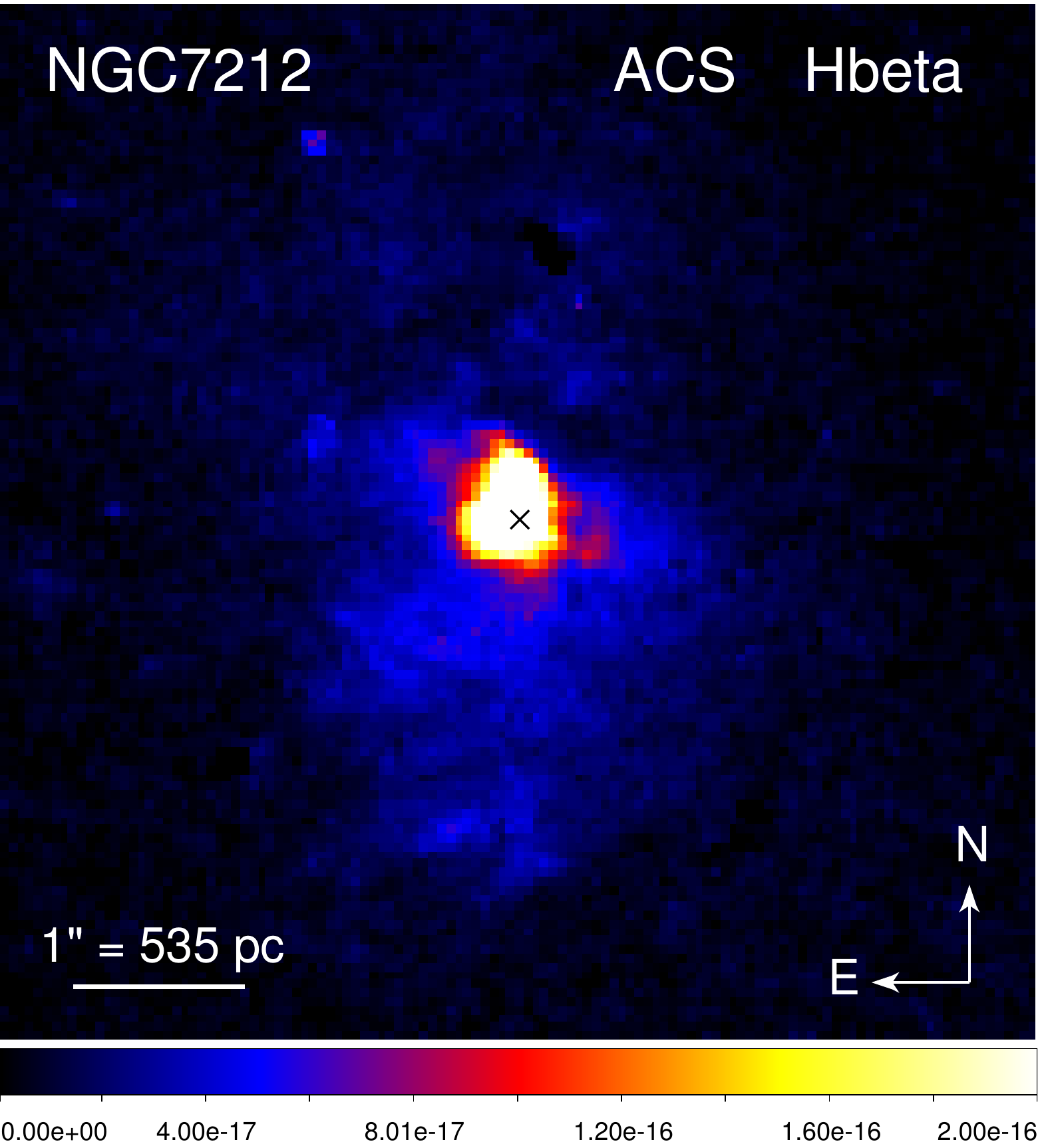}
\caption{NGC 7212 6$\arcsec$$\times$6$\arcsec$ continuum-subtracted emission line maps centered on the nucleus (black cross). All the images have a pixel scale of 0.05$\arcsec$. The colorbar denotes the surface brightness in the units of erg cm$^{-2}$ s$^{-1}$ pixel$^{-1}$.  }
\label{NGC7212_linemap}
\end{figure*}

NGC 7212 ($z$ = 0.0266, $D$ = 116.2 Mpc, 1$\arcsec$ = 535 pc) is a spiral galaxy belonging to a group of three interacting galaxies (e.g., \citealt{Wasilewski1981,Falcke1998}). Figure \ref{NGC7212_linemap} shows the continuum-subtracted line maps in the central $\sim$3 kpc $\times$ 3 kpc region. The extended, asymmetric, diffuse narrow line region is elongated in the northwest-southeast direction with a strong central component. The WFPC2 observations of [O\III] and H$\alpha$ are limited by the small field of view of the LRF. \cite{Cracco2011} confirm the presence of an extended ionization bicone up to 3.6 kpc from the nucleus with ground-based observations with a large field of view. There are multiple dust lanes across the galaxy. The dark region northwest of the nucleus is due to a sharp dust lane. The VLA map of NGC 7212 in \cite{Falcke1998} reveals a compact double radio source separated by 0.7$\arcsec$ in the N-S direction, offset from the optical nuclear emission peak. 

\begin{figure*} 
\centering
\includegraphics[width=9.5cm]{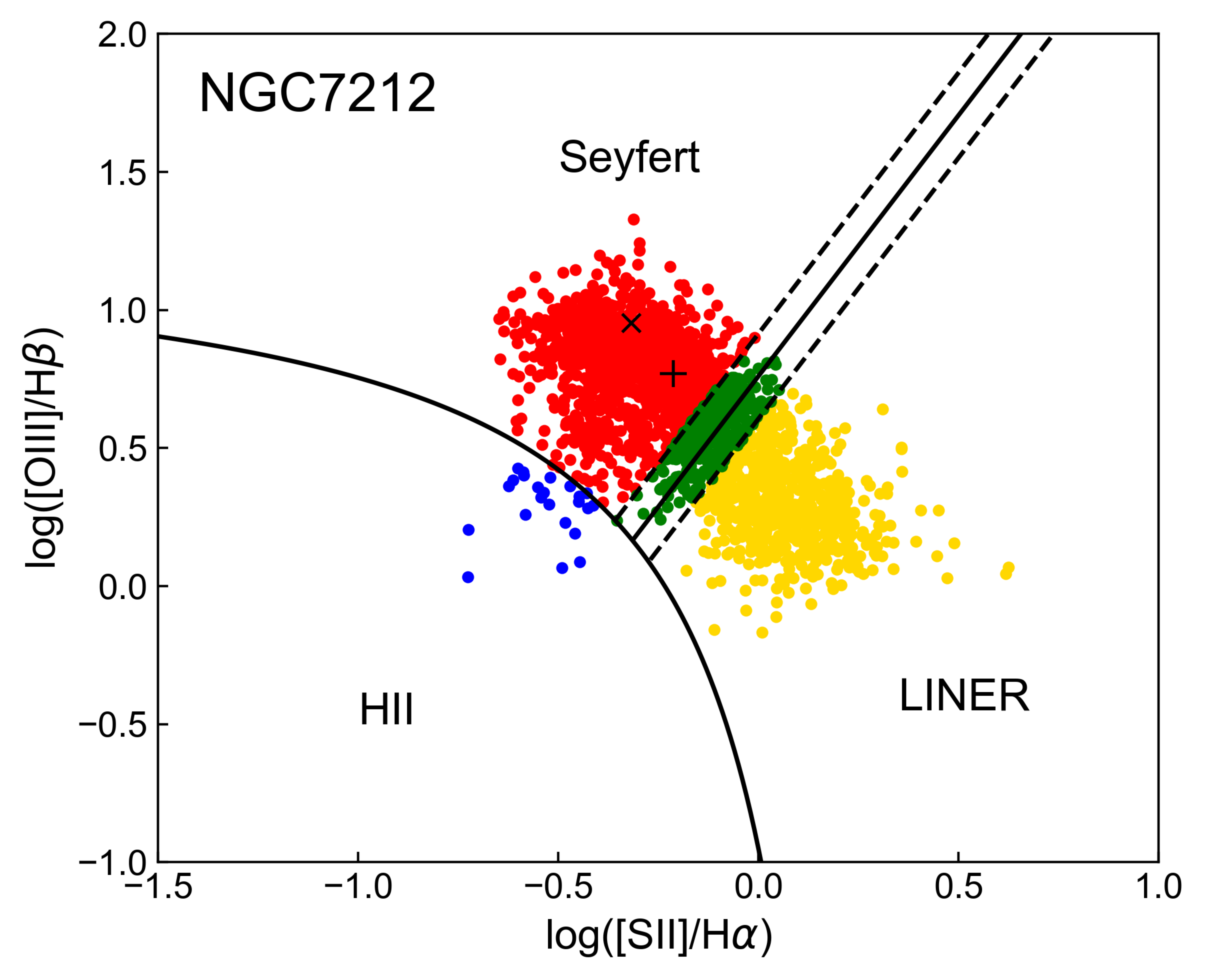} 
\includegraphics[width=7.7cm]{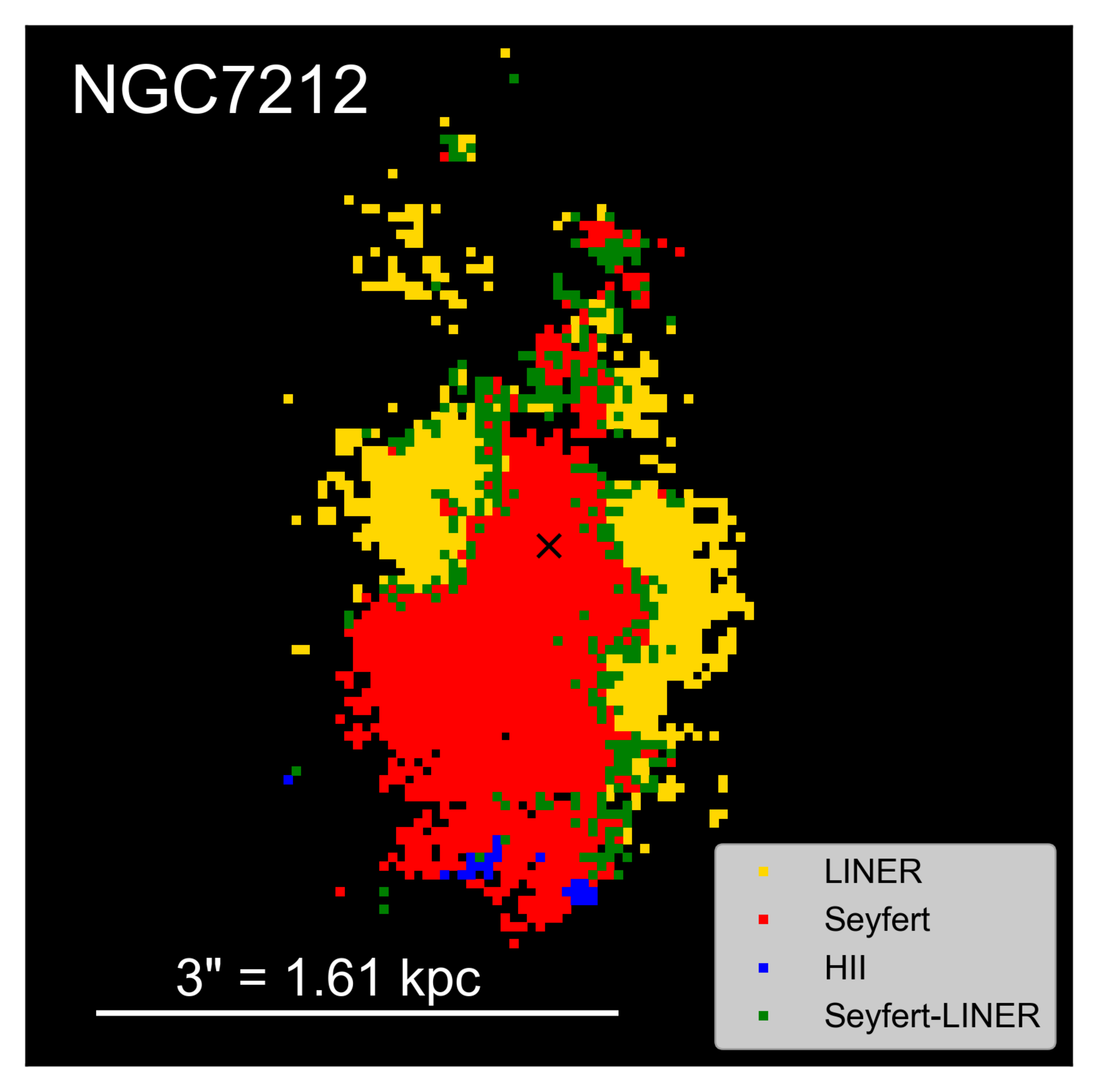}
\caption{{\bf Left}: BPT diagram of NGC 7212. The dividing lines/curves between different excitation mechanisms are defined in \cite{Kewley2006}. Red corresponds to Seyfert-like activity, yellow denotes LINER-like activity, and the transition zone is coded as green. Blue represents line ratios typical of H\II{} regions. Only pixels detected above 3$\sigma$ in all lines are used. Most H\II{} pixels are excluded by this criterion. The black cross marks the line flux ratios measured for the nuclear region ($r \leq 0.3\arcsec$). {\bf Right}: Spatially resolved BPT map (6$\arcsec$$\times$6$\arcsec$) with each pixel color-coded according to the BPT type as shown in the left panel. The black cross marks the nucleus. Black pixels have at least one line with $F_{\rm line}$ $<$ 3$\sigma$. The 3$\arcsec$ white bar represents the diameter of the SDSS 3$\arcsec$ fiber. We calculate the BPT line ratios ('+' in the left panel) using the integrated line fluxes within a 3$\arcsec$ diameter circular aperture centered on the nucleus to mimic SDSS observations. }
\label{NGC7212_BPT}
\end{figure*}

The ionization cone is again dominated by red Seyfert activity as shown in the resolved BPT map (Figure \ref{NGC7212_BPT}). A LINER cocoon is enveloping the Seyfert ionization cone along the eastern and western sides. A small number of pixels in the H\II{} region of the BPT diagram (left panel) are located in the southern end of the ionization cone (right panel), where stars can form. These star-forming clumps appear to be adjacent to the red Seyfert regions without a LINER layer in between. Such star-forming clumps have also been found in NGC 5643 \citep{Cresci2015}, which are H$\alpha$ bright clumps. Most Seyfert regions are still surrounded by a LINER cocoon, and star-forming regions are prevalent in the outer part \citep{Cresci2015}. 

Using the Magellan Echellette Spectrograph on the Clay telescope, \cite{Congiu2017} extracted spectra from different regions along the slit (PA = 16$^{\circ}$) and constructed BPT diagrams to diagnose ionization mechanisms in each region. Most points are clustered in the Seyfert region of the BPT diagrams. The main ionization mechanism in the NLR/ENLR is photoionization by the AGN. Some points are located close to the LINER region, which they explain as a consequence of shocks combined with velocity information. We cannot directly compare our BPT results region by region to theirs though, because their slit only covers a tiny portion of the ionization structures and their region bin size is much larger ($>$ 1$\arcsec$). The overall dominance of Seyfert over LINER excitation is consistent with that derived from our maps.

\subsection{NGC 7674}

\begin{figure*} 
\centering
\includegraphics[width=8cm]{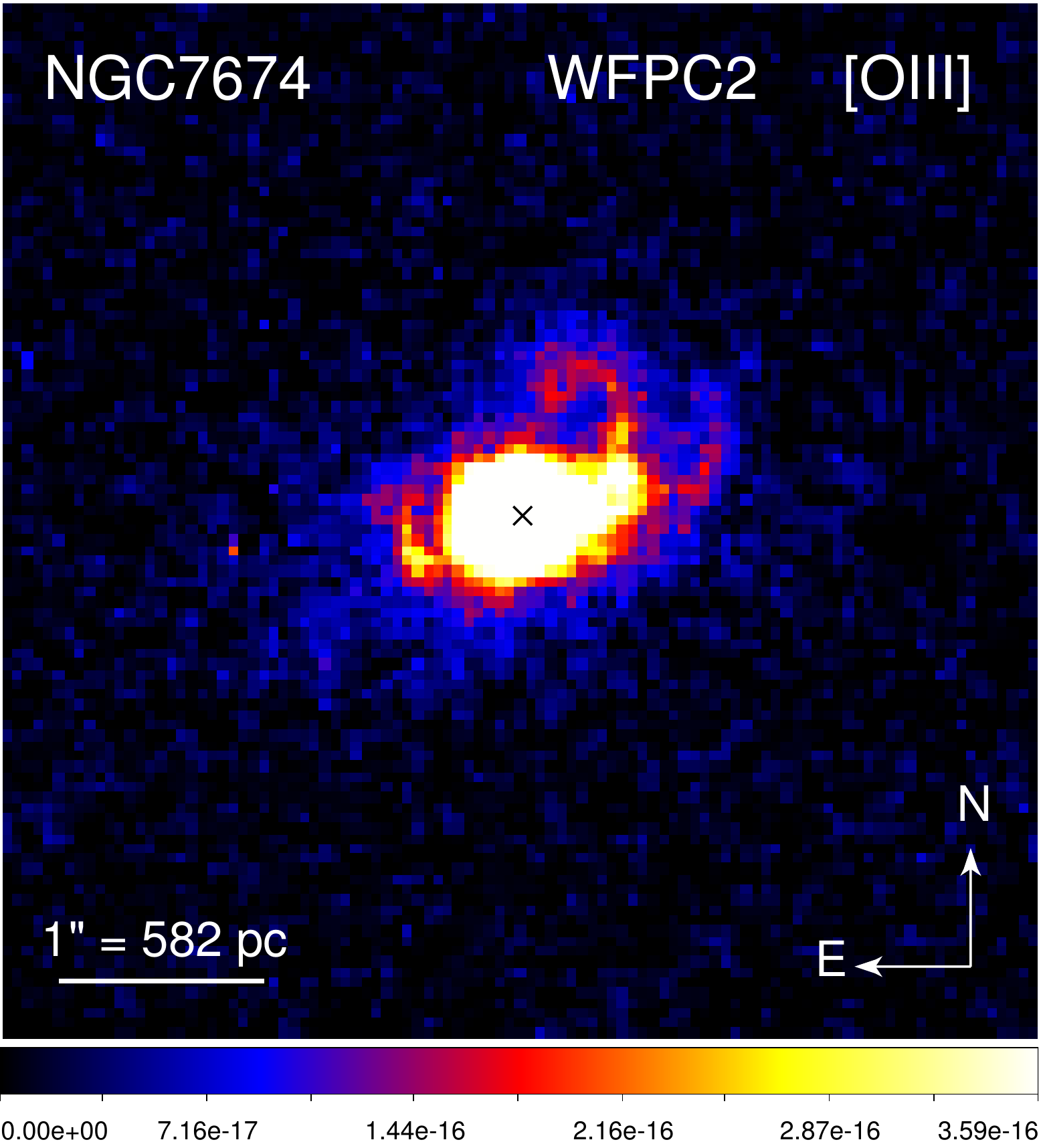}
\includegraphics[width=8cm]{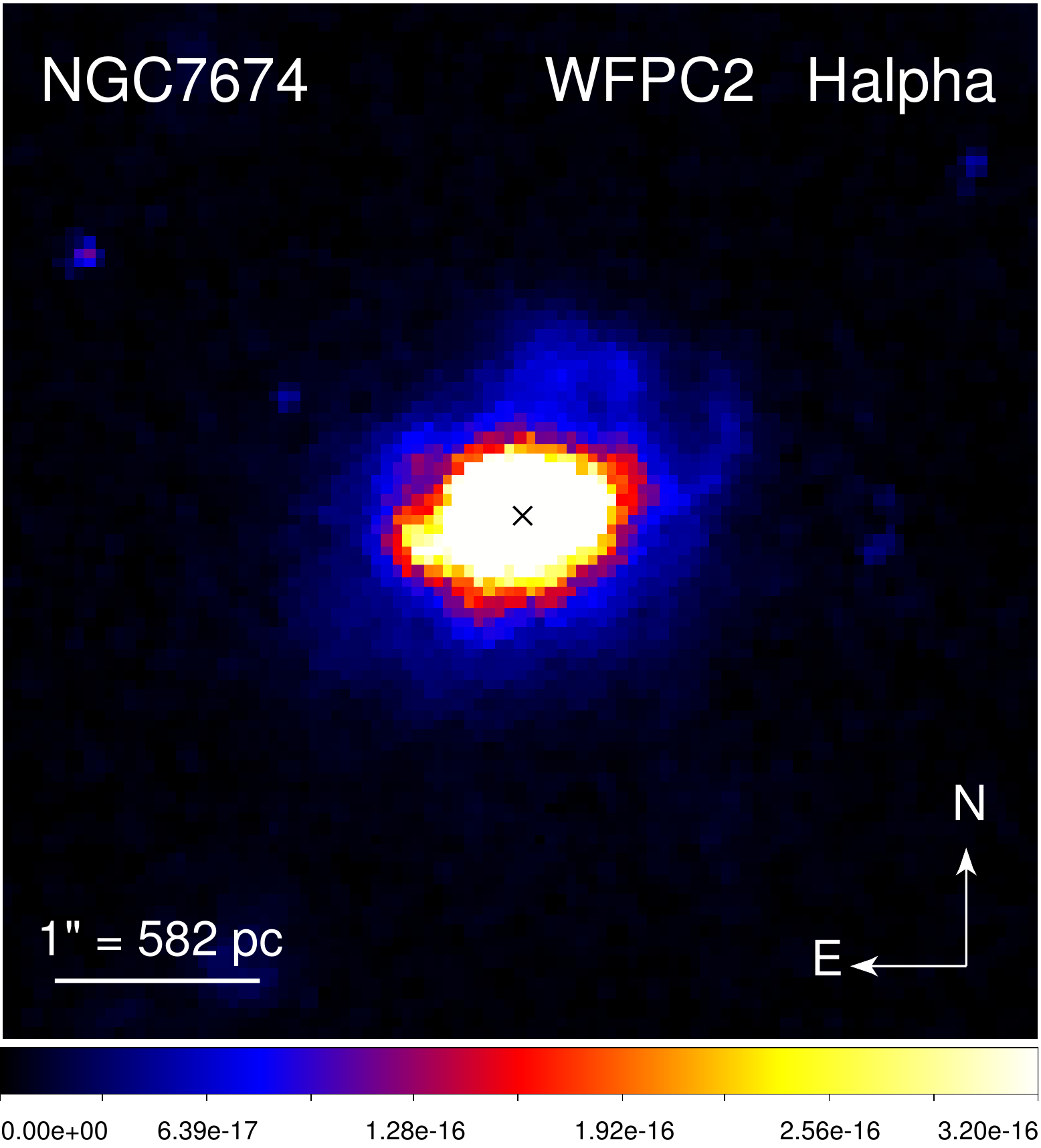}
\includegraphics[width=8cm]{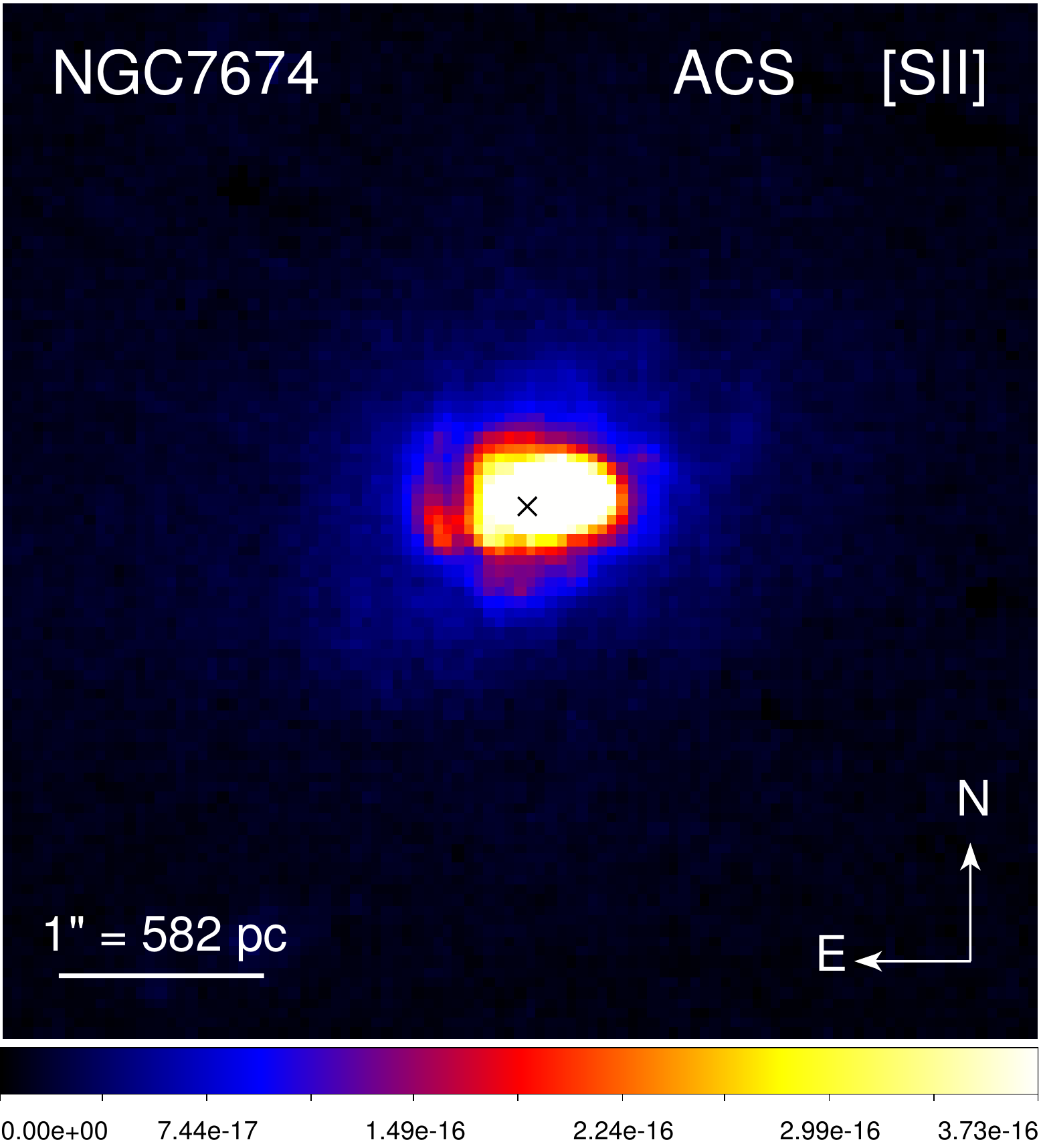}
\includegraphics[width=8cm]{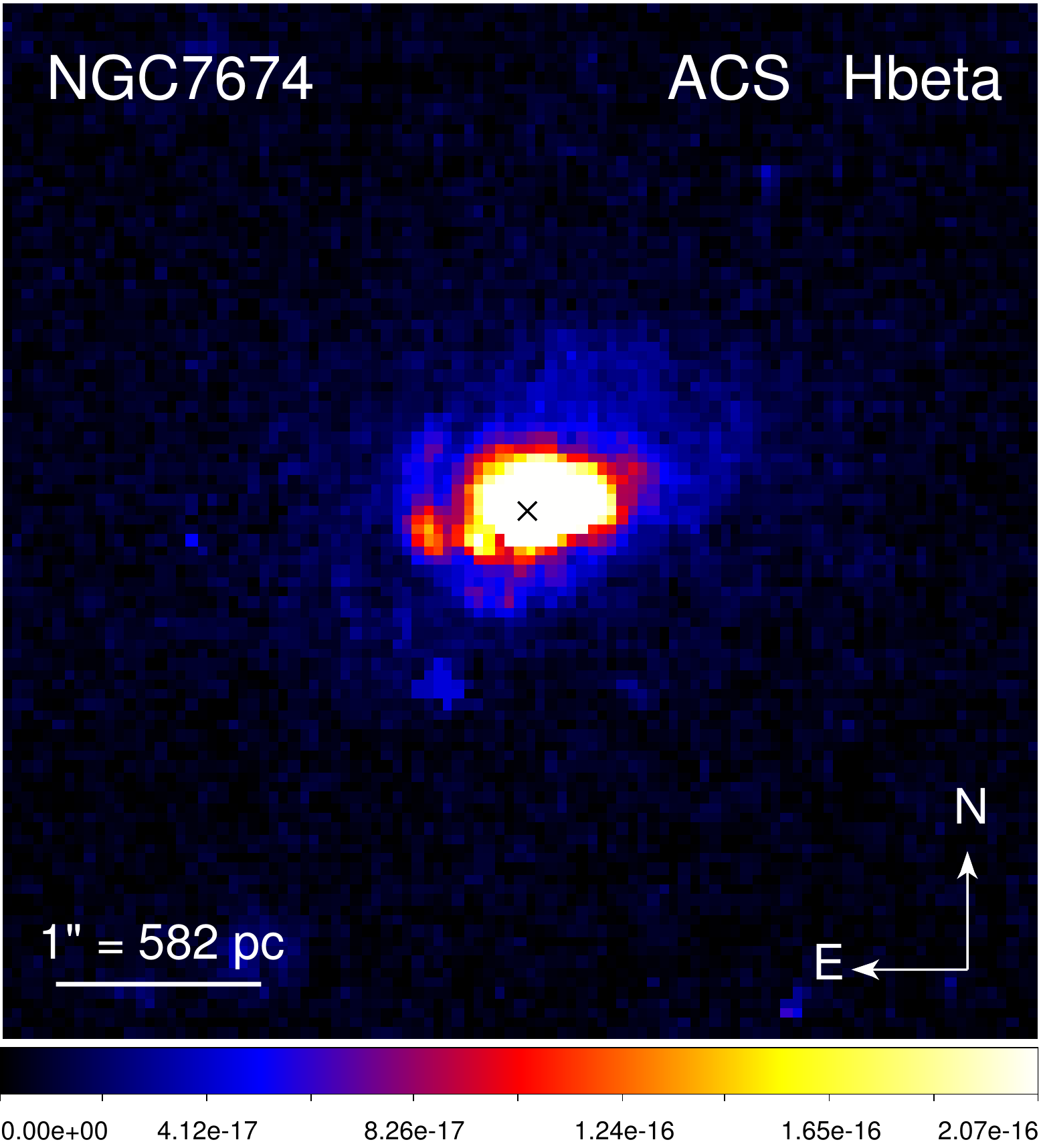}
\caption{NGC 7674 5$\arcsec$$\times$5$\arcsec$ continuum-subtracted emission line maps centered on the nucleus (black cross). All the images have a pixel scale of 0.05$\arcsec$. The colorbar denotes the surface brightness in the units of erg cm$^{-2}$ s$^{-1}$ pixel$^{-1}$.  }
\label{NGC7674_linemap}
\end{figure*}

NGC 7674 ($z$ = 0.02903, $D$ = 127.1 Mpc, 1$\arcsec$ = 582 pc) is the brightest member of a compact galaxy group \citep{Verdes1997,Schmitt2003}. Figure \ref{NGC7674_linemap} shows the continuum-subtracted line maps in the central $\sim$2.9 kpc $\times$ 2.9 kpc region. They exhibit a loop feature with a hole in emission northwest of the nucleus ($\sim$55$^{\circ}$ away from the cone axis), which is most prominent in the [O\III] image. One can also identify `arms' extending out from the nuclear region on both the eastern and western sides. NGC 7674 hosts a $\sim$0.7 kpc S-shaped radio jet \citep{Momijan2003}. Later \cite{Kharb2017} find evidence for the formation of the S-shaped radio structure as due to a black hole merger. 

\begin{figure*} 
\centering
\includegraphics[width=9.5cm]{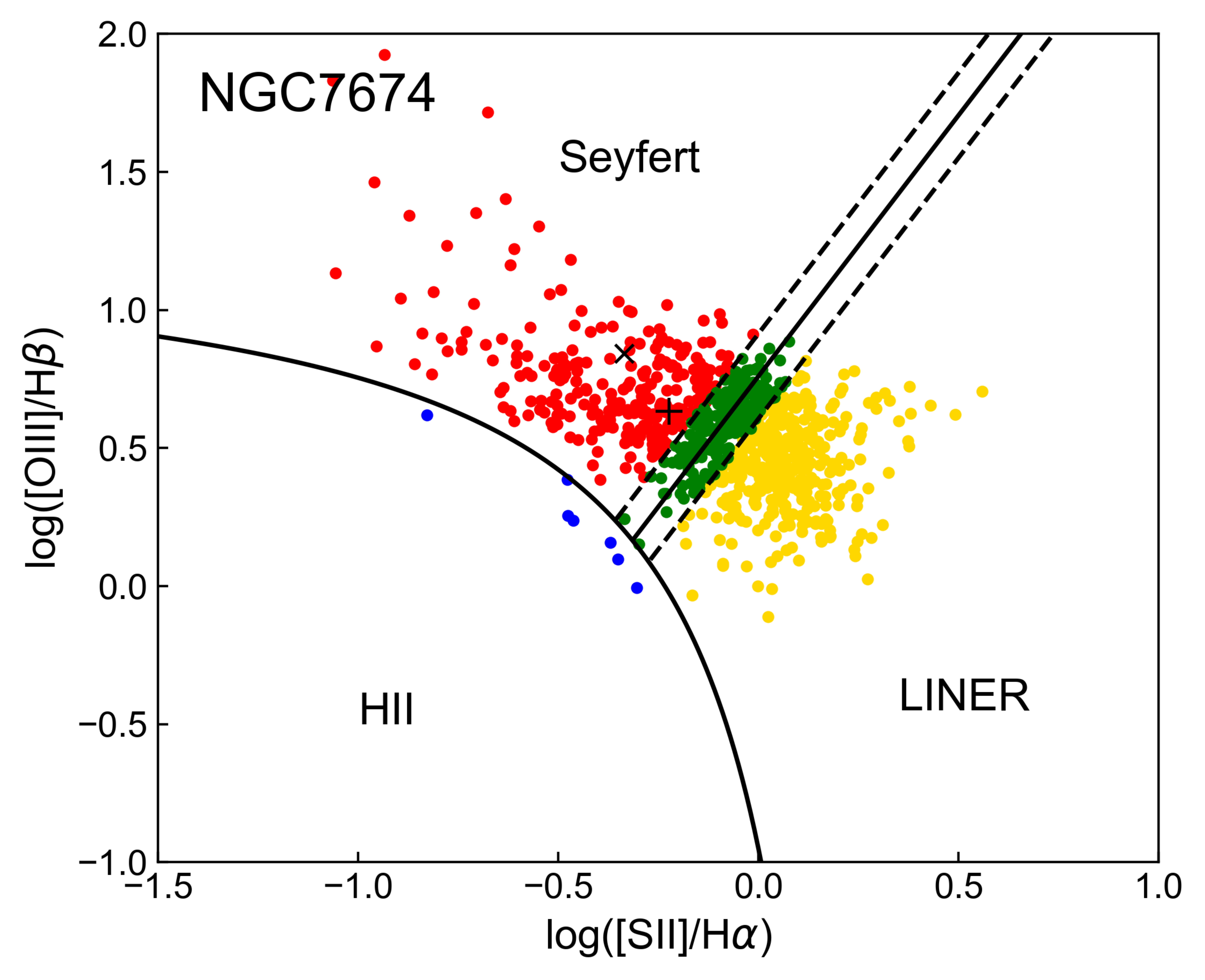} 
\includegraphics[width=7.7cm]{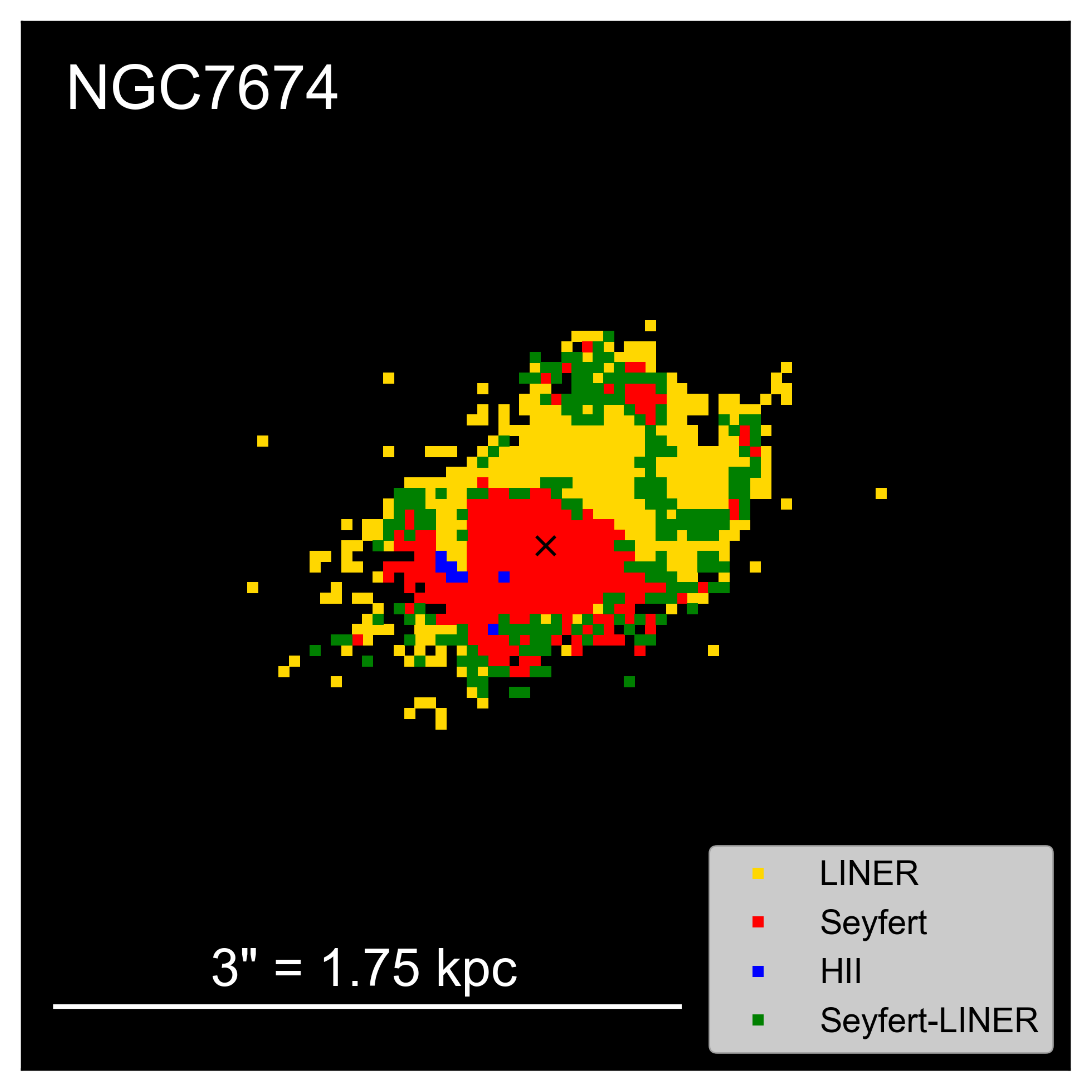} 
\caption{{\bf Left}: BPT diagram of NGC 7674. The dividing lines/curves between different excitation mechanisms are defined in \cite{Kewley2006}. Red corresponds to Seyfert-like activity, yellow denotes LINER-like activity, and the transition zone is coded as green. Blue represents line ratios typical of H\II{} regions. Only pixels detected above 3$\sigma$ in all lines are used. Most H\II{} pixels are excluded by this criterion. The black cross marks the line flux ratios measured for the nuclear region ($r \leq 0.35\arcsec$). {\bf Right}: Spatially resolved BPT map (5$\arcsec$ $\times$ 5$\arcsec$) with each pixel color-coded according to the BPT type as shown in the left panel. The black cross marks the nucleus. Black pixels have at least one line with $F_{\rm line}$ $<$ 3$\sigma$. The 3$\arcsec$ white bar represents the diameter of the SDSS 3$\arcsec$ fiber. We calculate the BPT line ratios ('+' in the left panel) using the integrated line fluxes within a 3$\arcsec$ diameter circular aperture centered on the nucleus to mimic SDSS observations. }
\label{NGC7674_BPT}
\end{figure*}

Based on the [O\III], H$\alpha$, [S\II], and H$\beta$ images, we created the BPT diagram and the resolved map (Figure \ref{NGC7674_BPT}). The nucleus and the eastern arm are dominated by red Seyfert activity. The loop and western arm show Seyfert or Seyfert-to-LINER type. The yellow LINER type permeates and surrounds the nucleus and those features. The blue H\II{} pixels are likely due to imperfections in the continuum subtraction rather than real star-forming regions, e.g., over-subtraction of H$\beta$ due to cosmic rays on top of the nuclear region of the continuum image.   

\subsection{NGC 2273}

\begin{figure*} 
\centering
\includegraphics[width=8cm]{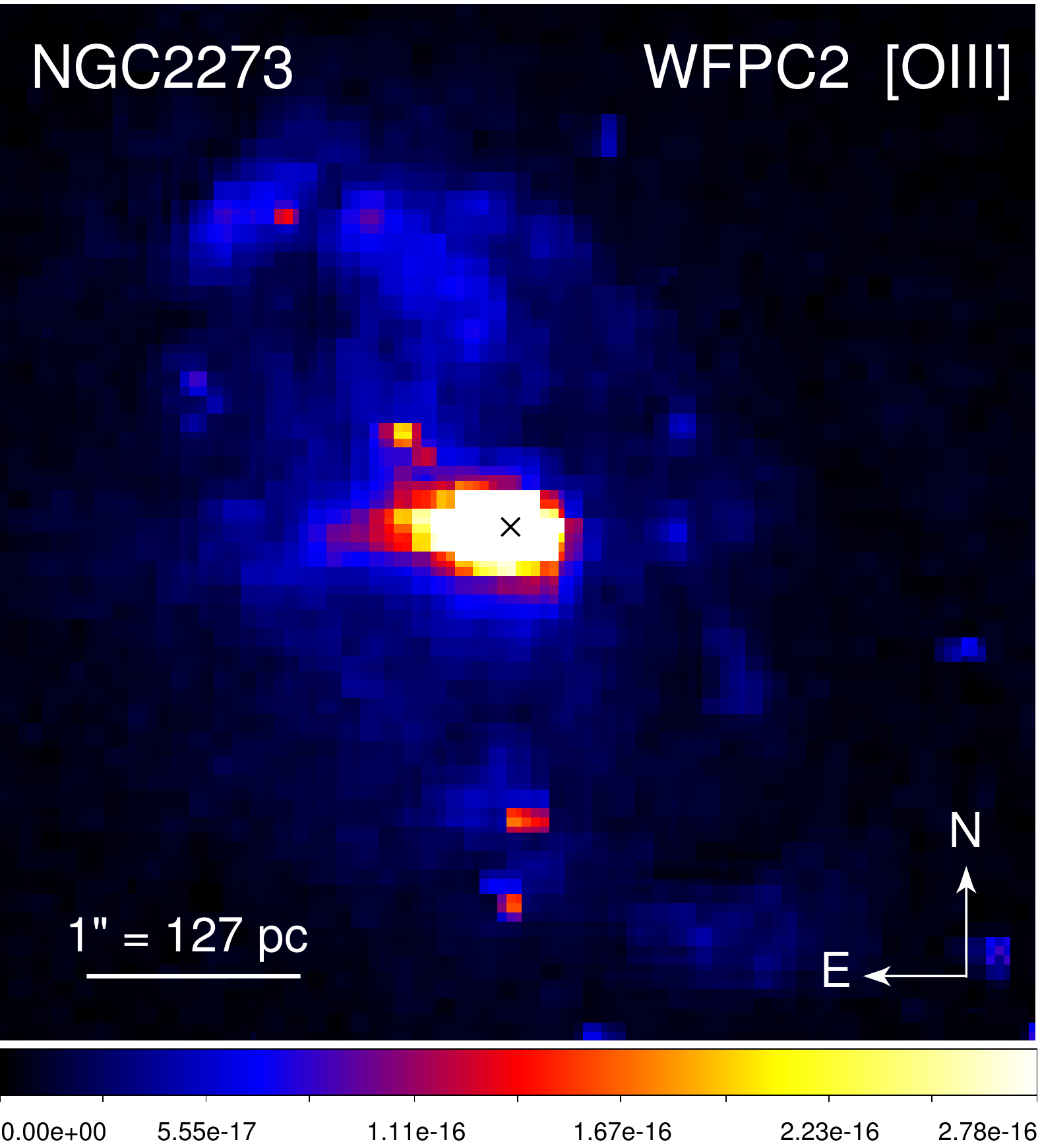}
\includegraphics[width=8cm]{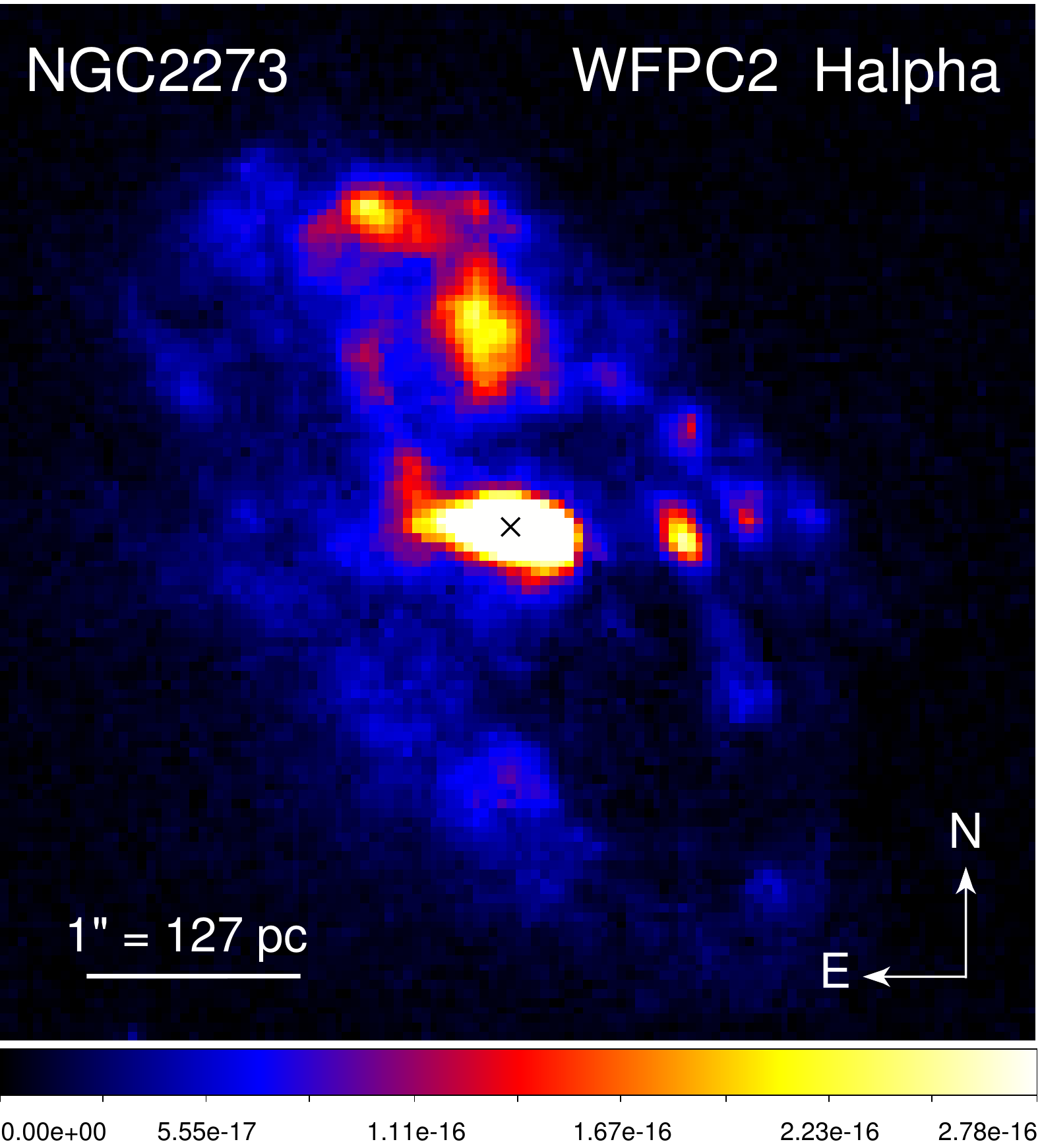}
\includegraphics[width=8cm]{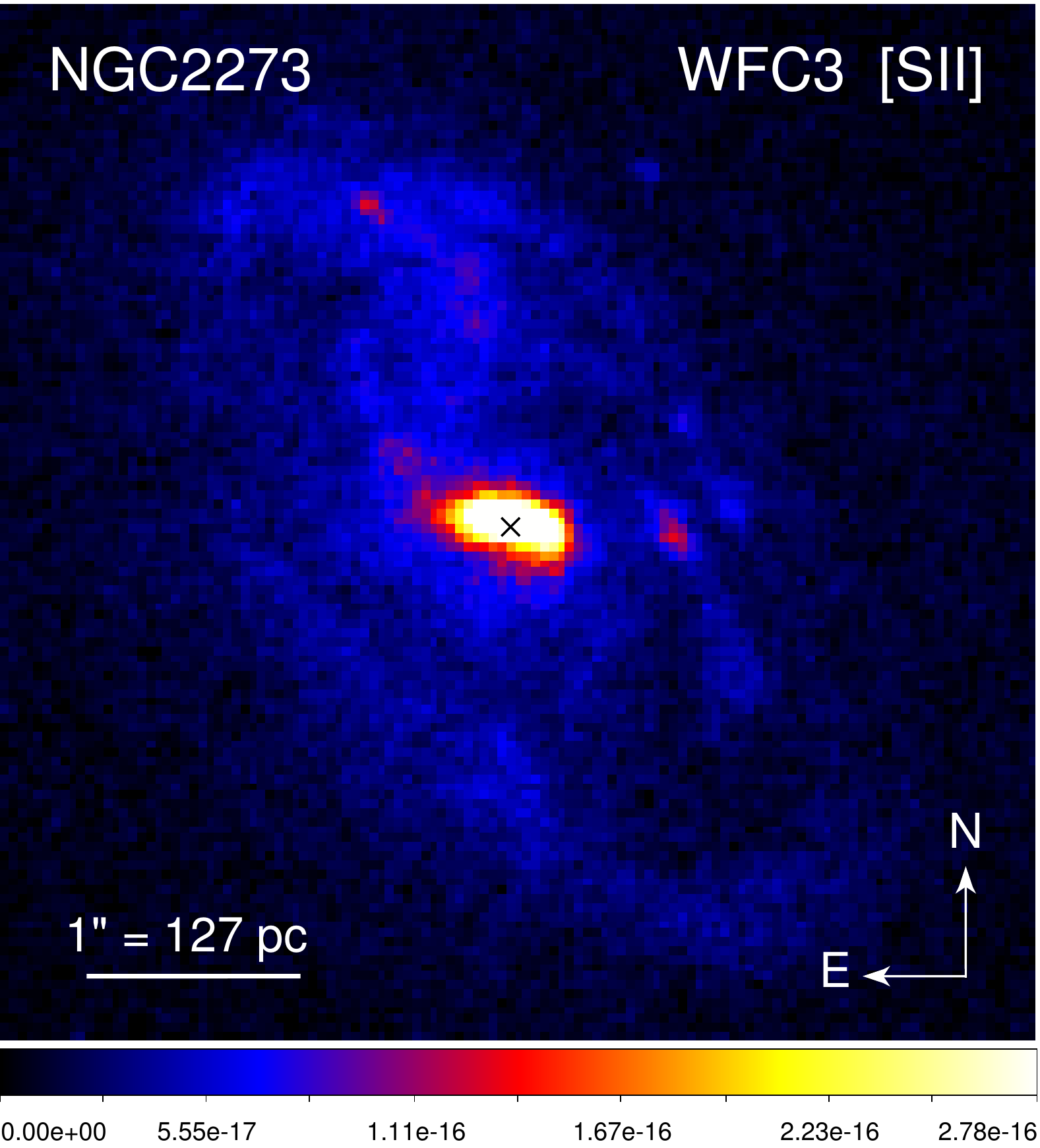}
\includegraphics[width=8cm]{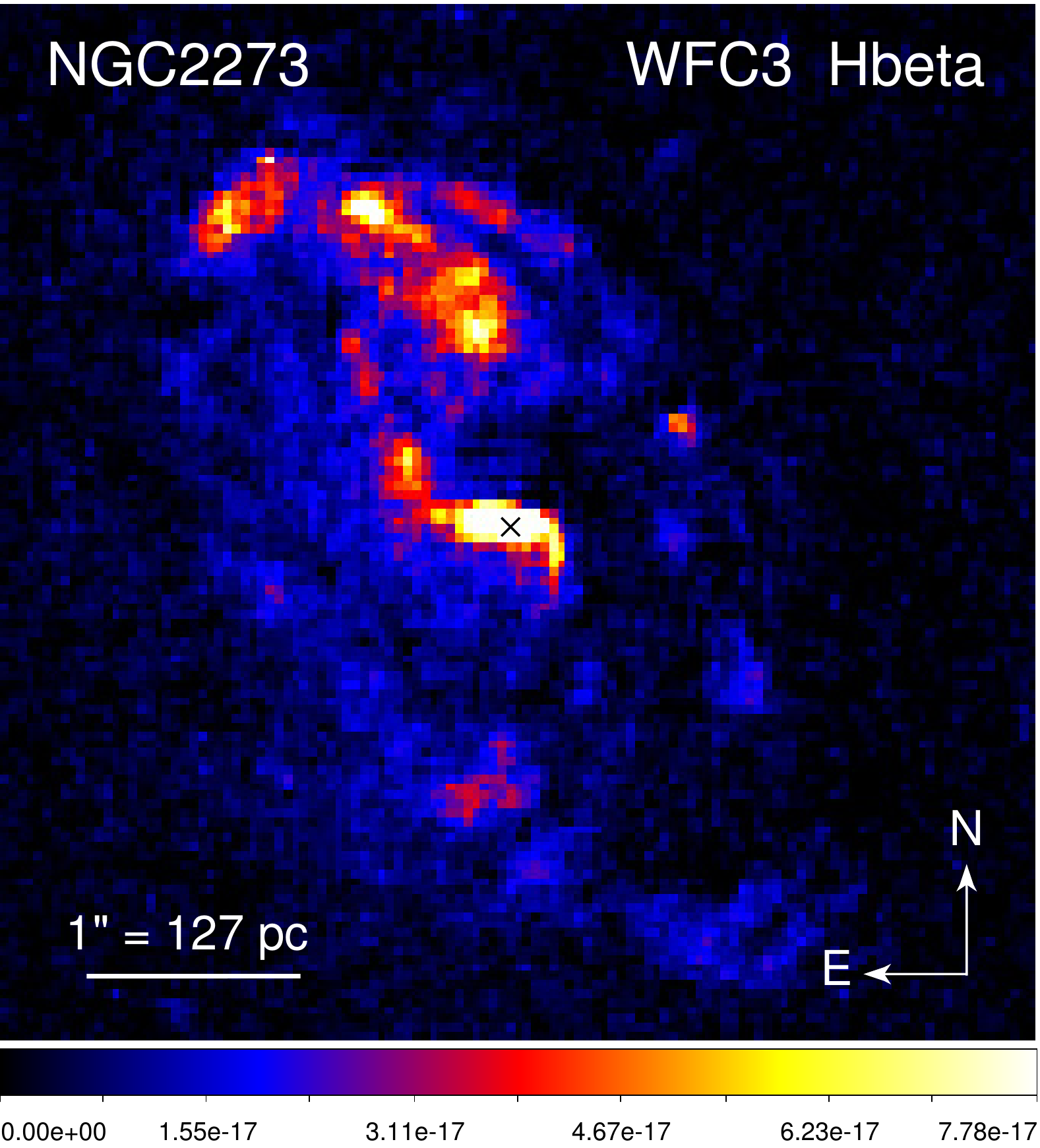}
\caption{NGC 2273 4.8$\arcsec$$\times$4.8$\arcsec$ continuum-subtracted emission line maps centered on the nucleus (black cross). All the images have a pixel scale of 0.04$\arcsec$. The colorbar denotes the surface brightness in the units of erg cm$^{-2}$ s$^{-1}$ pixel$^{-1}$. }
\label{NGC2273_linemap}
\end{figure*}

NGC 2273 ($z$ = 0.00614, $D$ = 26.4 Mpc, 1$\arcsec$ = 127 pc) is a ringed galaxy with an inner ring and two pseudo-rings in the outskirt  \citep{Aguerri1998,Ferruit2000}. Figure \ref{NGC2273_linemap} shows the continuum-subtracted line maps covering line-emitting regions out to the inner ring. The central region shows a jet-like bar structure. The H$\alpha$ and H$\beta$ images exhibit bright emission components in the north. \cite{Mundell2009} show that the radio emission from NGC 2273 consists of an E-W double source and fainter, extended emission to the northeast and southwest. The jet-like structure in [O\III] extends along the same E-W direction as the radio emission, suggesting interactions between the radio jet and the NLR gas. The diffuse radio emission could be a poorly collimated outflow wind from the nucleus \citep{Christopoulou1997,Mundell2009}.

\begin{figure*} 
\centering
\includegraphics[width=9.5cm]{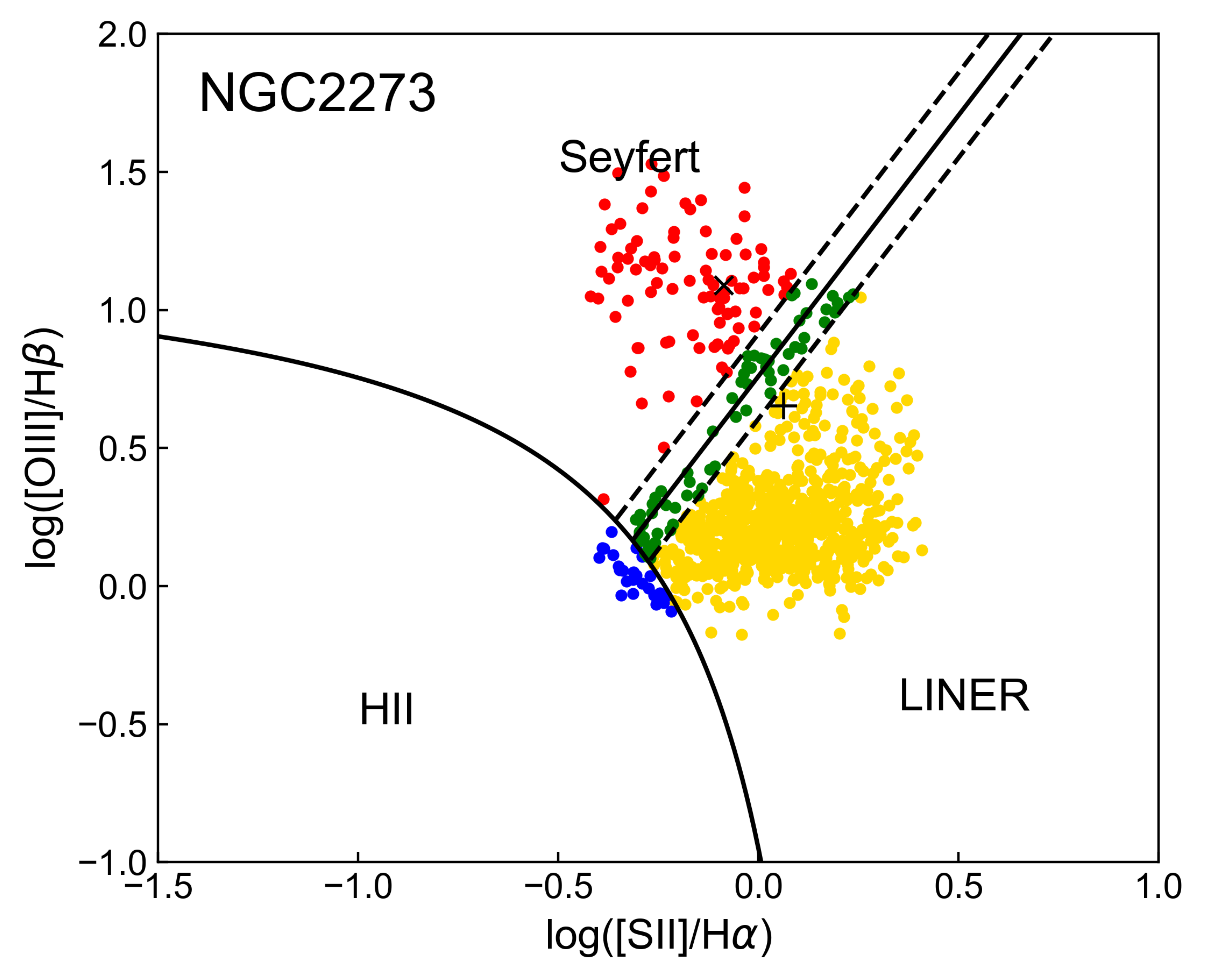} 
\includegraphics[width=7.7cm]{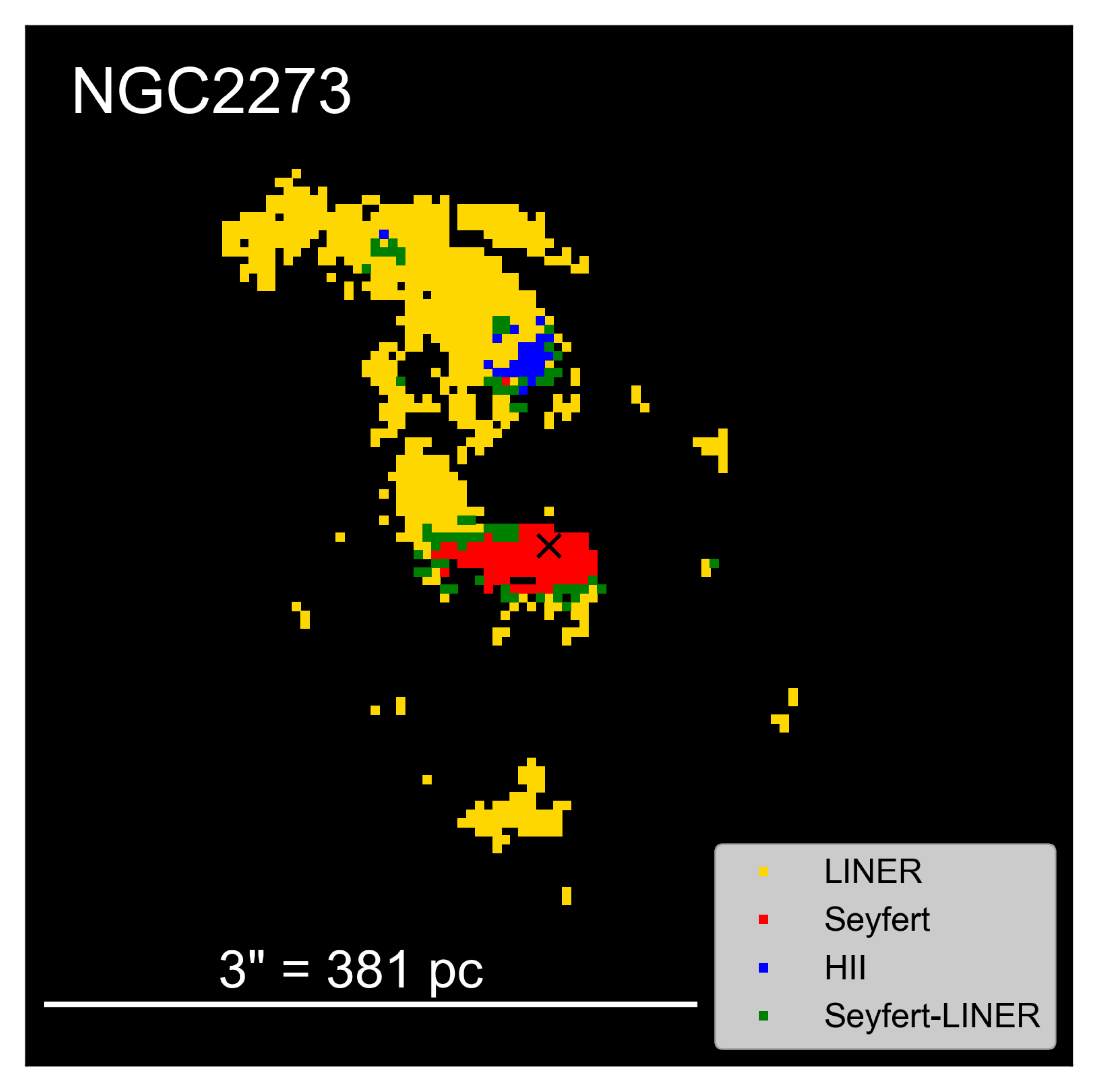}

\includegraphics[width=9.5cm]{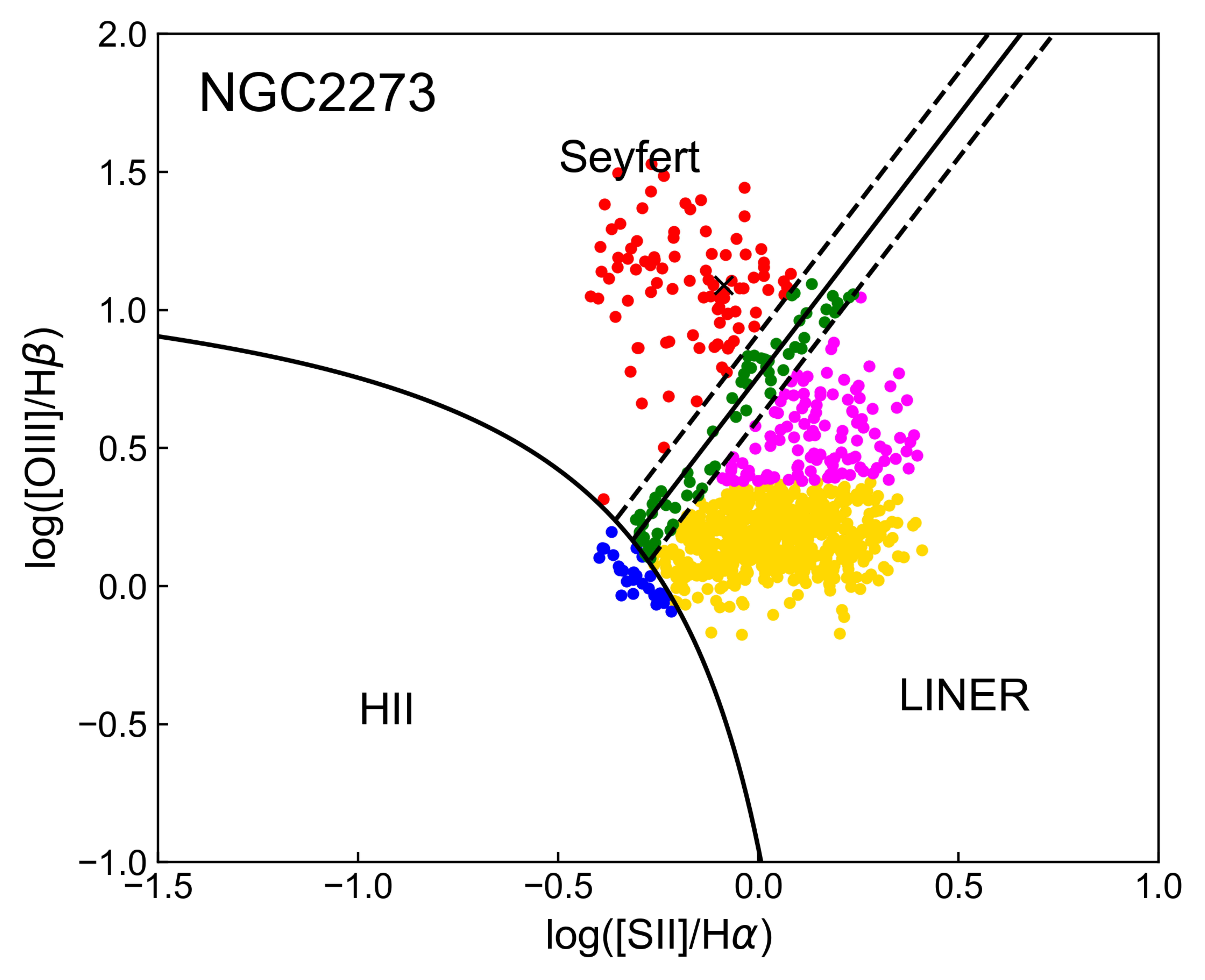} 
\includegraphics[width=7.7cm]{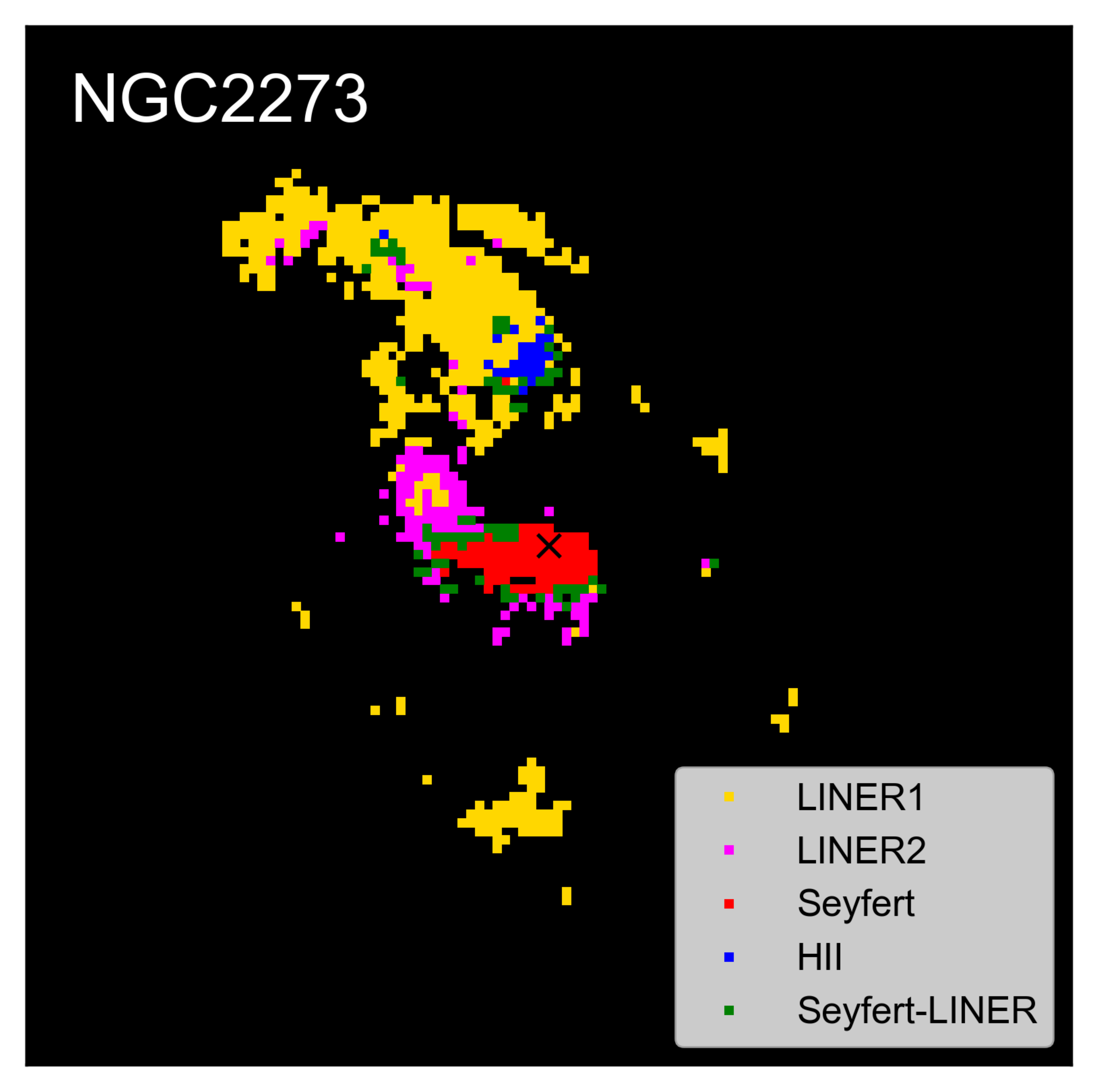}

\caption{{\bf Top Left}: BPT diagram of NGC 2273. The dividing lines/curves between different excitation mechanisms are defined in \cite{Kewley2006}. Red corresponds to Seyfert-like activity, yellow denotes LINER-like activity, and the transition zone is coded as green. Blue represents line ratios typical of H\II{} regions. Only pixels detected above 3$\sigma$ in all lines are used. Most H\II{} pixels are excluded by this criterion. The black cross marks the line flux ratios measured for the nuclear region ($r \leq 0.4\arcsec$). {\bf Top Right}: Spatially resolved BPT map (4.8$\arcsec$ $\times$ 4.8$\arcsec$) with each pixel color-coded according to the BPT type as shown in the left panel. The black cross marks the nucleus. Black pixels have at least one line with $F_{\rm line}$ $<$ 3$\sigma$. The 3$\arcsec$ white bar represents the diameter of the SDSS 3$\arcsec$ fiber. We calculate the BPT line ratios ('+' in the left panel) using the integrated line fluxes within a 3$\arcsec$ diameter circular aperture centered on the nucleus to mimic SDSS observations. {\bf Bottom}: We separate two groups of LINER regions by using different colors.  }
\label{NGC2273_BPT}
\end{figure*}

Figure \ref{NGC2273_BPT} displays the BPT diagram and the spatially resolved map. The nuclear region is dominated by red Seyfert activity and transitions to LINER type. A blue H\II{} region appears $\sim$0.8$\arcsec$ to the north of the nucleus. NGC 2273 has been found to host nuclear starburst activity in the central 2$\arcsec$ $\times$ 2$\arcsec$ region based on spectral signatures \citep{Gu2003}. 

The BPT diagram of NGC 2273 displays a special feature that is not seen in other sources. The pixels that satisfy the strong line criterion are distributed in two groups on the BPT diagram, a horizontal group and a diagonal group, suggesting that they may be associated with distinct structures in the resolved map. We further separate the LINER groups in the BPT diagram by using different colors (Figure \ref{NGC2273_BPT} bottom panels). Magenta points denote the pixels with $log$([O\III]/H$\beta$) $\geq$ 0.38, and yellow points belong to the horizontal group with $log$([O\III]/H$\beta$) $<$ 0.38. Indeed, the two groups are associated with different structures in the line-emitting region. Most magenta LINER-type pixels are adjacent to the central nuclear region, while most yellow LINER pixels are distributed farther out to the ring structure. We will discuss potentially different excitation mechanisms associated with these two distinct LINER groups in Section \ref{discussion}.

\subsection{NGC 5643}

\begin{figure*} 
\centering
\includegraphics[width=8cm]{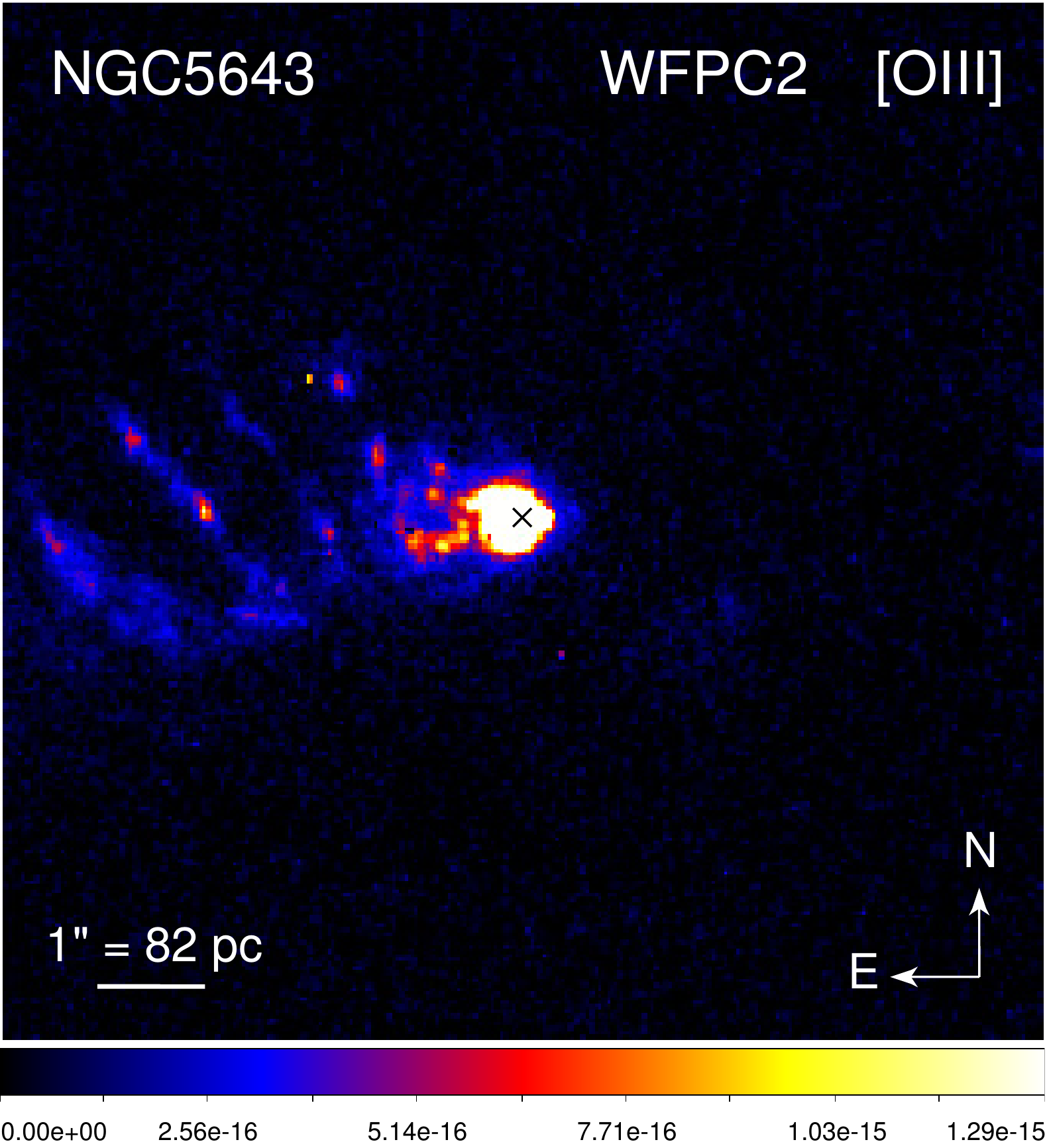}
\includegraphics[width=8cm]{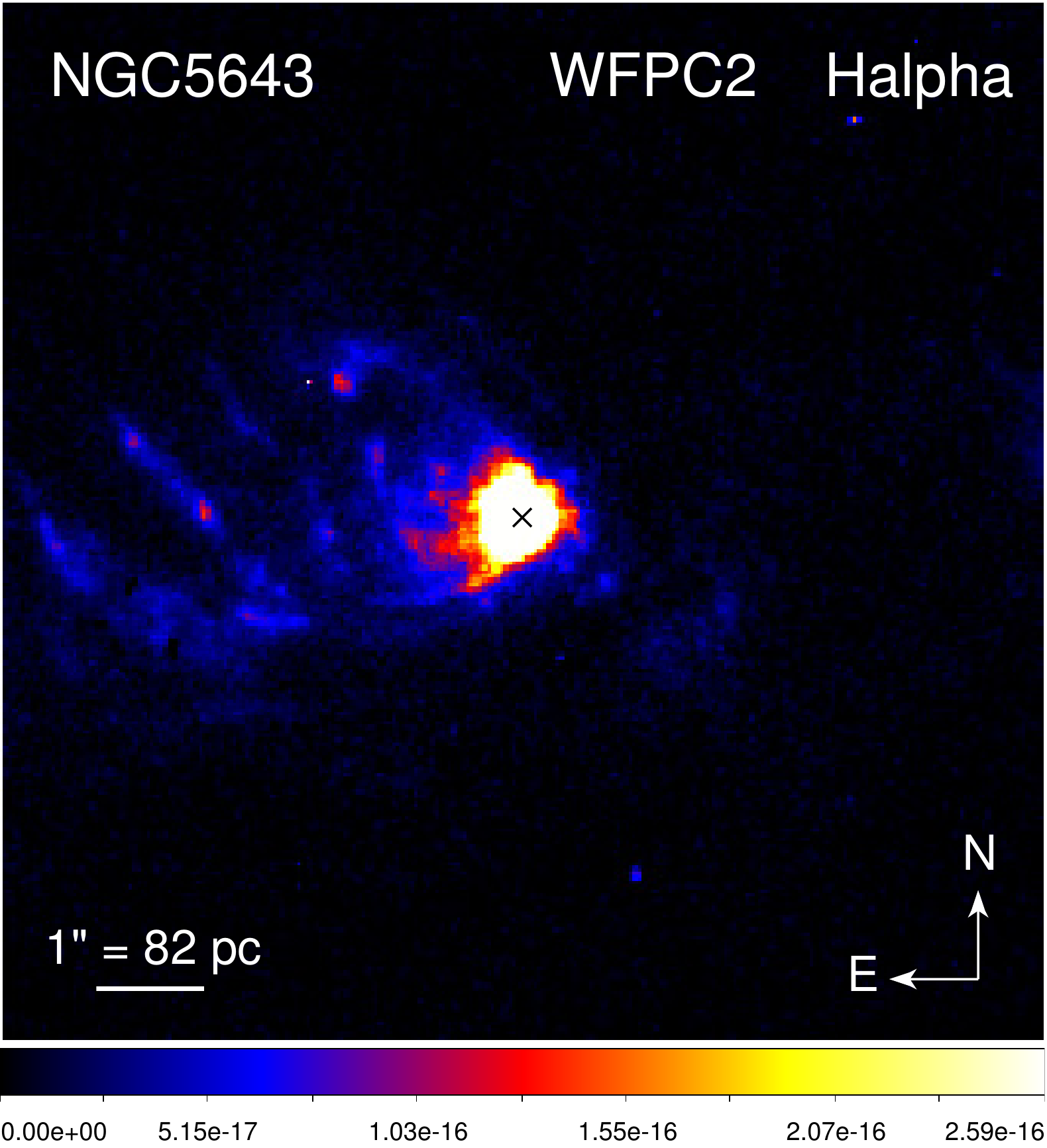}
\includegraphics[width=8cm]{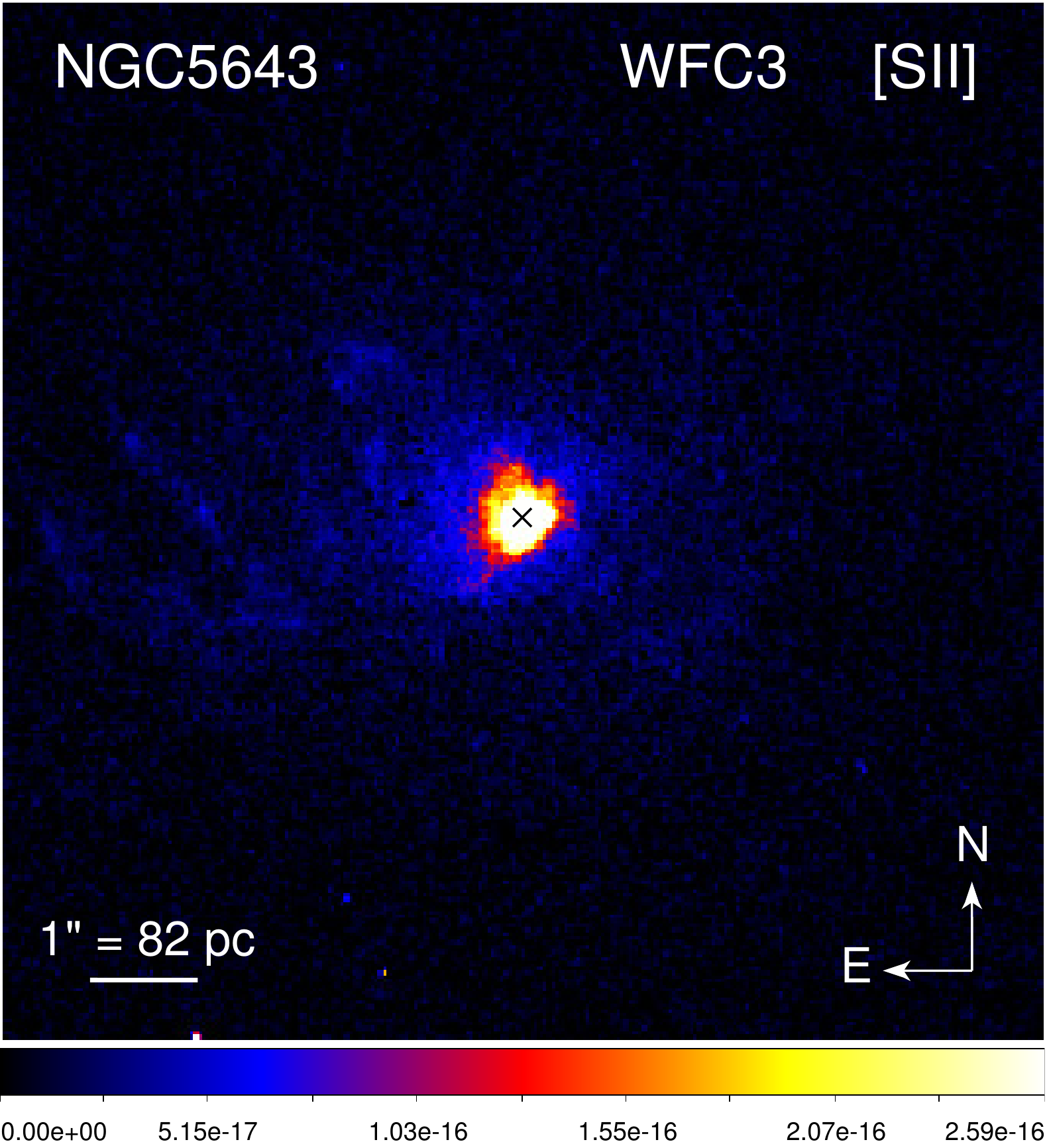}
\includegraphics[width=8cm]{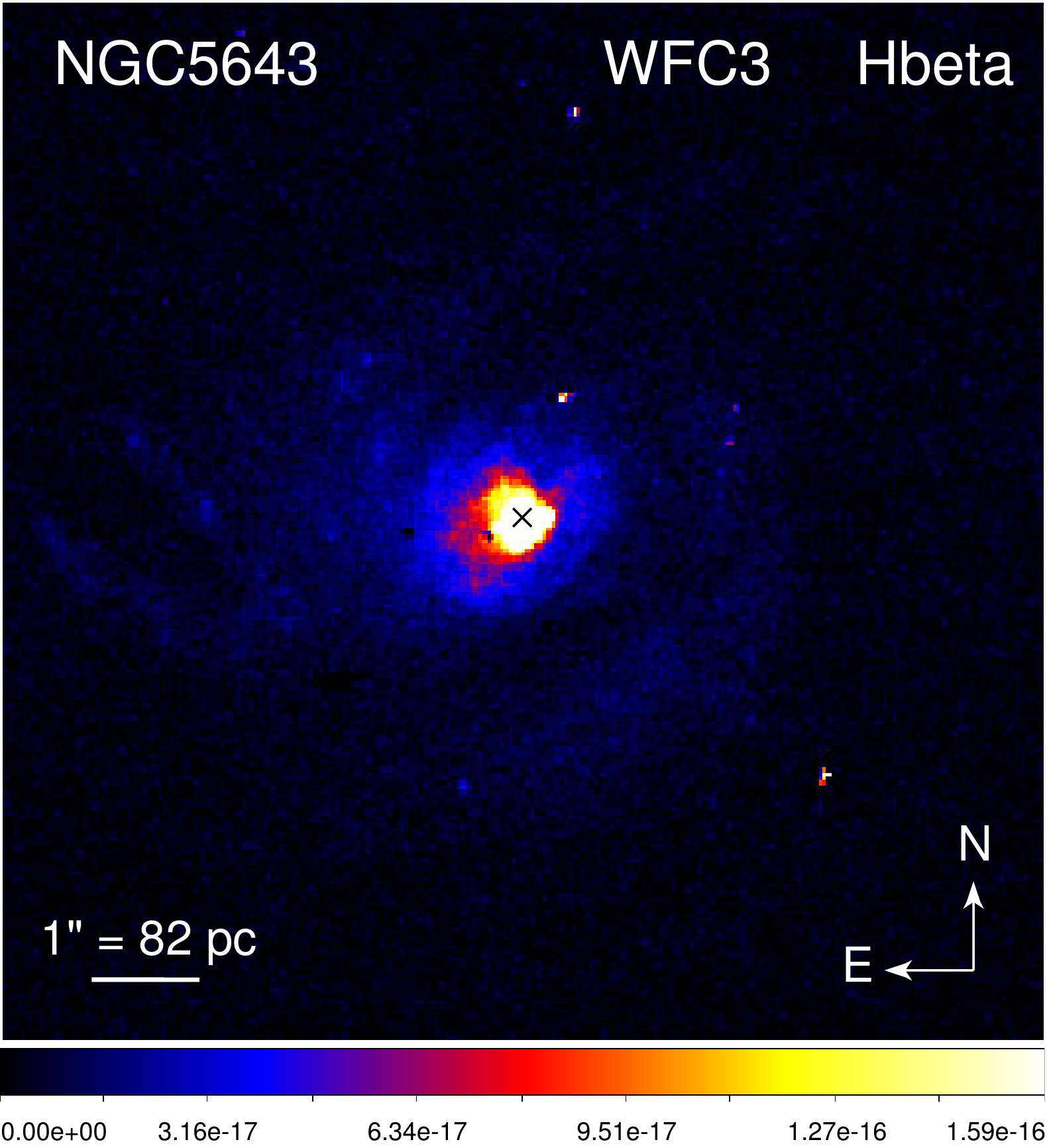}
\caption{NGC 5643 9.6$\arcsec$$\times$9.6$\arcsec$ continuum-subtracted emission line maps centered on the nucleus (black cross). All the images have a pixel scale of 0.04$\arcsec$. The colorbar denotes the surface brightness in the units of erg cm$^{-2}$ s$^{-1}$ pixel$^{-1}$. }
\label{NGC5643_linemap}
\end{figure*}

NGC 5643 ($z$ = 0.00399, $D$ = 17.1 Mpc, 1$\arcsec$ = 82 pc) has long been known to exhibit a one-sided ionization cone with filamentary structures \citep{Simpson1997}. Figure \ref{NGC5643_linemap} shows the continuum-subtracted line maps in the inner $r <$ 5$\arcsec$ (410 pc) region from the nucleus. Both the [O\III] and H$\alpha$ maps display well-resolved filamentary structures to the east of the nucleus, while barely any emission is detected west of the nucleus. The filaments are also present in the [S\II] and H$\beta$ maps but are much fainter. The H$\beta$ image scale is stretched such that a dark region (nuclear dust structure; \citealt{Cresci2015}) west of the nucleus in the shape of a crescent moon can be recognized. Faint emission appears southwest of the dark region. \cite{Cresci2015} (see also \citealt{Mingozzi2019}) reveal a more complete, double-sided ionization structure using deep VLT/MUSE integral field observations, although the spatial resolution is lower (0.88$\arcsec$) than our {\it HST} observations. 

\begin{figure*} 
\centering
\includegraphics[width=9.5cm]{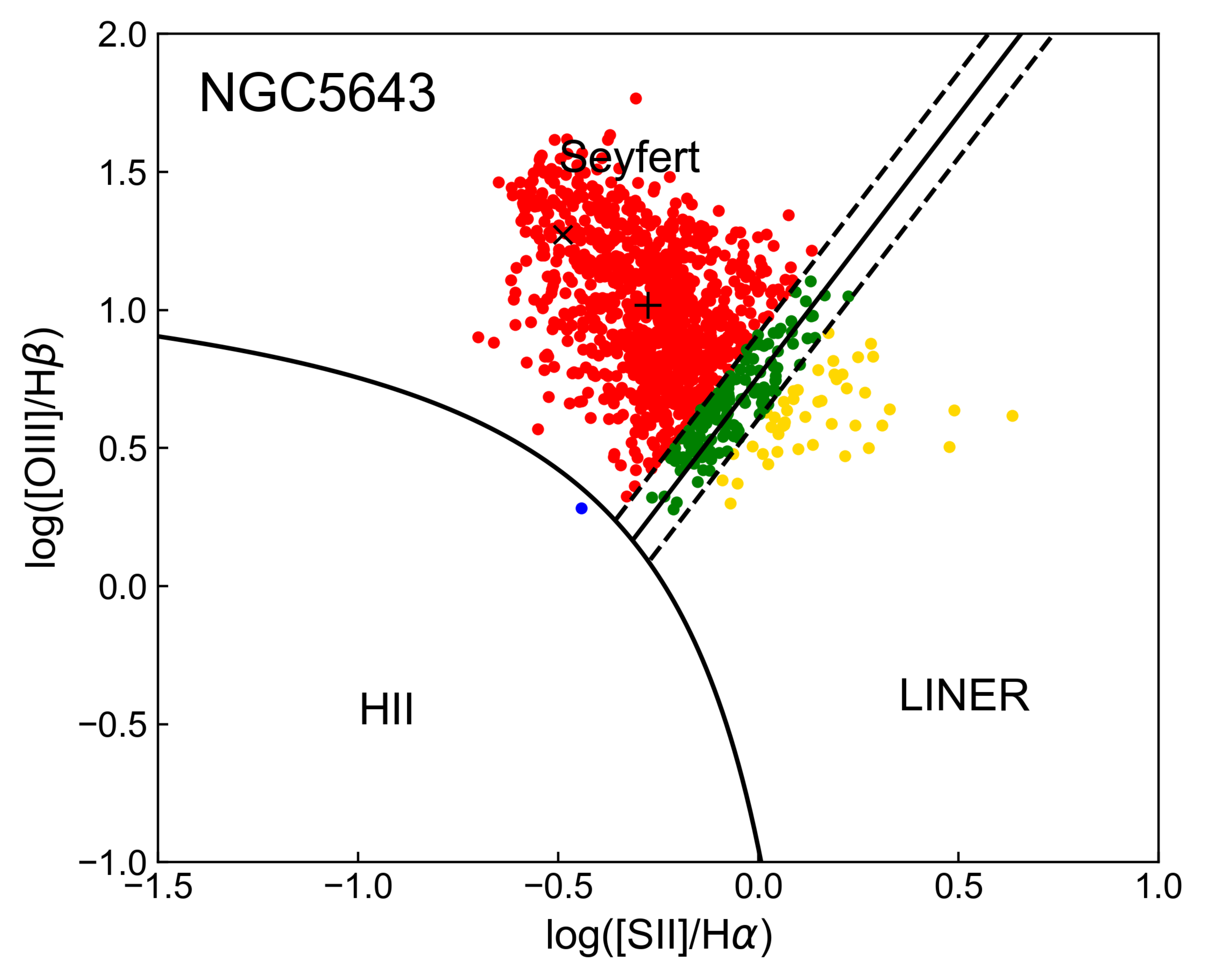} 
\includegraphics[width=7.7cm]{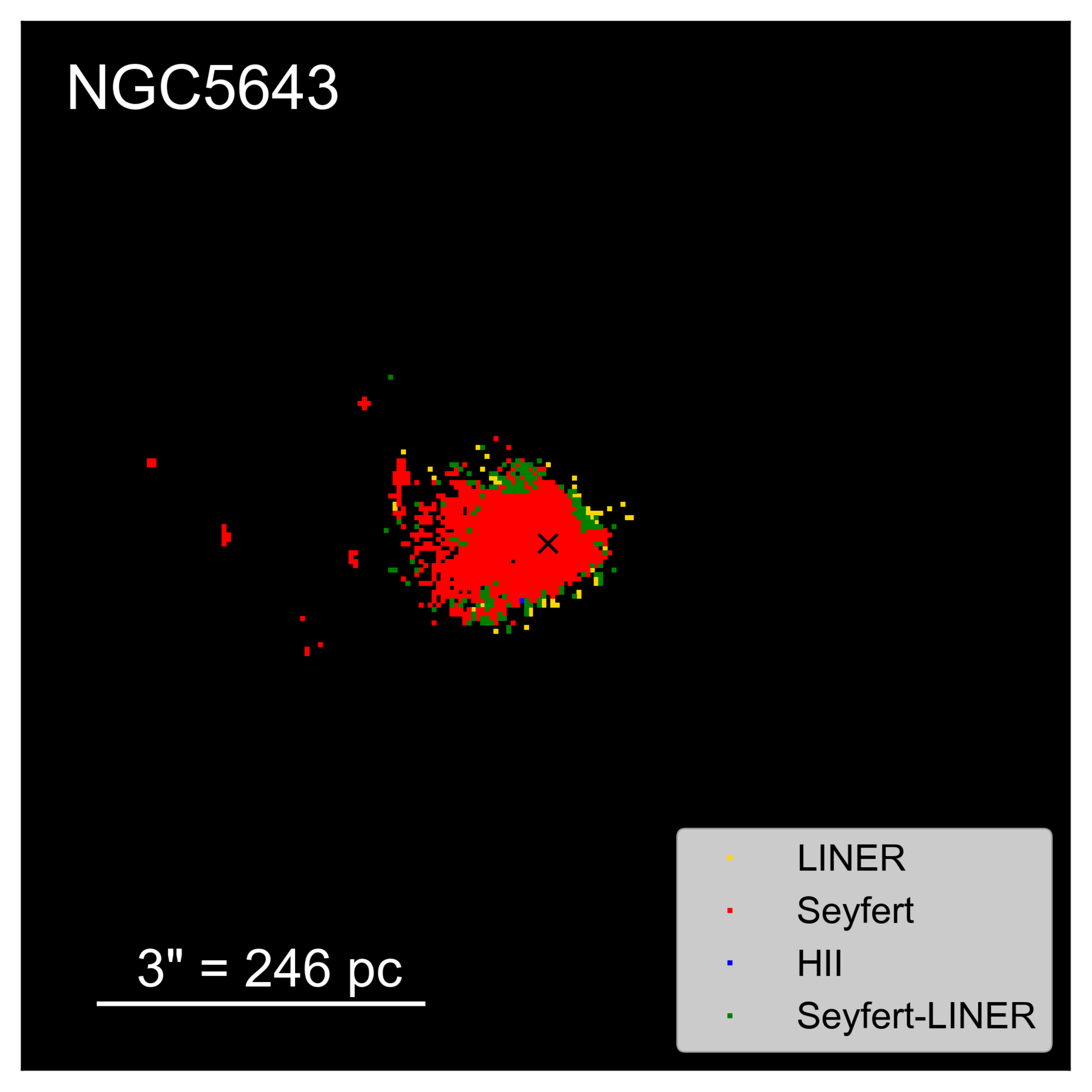}
\caption{{\bf Left}: BPT diagram of NGC 5643. The dividing lines/curves between different excitation mechanisms are defined in \cite{Kewley2006}. Red corresponds to Seyfert-like activity, yellow denotes LINER-like activity, and the transition zone is coded as green. Blue represents line ratios typical of H\II{} regions. Only pixels detected above 3$\sigma$ in all lines are used. Most H\II{} pixels are excluded by this criterion. The black cross marks the line flux ratios measured for the nuclear region ($r \leq 0.35\arcsec$). {\bf Right}: Spatially resolved BPT map (9.6$\arcsec$ $\times$ 9.6$\arcsec$) with each pixel color-coded according to the BPT type as shown in the left panel. The black cross marks the nucleus. Black pixels have at least one line with $F_{\rm line}$ $<$ 3$\sigma$. The 3$\arcsec$ white bar represents the diameter of the SDSS 3$\arcsec$ fiber. We calculate the BPT line ratios ('+' in the left panel) using the integrated line fluxes within a 3$\arcsec$ diameter circular aperture centered on the nucleus to mimic SDSS observations.  }
\label{NGC5643_BPT}
\end{figure*}

Figure \ref{NGC5643_BPT} shows our {\it HST} BPT diagram and color-coded map. Most pixels that meet the 3$\sigma$ criterion are located in the nuclear region, and are dominated by Seyfert activity. The relatively bright spots in the filamentary structures east of the nucleus also show red Seyfert-type excitation. The central region tends to transition from Seyfert to LINER, but there are no more pixels to reveal a complete LINER cocoon structure. \cite{Cresci2015} also present BPT maps based on their VLT/MUSE data. They not only observed the four lines required in the S-BPT diagram, but also obtained [N\II] and [O\I] to make the N-BPT and O-BPT maps. The MUSE data are deep enough and have a large enough field of view to allow them to map the entire ionization bicone, which shows Seyfert activity. The ionization bicone is surrounded by LINER regions, which then transition farther out into the star-forming regions in the host galaxy.   

NGC 5643 possesses a diffuse, low-luminosity radio jet on both sides of the nucleus, extending $\sim$30$\arcsec$ long \citep{Leipski2006}. A central radio component overlaps with the optical nuclear emission shown in Figure \ref{NGC5643_linemap}, and the whole radio structure matches well with the full ionization cones shown in \cite{Cresci2015}. They also confirm the presence of outflowing gas from the nuclear region based on the high velocity ionized gas kinematics (more discussion in Section \ref{discussion}).

\section{Discussion}
\label{discussion}

Based on the BPT mapping results, a coherent and consistent picture is emerging that while the nucleus and ionization cones are dominated by Seyfert-type emission, they transition to and are surrounded by a LINER cocoon extending up to $\sim$250 pc in thickness. This is the first AGN sample to clearly resolve LINER regions and where Seyfert excitation ends and LINER begins within their host galaxies. This common LINER cocoon within traditionally classified Seyfert galaxies is a new feature that needs an explanation. We have also identified a loop feature in two galaxies (NGC 3081 and NGC 7674), the formation of which is to be investigated. 

Since dust features are quite common and cannot be neglected in these Seyfert 2 galaxies, in this section we will first discuss dust extinction effects on the NLR/ENLR morphology, the detectability of the BPT lines, as well as the line ratios and BPT classification. We will then discuss possible origins of the Seyfert ionization cones, LINER cocoon features, and the loops.

\subsection{Dust extinction effects}
\subsubsection{BPT line ratios insensitive to dust extinction}

The line ratios in our BPT diagrams are based on the observed line fluxes in each pixel. In the definitional BPT paper by \cite{Kewley2006}, they corrected the emission line fluxes for dust extinction using the Balmer decrement and assumed a reddening curve. They used an intrinsic H$\alpha$/H$\beta$ ratio of 2.85 for galaxies dominated by star formation and H$\alpha$/H$\beta$ = 3.1 for galaxies dominated by AGN. However, we are dissecting regions within galaxies rather than classifying galaxies and correcting for dust extinction globally. We cannot pre-assume the excitation mechanism of a local pixel to be intrinsically star formation driven or AGN driven. Although the measured line fluxes are subject to dust extinction correction, the pairs of lines in the BPT line ratios, e.g., [O\III]$\lambda$5007 and H$\beta$ or [S\II] and H$\alpha$, are by design close in wavelength thus the line ratios are largely immune to dust extinction correction in terms of BPT classification \citep{Baldwin1981}. A small fraction of pixels close to the boundary curves/lines between H\II{}, LINER, and Seyfert regions on the BPT diagram may be sensitive to small changes in the line ratios if corrected for dust extinction, but this would not change the Seyfert and LINER overall structures seen in our resolved maps. 

\subsubsection{Dust distribution affecting NLR/ENLR morphology and detectability}
 
It has been known that Seyfert 2 galaxies are significantly more likely to show nuclear dust absorption, for example in the form of dust lanes and patches close to the nucleus, compared to Seyfert 1 galaxies \citep{Malkan1998}. We have seen multiple strong dust lanes in this sample of Seyfert 2 galaxies, e.g., in NGC 1386, NGC 7212, and NGC 5643. Some Seyfert regions exhibit clear bicones while others look highly asymmetric and conical, where dust distribution may play a key role and obscures one side of the ionization cones.

Dust attenuation can significantly impact the inferred spatial distributions of the emission associated with each of the ionization mechanisms \citep{Davies2017}. Some or all of the emission lines may not pass the detection threshold due to strong dust attenuation, which prevents the underlying emission associated with star formation, LINER, or AGN activity from being recovered. Deep observations would help alleviate the problem. For example in the case of NGC 5643, what appears to be a one-sided ionization cone in the {\it HST} narrow band images turns out to be a double-sided ionization structure revealed by the MUSE IFU \citep{Cresci2015}. The emission lines all meet the detection criterion and the excitation mechanisms can be derived even in the regions with prominent nuclear dust structures.

\subsection{The origins of Seyfert ionization cones and LINER cocoon}

Since the ionization cone interiors of our sources are almost entirely Seyfert-type emission, one can simply draw the conclusion that the origin is predominantly due to AGN photoionization. However, fast shocks ($v >$ 500 km s$^{-1}$), which have a photoionizing precursor, can also produce the Seyfert AGN photoionization-type line ratios on the BPT diagram \citep{Allen2008,Kewley2019}. For example, the S-shaped ionization structure in NGC 3393 is thought to be predominantly by AGN photoionization, but the S-shape could also be formed by fast shocks with photoionized precursors \citep{Maksym2016,Maksym2017}. 

LINERs are more complicated in terms of excitation mechanism or power source (e.g., \citealt{Ho2008,Yan2012}). Some LINER emission is due to gas ionized by low-luminosity AGN (e.g., \citealt{Ho1996,Constantin2006}) or regions exposed to absorbed or filtered nuclear radiation (e.g., \citealt{Kraemer2008,Heckman2014}). Further evidence of AGN in LINERs is supported by radio emission. More than half of nuclear LINERs have pc-scale radio nuclei and sub-pc radio jets \citep{Heckman1980,Nagar2005}. In some other cases, shocks by jets or outflows are required to power the LINER emission. Extended LINER emission (i.e., $>$ 1 kpc) can be produced by shocks resulting from galactic-scale outflows (e.g., \citealt{Dopita1995,Ho2014,Ho2016}). Slow shocks, which are unable to drive a precursor, can produce LINER-type line ratios on the BPT diagram \citep{Rich2011,Sutherland2017,Kewley2019}. 

The green Seyfert-LINER transition zones at the edge of the ionization cones shown in our resolved maps in Section \ref{results} are likely due to absorbed or filtered AGN ionization. Almost all of the Seyfert 2 galaxies in our sample host radio sources that overlap with the NLR structures, indicating interactions between the optical and radio components. The optical or radio emission could be alternatively explained as the interaction between an AGN disk wind and the surrounding ISM (e.g., \citealt{Venturi2017,Panessa2019}). \cite{Fischer2013} studied the NLR kinematics of nearby AGN, including the 7 Seyfert 2 galaxies in this work, using {\it HST}/STIS optical spectroscopy. They find that Mrk 573 (see also \citealt{Revalski2018}), NGC 5643 (see also \citealt{Cresci2015}), and NGC 7674 exhibit clear outflow kinematics, while NGC 1386, NGC 2273, NGC 3081, and NGC 7212 show either complex or ambiguous kinematics. The LINER regions in the sources associated with clear radio jets or outflow signatures are most likely due to jet/outflow induced shocks. Using long-slit spectroscopy, \cite{Storchi1996a,Storchi1996b} find a correlation between shock-sensitive line ratios [N\II]/H$\alpha$, [S\II]/H$\alpha$ and the FWHM line width of [N\II] in a sample of nearby AGN including NGC 3081, which indicates shocks contributing to the ionization of the gas in the circumnuclear region, covering the LINER cocoon region that we have identified. 

In different regions of the same galaxy, the LINER emission can be attributed to totally different mechanisms. For example, NGC 2273 is the only galaxy in this sample that displays two distinct groups on the BPT diagram. The relatively high [O\III]/H$\beta$ ratio LINER group is associated with the central optical emission and radio jet, and therefore the excitation mechanism is likely due to the jet-ISM interaction. The flat, relatively low [O\III]/H$\beta$ ratio LINER group is located farther from the nucleus, associated with the ring structure, and not associated with radio jets or outflows, which could be explained as photoionization by absorbed or filtered AGN radiation or regions with high gas densities relative to the AGN ionizing flux. Since NGC 2273 has been found to host nuclear starburst activity, the starburst could also power the LINER emission.

However, one cannot separate excitation sources just based on optical diagnostics when shock excitation is involved \citep{Kewley2019}. In order to verify the regions we suspected to be shock excited, we would need high resolution IFU spectroscopy and the following criteria must be satisfied as summarized by \cite{Kewley2019}: 1. The velocity dispersion distribution is bimodal with the broad component being greater than 80 km s$^{-1}$; 2. The velocity dispersions correlate with shock-sensitive emission line ratios such as [N\II]/H$\alpha$, [S\II]/H$\alpha$, or [O\I]/H$\alpha$; 3. The line ratios are consistent with the predictions from shock models. A 3D diagnostic diagram that combines radius as the x-axis, velocity dispersion as the y-axis, and line ratio as the z-axis is able to separate all three power sources, i.e., star formation, AGN, and shocks \citep{DAgostino2019}. 

Currently, such 3D diagnostics are not available for our sources. {\it Chandra} X-ray observations, in addition to radio observations, can provide complementary and critical direct measurements of AGN strength and spectroscopic diagnostics of shocks and photoionization, which can be directly compared to narrow line images (e.g., \citealt{Wang2009,Paggi2012,Maksym2017,Jones2020}). For example in Mrk 573, \cite{Paggi2012} also find evidence for shocks at the jet-ISM interface based on {\it Chandra} X-ray data.
A detailed comparison of optical, radio, and X-ray emission of the resolved structures in individual sources is beyond the scope of this paper, and will be presented in future publications. 

\subsection{The origin of the loop features}

With {\it HST}'s superior spatial resolution, we have identified a loop feature in at least two sources, NGC 3081 (Figure \ref{NGC3081_linemap}) and NGC 7674 (Figure \ref{NGC7674_linemap}). For NGC 3081, the loop feature is bright in [O\III] but is better resolved into small emission line knots in H$\alpha$. It also appears in [S\II] and H$\beta$ although less prominently.  The loop feature in NGC 7674 is prominent in [O\III] and also appears in H$\alpha$, although the resolution of the H$\alpha$ image is lower and the loop is not well resolved. Both loops align more along the perpendicular direction of the ionization bicone than the bicone direction. The BPT classifications of these two loops are either Seyfert (NGC 3081) or a mixed of Seyfert, Seyfert-LINER transition, and LINER types (NGC 7674). The simplest explanation for the Seyfert dominated regions is AGN photoionization. However, as discussed above the origin of LINER emission is complicated and other mechanisms must play a role in shaping this special loop structure.

A very similar loop feature was identified in Seyfert galaxy IC 5063 with {\it HST} in the [S\II] and [N\II] line maps perpendicular to the main emission line structure (Maksym et al. in prep). This loop could result from a plume of hot gas ablated by jet-ISM interactions from the galactic plane before escaping laterally. The loops in NGC 3081 and NGC 7674 share the same trend of arising more perpendicular to than along the ionization bicone axis. The loop in NGC 3081 also happens to coincide with the main bipolar outflow described by \cite{Schnorr2016} in their GMOS IFU study. \cite{Fischer2013} also show a bipolar outflow in NGC 7674 along the N-S direction, possibly intersecting the loop region. So there is probably some outflow-disk gas interaction that is exciting the loops in NGC 3081 and NGC 7674, although the interaction may not be as direct as for IC 5063. We would need additional diagnostics including kinematics in and around the loop region to investigate whether the loops in NGC 3081 and NGC 7674 are due to outflowing gas from jet-ISM interactions or AGN winds. Although less likely given the scales, we cannot rule out the possibility that these loop features are the consequences of post-merger events or tidal forces (e.g., \citealt{Ji2014,Keel2015}).

\subsection{Galaxy survey classifications}

Traditional optical spectroscopic observations, by which these galaxies were classified, place narrow ($\sim$ 1$\arcsec$) slits on their nuclei, and their nuclear spectra are classified as Seyfert 2 type. Long-slit spectra aligned with the bicone axes will tend to systematically miss the LINER regions surrounding the bicones. Given the common LINER emission in these traditionally classified Seyfert 2 galaxies, we further investigate whether this would affect BPT classification in galaxy surveys. Since we use the \cite{Kewley2006} BPT classification scheme, especially the Seyfert-LINER separation calibrated and validated on emission line galaxies selected from the Sloan Digital Sky Survey (SDSS), here we conduct a simple experiment to mimic the SDSS's 3$\arcsec$ diameter fiber and calculate the BPT line ratios ('+') using the integrated line fluxes measured within this 3$\arcsec$ diameter circular aperture centered on the nucleus. 

The SDSS BPT classifications based on this experiment would be: NGC 5643, NGC 7212, and NGC 7674 remain Seyfert type; Mrk 573, NGC 1386, and NGC 3081 are in the Seyfert-LINER transition zone; Only NGC 2273 falls in the LINER region of the BPT diagram.

This simple exercise demonstrates that when we talk about BPT galaxy classifications, it is important to specify the spatial scales (and resolution) associated with the measured BPT line ratios. Only spatially resolved observations like ours or with high resolution IFUs can disentangle the emission types in different regions of the same galaxy, which are not necessarily in agreement with survey classifications. Admittedly, this simple exercise may not represent real SDSS spectroscopic observations, which requires a more detailed simulation that is beyond the scope of this paper.

\section{Summary and Conclusions}
\label{conclusions}

We have used {\it HST} narrow band imaging of optical emission lines in seven nearby Seyfert 2 galaxies to construct spatially resolved BPT diagrams and line maps, in order to diagnose different excitation mechanisms in the same galaxy. 

The BPT mapping results of eight Seyfert 2 galaxies (including NGC 3393) have informed a coherent and consistent picture that the nucleus and ionization cones are almost entirely Seyfert-type emission, which can be interpreted as predominantly photoionization by the AGN. The Seyfert nucleus and ionization cones transition to and are surrounded by a LINER cocoon. We resolved where Seyfert excitation ends and LINER begins; the intermediate Seyfert-LINER transition zone is resolved as a thin layer closely wrapping around the ionization cones. We detected LINER cocoons extending up to $\sim$250 pc in thickness. Deeper observations are expected to reveal the full extent of the LINER cocoon. We discussed possible origins of the LINER regions. The LINER regions associated with radio jets or AGN outflows are likely due to shock excitation induced by the jet-ISM interaction or the AGN wind pushing into the surrounding ISM. Some LINER regions could be photoionized by absorbed or filtered AGN radiation.

The star-forming regions bounding the LINER regions and permeating in the outskirt of NGC 5643 \citep{Cresci2015} are not seen in our {\it HST} resolved maps, simply because most H\II{} pixels do not meet the 3$\sigma$ criterion in all lines and therefore are not used in our BPT mapping. We do see some H\II{} clumps in the circumnuclear regions of NGC 2273, NGC 3081, and NGC 7212, indicating that the local conditions permit star formation activity. 

The ubiquity of the LINER ``cocoons" suggests that although they are classified as Seyfert 2 galaxies, the circumnuclear regions are not necessarily Seyfert type, and LINER regions play an important role in Seyfert 2 galaxies. We have demonstrated that these traditionally classified Seyfert 2 galaxies may not necessarily be classified as Seyfert type in galaxy surveys like the SDSS in a 3$\arcsec$ fiber. We have also identified in NGC 2273 two groups of LINERs on the BPT diagram, associated with different structures of the emission line map. Spatially resolved observations can disentangle excitation types in different regions of the same galaxy, and corresponding spatially resolved diagnostics are required to understand the excitation mechanisms in different regions. 

Since circumnuclear dust features are quite common in these Seyfert 2 galaxies, we discussed dust extinction effects on our BPT mapping results. Our spatially resolved BPT maps were constructed based on the observed line fluxes without dust extinction correction due to the difficulty and uncertainty of properly correcting for dust extinction in each local pixel. Nevertheless, our conclusions would not change given that the BPT lines in each pair are chosen to be largely immune to dust extinction. The dust distribution does strongly shape the NLR/ENLR morphology and affects the detectability of emission lines. Some or all of the emission lines may not pass the detection threshold due to strong dust attenuation. 

We have also discovered line-emitting loop features in the NLRs of two galaxies, NGC 3081 and NGC 7674. The BPT classifications of the loops are either Seyfert or a mix of Seyfert and LINER types. The formation mechanisms of these loops need further investigation.

To fully understand the excitation mechanisms region by region, especially when shocks are involved, we would need kinematic information with matched spatial resolution, i.e., high resolution IFU spectroscopy. A combined analysis of multi-wavelength data, e.g., optical, radio and X-ray with matched resolution, will provide unambiguous diagnostics of various power sources and help understand the AGN-host galaxy interactions, i.e., feedback processes.

\section*{acknowledgments}

We thank the anonymous referee for the constructive comments that helped improve the manuscript. Support for this work was provided by NASA through {\it HST} grant GO-15350 from the Space Telescope Science Institute. WPM acknowledges support by {\it Chandra} grants GO8-19096X, GO5-16101X, GO7-18112X, GO8-19099X. JW acknowledges support from the National Key R\&D Program of China (2016YFA0400702) and NSFC grant U1831205. This work is based on observations made with the NASA/ESA Hubble Space Telescope, obtained from the data archive at the Space Telescope Science Institute. STScI is operated by the Association of Universities for Research in Astronomy, Inc. under NASA contract NAS 5-26555. This work makes use of the NASA-IPAC Extragalactic Database (NED).

\end{document}